\documentclass[%
reprint,
 amsmath,amssymb,
 aps,
]{revtex4-2}

\usepackage{graphicx}% Include figure files
\usepackage{dcolumn}% Align table columns on decimal point
\usepackage{bm}% bold math
\usepackage{multirow}% bold math
\usepackage{color}
\usepackage{lipsum}
\usepackage{longtable}
\usepackage{dcolumn}   % Align table columns on decimal point
\usepackage{bm} % bold math
\usepackage{amsfonts}  % Common math fonts
\usepackage{amsmath} % Common math functions
\usepackage{amssymb} % Common math symbols
\usepackage{graphicx}% Include figure files
\usepackage{multirow}
\usepackage{tabularx}
\usepackage{lipsum}
\usepackage{float}
\usepackage{blindtext}
\begin{document}

\preprint{Synth of SHE/Feb 2021}
\title{Exploring an experimental route of synthesizing superheavy elements beyond Z $>$ 118 }
%title{Synthesis of heavy and superheavy elements in perspectives of experiments}
%\title{INFLUENCE OF ANGULAR MOMENTUM AND NUCLEAR DEFORMATION IN SYNTHESIS OF SUPERHEAVY ELEMENTS: PRESENT AND FUTURE EXPERIMENTAL SCENARIO}
\author{H.C. Manjunatha$^{1}$, Y.S.Vidya$^{2}$, P.S.Damodara Gupta$^{1}$,  N.Manjunatha$^{1}$, N.Sowmya$^{1}$, L.Seenappa$^{1}$ and T. Nandi$^{3\ddagger}$}
\affiliation {\it $^1$Department of Physics, Government College for women, Kolar-563101, Karnataka, India}
\affiliation {\it $^2$Department of Physics, LBS Govt. First Grade College, RT Nagar, Bangalore–560032, India}%LBSal Bahadur Shastri
\affiliation{$^{1}$1003 Regal, Mapsko Royal Ville, Sector 82, Gurugram-122004, Haryana, India.}
\thanks {{Superannuated from Inter-University Accelerator Centre, Aruna Asaf Ali Marg, New Delhi-110067, India.}\\ $^\dagger$nanditapan@gmail.com }%and  $^\ddagger$manjunathhc@rediffmail.com}
%\thanks{For correspondences: $^\ddagger$nanditapan@gmail.com}  
\date{\today}
\begin{abstract}
Role of the Coulomb interaction, mean fissility, mass asymmetry, and charge asymmetry parameters on the synthesis of heavy and superheavy elements has been examined with respect to the deformation parameters of the projectile and target nuclei explicitly in light of the experimental results. The observed facts are classified into four categories and are then used to study several unsuccessful as well as planned reactions to synthesize the new superheavy elements $Z= 119, 120$. Concrete inference is too difficult to draw from these results because of excessive deviations in evaporation residue cross-section data. It is found that the arbitrary choice of excitation energy for the experiments studied was the root cause of such large deviations. Such a complex issue can be resolved well by theoretical excitation function studies using the advanced statistical model or the dinuclear system model and choosing the excitation energy corresponding to the energy where the excitation function curve shows the maximum. We believe this method may help us to predict whether the estimated evaporation residue cross section can be measurable within the experimental limit of the existing facilities for the future reactions planned.
\end{abstract}
%%%%%%%%%%%%%%%%%%%%%%%%%%%%%%%%%%%%%%%%%%%%%%%%
\maketitle
\begin{figure}
    \centering
    \includegraphics[width=\linewidth]{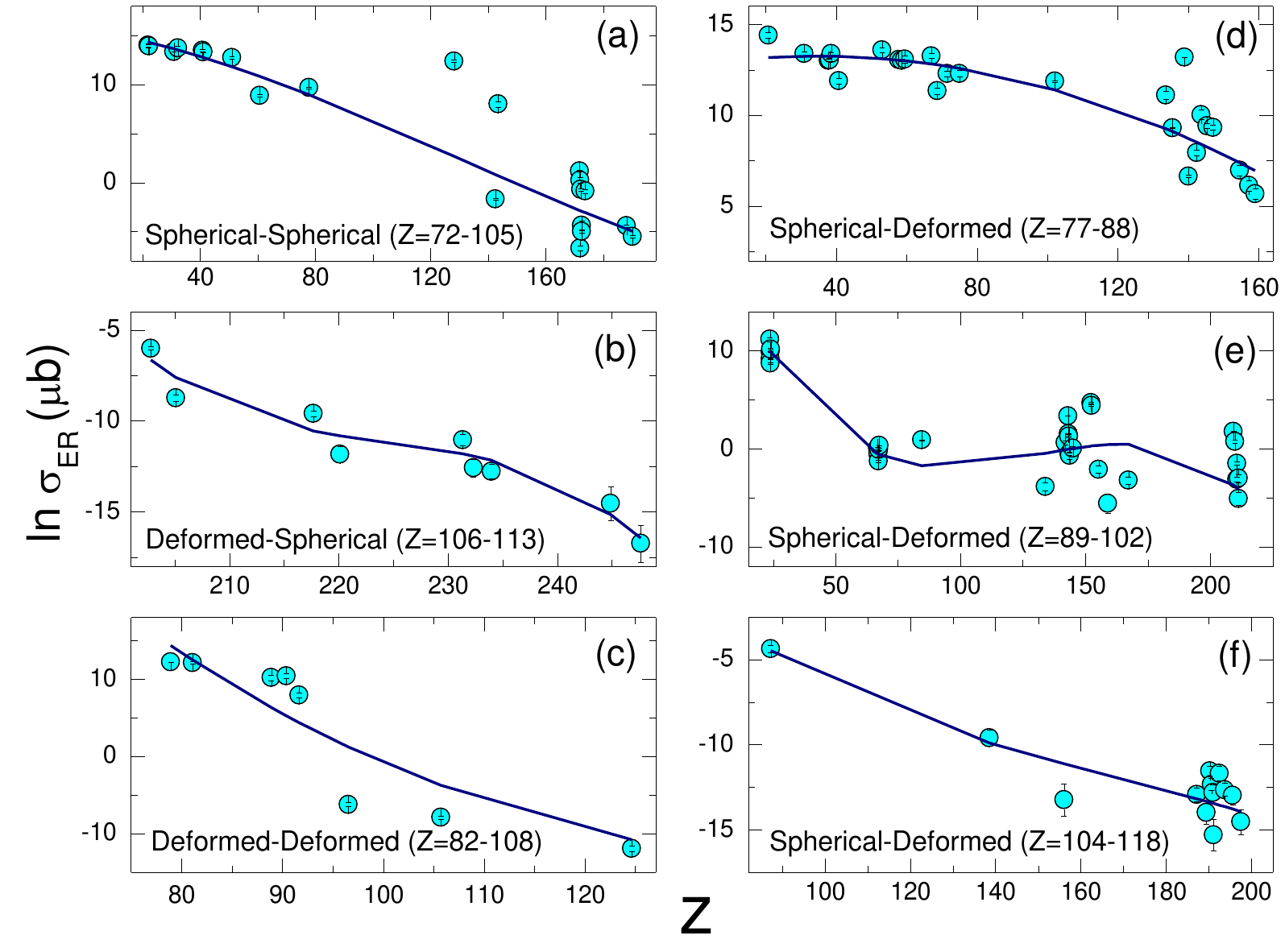}
    \caption{Variation of measured production cross-section as a function of Coulomb interaction parameter for different projectile-target combinations: (a) both the projectile and target are spherical \cite{lukyanov2009study,ramler1959excitation,vinodkumar2009fusion,penionzhkevich2008complete,dasgupta2010suppression,signorini2004subbarrier,gasques2009suppression,LEBEYEC1972405,PhysRevC.52.243,PhysRevC.64.054606,khuyagbaatar2010new,gaggeler1989cold,PhysRevC.64.054606,PhysRevC.64.054606,PhysRevC.64.054606,belozerov2003spontaneous,oganessian2001measurements,gaggeler1989cold,hessberger2001decay,hessberger2001decay}, (b) deformed projectile and spherical target \cite{andreev1989measurement,hofmann2004properties,munzenberg1981identification,hofmann2004properties,munzenberg1984evidence,morita2004status,hofmann1995production,hofmann1995new,hofmann1996new,morita2004experiment}, (c) both  theprojectile and target are deformed \cite{hinde1982fission,mahata2003fusion,andreyev1997statistical,andreyev1997decay,andreev1989measurement,haba2014production,haba2012production,dvorak2008observation} and (d-f) spherical projectile and deformed target in specified atomic number regions of the compound nuclei \cite{chakrabarty2000complete,kumar2013low,zhang2014complete,hinde1982fission,sharma2015systematic,scholz2014measurement,fang2013fusion,fang2015complete,penionzhkevich2007excitation,penionzhkevich2008complete,singh2009investigation,shrivastava1999shell,sikkeland1970study,mayorov2014production,baba1988evaporation,maiti2011production,mayorov2014production,vermeulen1984cross,mayorov2014production,mayorov2015evaporation,vermeulen1984cross,LEBEYEC1972405,corradi2005excitation,mayorov2014production,mayorov2015evaporation,vermeulen1984cross,mayorov2015evaporation,vermeulen1984cross,sahm1985fusion,sahm1985fusion,vermeulen1984cross,SAHM1985316,vermeulen1984cross,vermeulen1984cross,schmidt1981barrier,vermeulen1984cross,SAHM1985316,SAHM1985316,SAHM1985316,PhysRevC.62.054603,SAHM1985316,vermeulen1984cross,gaggeler1989cold,PhysRev.111.1358,PhysRevC.17.1706,PhysRevC.17.1706,PhysRev.111.1358,gaggeler1989cold,fleury1973excitation,gaggeler1989cold,PhysRevLett.93.162701,PhysRev.172.1232,PhysRev.172.1232,PhysRev.172.1232,PhysRev.172.1232,PhysRev.172.1232,nagame2002status,nishio2006measurement,nishio2010nuclear,oganessian2004measurements,oganessian2015super,oganessian2004measurements,oganessian2004experiments,oganessian2015super,oganessian2015super,oganessian2015super,oganessian2004measurements,oganessian2015super,oganessian2015super}.}
    \label{Spe-DefandDer-Spe}
\end{figure}
\begin{figure}
    \centering
    \includegraphics[width=\linewidth]{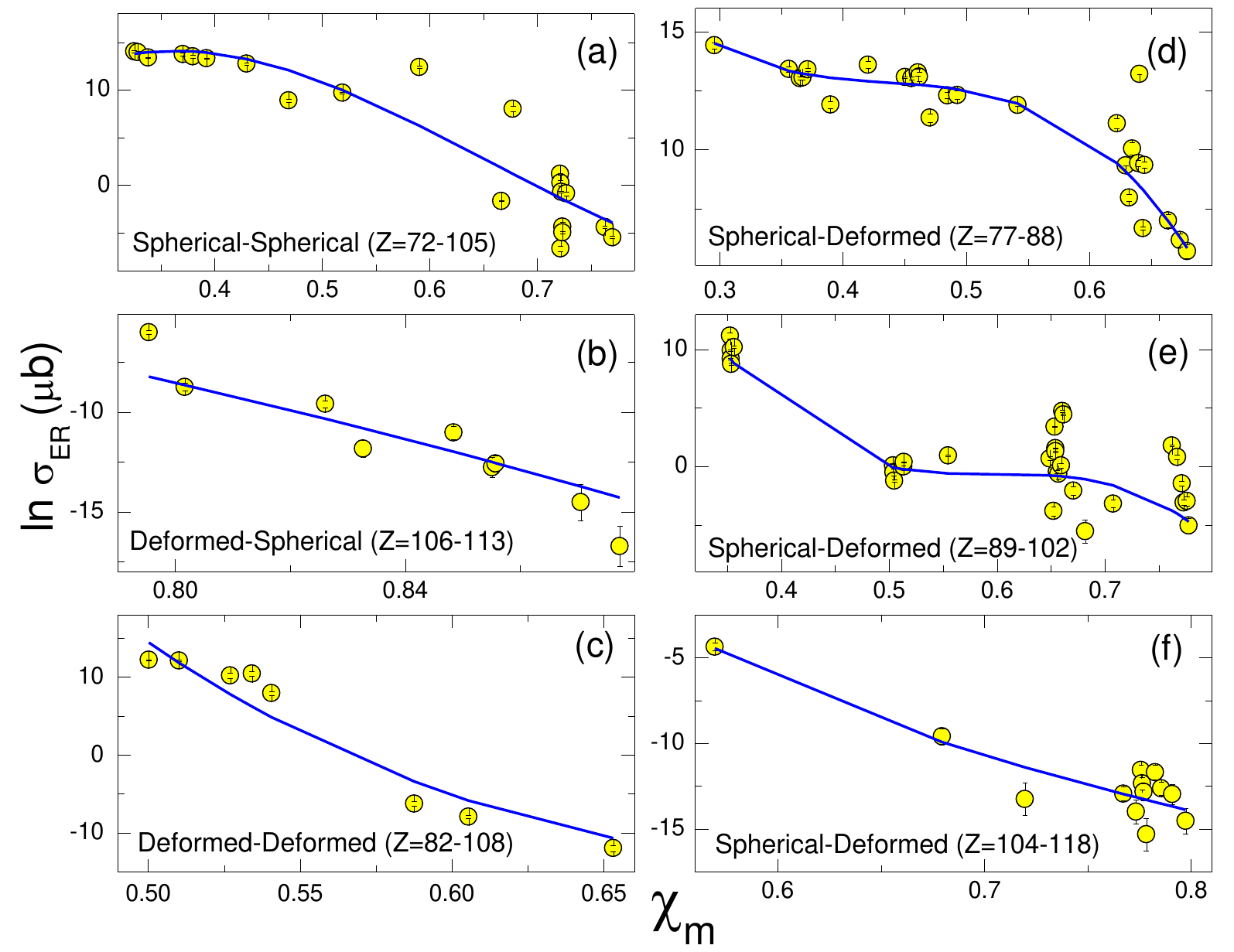}
    \caption{Variation of measured production cross-section as a function of mean fissility parameter for different projectile-target combinations as mentioned in Fig.\ref{Spe-DefandDer-Spe} \cite{lukyanov2009study,ramler1959excitation,vinodkumar2009fusion,penionzhkevich2008complete,dasgupta2010suppression,signorini2004subbarrier,gasques2009suppression,LEBEYEC1972405,PhysRevC.52.243,PhysRevC.64.054606,khuyagbaatar2010new,gaggeler1989cold,PhysRevC.64.054606,PhysRevC.64.054606,PhysRevC.64.054606,belozerov2003spontaneous,oganessian2001measurements,gaggeler1989cold,hessberger2001decay,hessberger2001decay,andreev1989measurement,hofmann2004properties,munzenberg1981identification,hofmann2004properties,munzenberg1984evidence,morita2004status,hofmann1995production,hofmann1995new,hofmann1996new,morita2004experiment,hinde1982fission,mahata2003fusion,andreyev1997statistical,andreyev1997decay,andreev1989measurement,haba2014production,haba2012production,dvorak2008observation,chakrabarty2000complete,kumar2013low,zhang2014complete,hinde1982fission,sharma2015systematic,scholz2014measurement,fang2013fusion,fang2015complete,penionzhkevich2007excitation,penionzhkevich2008complete,singh2009investigation,shrivastava1999shell,sikkeland1970study,mayorov2014production,baba1988evaporation,maiti2011production,mayorov2014production,vermeulen1984cross,mayorov2014production,mayorov2015evaporation,vermeulen1984cross,LEBEYEC1972405,corradi2005excitation,mayorov2014production,mayorov2015evaporation,vermeulen1984cross,mayorov2015evaporation,vermeulen1984cross,sahm1985fusion,sahm1985fusion,vermeulen1984cross,SAHM1985316,vermeulen1984cross,vermeulen1984cross,schmidt1981barrier,vermeulen1984cross,SAHM1985316,SAHM1985316,SAHM1985316,PhysRevC.62.054603,SAHM1985316,vermeulen1984cross,gaggeler1989cold,PhysRev.111.1358,PhysRevC.17.1706,PhysRevC.17.1706,PhysRev.111.1358,gaggeler1989cold,fleury1973excitation,gaggeler1989cold,PhysRevLett.93.162701,PhysRev.172.1232,PhysRev.172.1232,PhysRev.172.1232,PhysRev.172.1232,PhysRev.172.1232,nagame2002status,nishio2006measurement,nishio2010nuclear,oganessian2004measurements,oganessian2015super,oganessian2004measurements,oganessian2004experiments,oganessian2015super,oganessian2015super,oganessian2015super,oganessian2004measurements,oganessian2015super,oganessian2015super}.} 
    \label{meanfissility}
\end{figure}
\begin{figure}
    \centering
    \includegraphics[width=\linewidth]{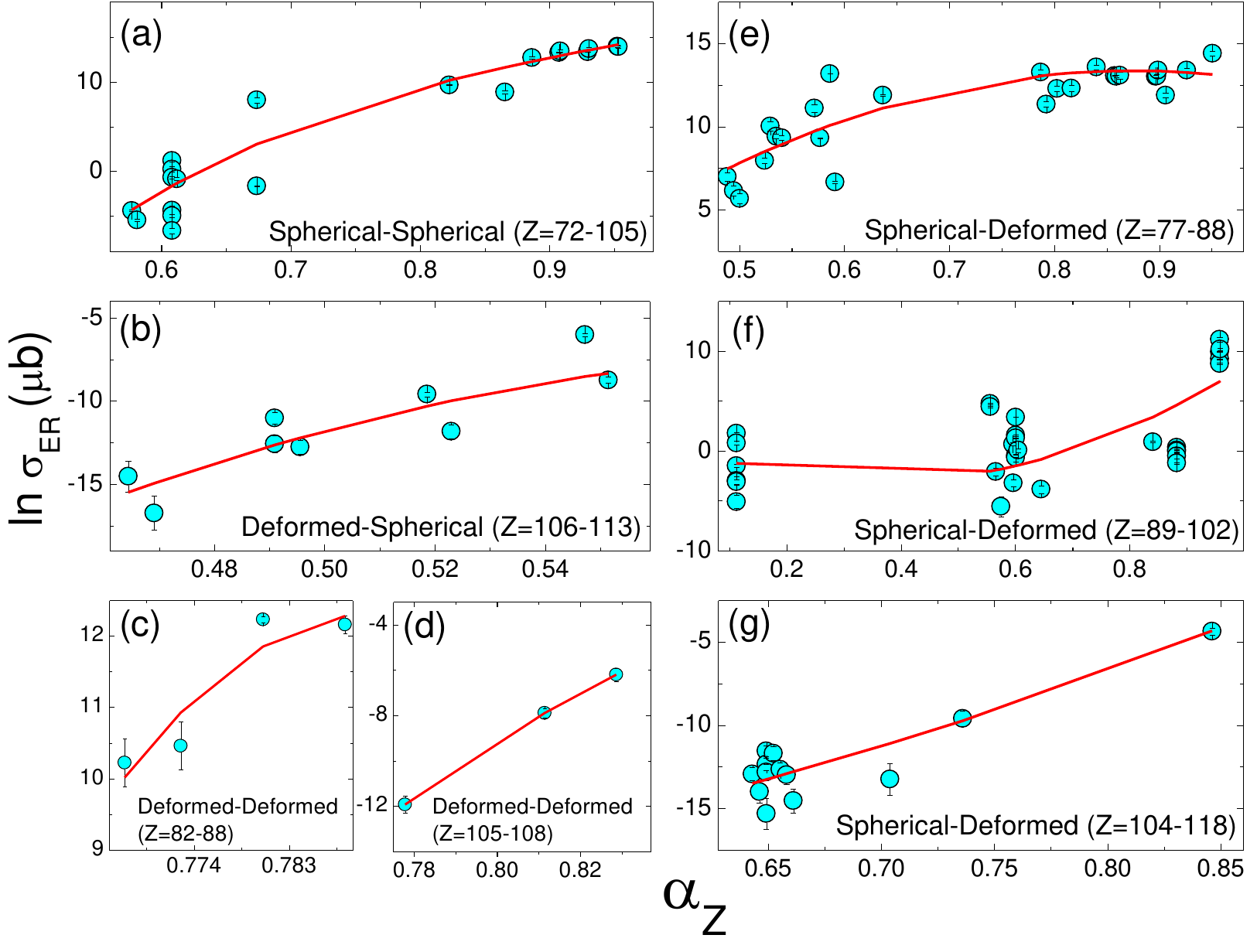}
    \caption{Variation of measured production cross-section as a function of charge asymmetry  parameter for different projectile-target combinations: (a) both projectile and target are spherical\cite{lukyanov2009study,ramler1959excitation,vinodkumar2009fusion,penionzhkevich2008complete,dasgupta2010suppression,signorini2004subbarrier,gasques2009suppression,LEBEYEC1972405,PhysRevC.52.243,PhysRevC.64.054606,khuyagbaatar2010new,gaggeler1989cold,PhysRevC.64.054606,PhysRevC.64.054606,PhysRevC.64.054606,belozerov2003spontaneous,oganessian2001measurements,gaggeler1989cold,hessberger2001decay,hessberger2001decay}, (b) deformed projectile and spherical target\cite{andreev1989measurement,hofmann2004properties,munzenberg1981identification,hofmann2004properties,munzenberg1984evidence,morita2004status,hofmann1995production,hofmann1995new,hofmann1996new,morita2004experiment}, (c-d) deformed projectile and deformed target\cite{hinde1982fission,mahata2003fusion,andreyev1997statistical,andreyev1997decay,andreev1989measurement,haba2014production,haba2012production,dvorak2008observation} and (e-g) spherical projectile and deformed target in specified atomic number regions of the compound nuclei\ref{Spe-DefandDer-Spe}\cite{chakrabarty2000complete,kumar2013low,zhang2014complete,hinde1982fission,sharma2015systematic,scholz2014measurement,fang2013fusion,fang2015complete,penionzhkevich2007excitation,penionzhkevich2008complete,singh2009investigation,shrivastava1999shell,sikkeland1970study,mayorov2014production,baba1988evaporation,maiti2011production,mayorov2014production,vermeulen1984cross,mayorov2014production,mayorov2015evaporation,vermeulen1984cross,LEBEYEC1972405,corradi2005excitation,mayorov2014production,mayorov2015evaporation,vermeulen1984cross,mayorov2015evaporation,vermeulen1984cross,sahm1985fusion,sahm1985fusion,vermeulen1984cross,SAHM1985316,vermeulen1984cross,vermeulen1984cross,schmidt1981barrier,vermeulen1984cross,SAHM1985316,SAHM1985316,SAHM1985316,PhysRevC.62.054603,SAHM1985316,vermeulen1984cross,gaggeler1989cold,PhysRev.111.1358,PhysRevC.17.1706,PhysRevC.17.1706,PhysRev.111.1358,gaggeler1989cold,fleury1973excitation,gaggeler1989cold,PhysRevLett.93.162701,PhysRev.172.1232,PhysRev.172.1232,PhysRev.172.1232,PhysRev.172.1232,PhysRev.172.1232,nagame2002status,nishio2006measurement,nishio2010nuclear,oganessian2004measurements,oganessian2015super,oganessian2004measurements,oganessian2004experiments,oganessian2015super,oganessian2015super,oganessian2015super,oganessian2004measurements,oganessian2015super,oganessian2015super}.}
    \label{chargeasymmetry}
\end{figure}
\begin{figure}
    \centering
    \includegraphics[width=\linewidth]{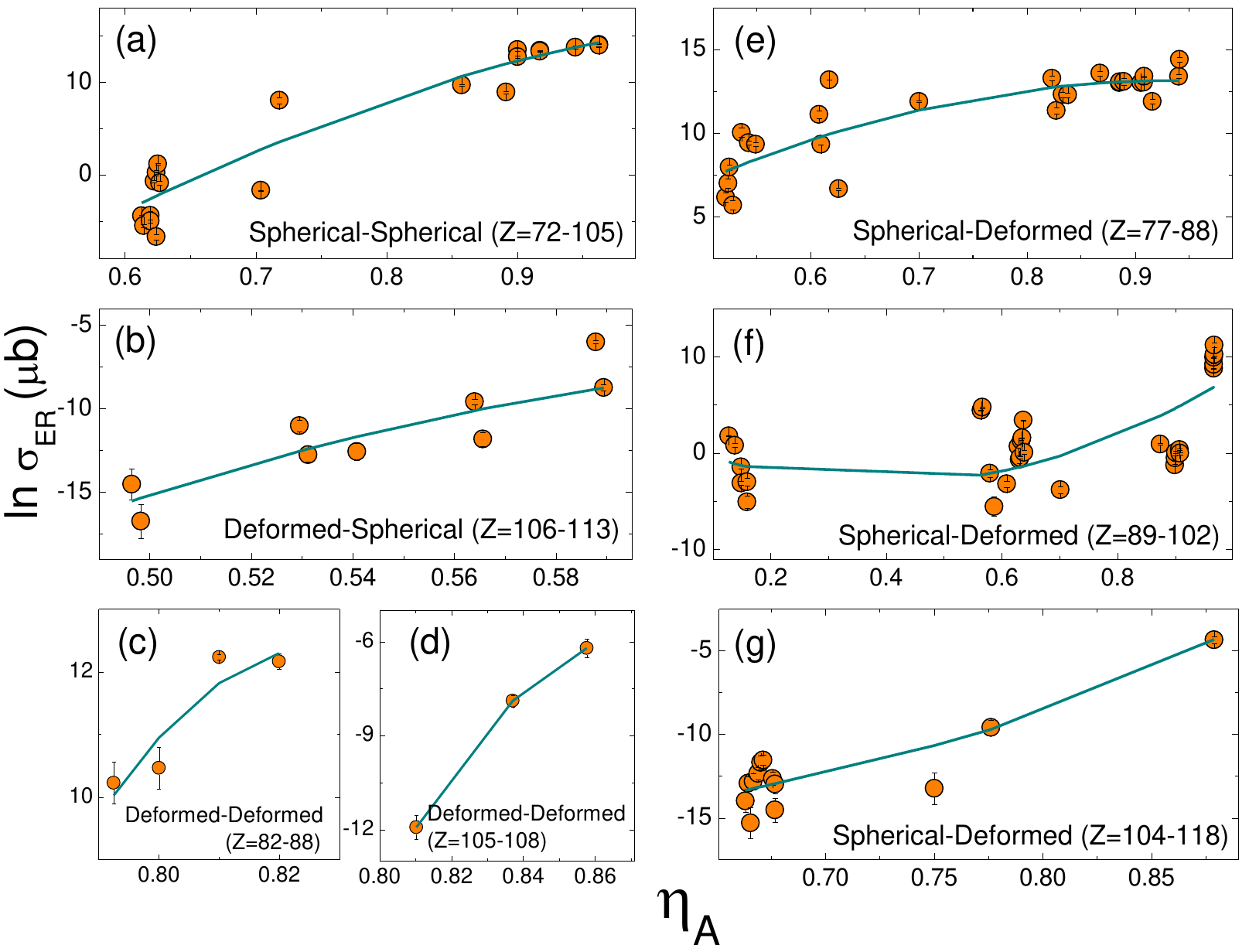}
    \caption{Variation of measured production cross-section as a function mass asymmetry  parameter for different projectile-target combinations \cite{lukyanov2009study,ramler1959excitation,vinodkumar2009fusion,penionzhkevich2008complete,dasgupta2010suppression,signorini2004subbarrier,gasques2009suppression,LEBEYEC1972405,PhysRevC.52.243,PhysRevC.64.054606,khuyagbaatar2010new,gaggeler1989cold,PhysRevC.64.054606,PhysRevC.64.054606,PhysRevC.64.054606,belozerov2003spontaneous,oganessian2001measurements,gaggeler1989cold,hessberger2001decay,hessberger2001decay,andreev1989measurement,hofmann2004properties,munzenberg1981identification,hofmann2004properties,munzenberg1984evidence,morita2004status,hofmann1995production,hofmann1995new,hofmann1996new,morita2004experiment,hinde1982fission,mahata2003fusion,andreyev1997statistical,andreyev1997decay,andreev1989measurement,haba2014production,haba2012production,dvorak2008observation,chakrabarty2000complete,kumar2013low,zhang2014complete,hinde1982fission,sharma2015systematic,scholz2014measurement,fang2013fusion,fang2015complete,penionzhkevich2007excitation,penionzhkevich2008complete,singh2009investigation,shrivastava1999shell,sikkeland1970study,mayorov2014production,baba1988evaporation,maiti2011production,mayorov2014production,vermeulen1984cross,mayorov2014production,mayorov2015evaporation,vermeulen1984cross,LEBEYEC1972405,corradi2005excitation,mayorov2014production,mayorov2015evaporation,vermeulen1984cross,mayorov2015evaporation,vermeulen1984cross,sahm1985fusion,sahm1985fusion,vermeulen1984cross,SAHM1985316,vermeulen1984cross,vermeulen1984cross,schmidt1981barrier,vermeulen1984cross,SAHM1985316,SAHM1985316,SAHM1985316,PhysRevC.62.054603,SAHM1985316,vermeulen1984cross,gaggeler1989cold,PhysRev.111.1358,PhysRevC.17.1706,PhysRevC.17.1706,PhysRev.111.1358,gaggeler1989cold,fleury1973excitation,gaggeler1989cold,PhysRevLett.93.162701,PhysRev.172.1232,PhysRev.172.1232,PhysRev.172.1232,PhysRev.172.1232,PhysRev.172.1232,nagame2002status,nishio2006measurement,nishio2010nuclear,oganessian2004measurements,oganessian2015super,oganessian2004measurements,oganessian2004experiments,oganessian2015super,oganessian2015super,oganessian2015super,oganessian2004measurements,oganessian2015super,oganessian2015super} as mentioned in Fig.\ref{chargeasymmetry}}%:(a) deformed projectile and spherical target, (b-d) spherical projectile and deformed target for making compound nucleus, (e-f) deformed projectile and deformed target and (g) both projectile and target are spherical in specified atomic number regions of the compound nuclei.}
    \label{massasymmetry}
\end{figure}

%%%%%%%%%%%%%%%%%%%%%%%%%%%%%%%%%%% Introduction begins
\indent Elements with atomic number $Z\ge104$ known as transactinides or superheavy elements (SHE) are produced using either cold \cite{hofmann2000discovery} or hot fusion reactions \cite{oganessian2007heaviest,oganessian2015super}. These methods have enabled us to synthesize the SHEs up to the element oganesson $Z=118$  \cite{lee2017elemental}. The cold fusion reactions have been applied for SHEs from $Z=104-113$ and the hot fusion reactions from $Z=110-118$. In cold fusion reactions, $^{208}Pb$ or $^{209}Bi$ target nuclei are bombarded with the projectiles heavier than $^{48}Ca$ and in hot fusion reactions the $^{48}Ca$ projectiles are impacted on an actinide target. To produce the elements heavier than oganesson, the projectiles heavier than $^{48}Ca$ are required as the target elements heavier than Californium $(Z = 98)$ are not possible yet \cite{hofmann2016review,giuliani2019colloquium} and all such experimental attempts have been failed till date \cite{haba2019new,novikov2020formation} irrespective of the fact that increasing stability was expected for superheavy elements Z$>$110 \cite{hamilton2013search,oganessian2015super}. Accordingly special effort is being invested to revamp the detection sensitivity close to fb. On the other side, we have taken up a project in pinning the possible reason of the said debacle.\\
%%%%%%%%%%%%%%%%%%%%%%%%%%%%%%%%%%%%%%%%%%%%%%%%%%%%%%%%%%%%%%%%%%%%%%%Introduction begins 
\indent A memory of entrance channel effects such as the Coulomb interaction parameter ($z$), compound nucleus fissility ($\chi_{CN}$), effective entrance channel fissility ($\chi_{eff}$), mean fissility ($\chi_{m}$), charge ($\alpha_Z$) and mass asymmetry ($\eta_A$) is often retained in the heavy-ion reactions \cite{lesko1983properties,fazio2004formation}, especially, producing compound nucleus (CN) of A$\approx$ 220 \cite{mandaglio2018effects}, A$\approx$ 230 \cite{ramamurthy1985interpretation}, actinide \cite{soheyli2012non} and superheavy elements \cite{nhan2019investigation}. The entrance channel dynamics for the cold fusion reactions have been studied in detail about two decades ago \cite{giardina2000effect}. While we have extended this study recently for the hot fusion reactions \cite{manjunath2020entrance}. Furthermore, other entrance channel effects such as target deformation .\cite{itkis2015fusion} and excitation energy \cite{itkis2008processes} determine the quasifission characteristics \cite{schmitt2019new}, which are the important issues for a reaction to synthesizing a superheavy element \cite{sekizawa2019time}. Role of all these entrance channel effects has been examined explicitly by the experimental evaporation residue cross-sections of the heavy-ion reactions used for synthesizing the SHEs $Z=104-118$ successfully. The same treatment has also been applied on the unsuccessful reactions which were employed for the SHEs $Z=119-120$. Such works help us to establish certain systematic trends, which in turn allow us to plan for a most suitable reaction. Note that the  beam energy required for this reaction is extremely important and it can be found from excitation function studies using either the advanced statistical model (ASM) \cite{sridhar2018search,sridhar2019studies,manjunatha2018investigations} or the dinuclear system (DNS) model \cite{giardina2000effect}.  In this letter, we describe the full survey with a happy note that present ideas may lead to the synthesis of the SHEs beyond $Z=118$.\\
%%%%%%%%%%%%%%%%%%%%%%%%%%%%%%%%%%%%%%%%%%%%%%%%%%%%Discussion on entrance channel parameter effects begin:
\indent As the projectile is made to incident on the target, a compound nucleus is formed at certain excitation energy, which cools down by evaporation of neutrons or any other light particles and the evaporation residue cross-section ($\sigma_{ER}$) is the measure of the formation of the superheavy nuclei. This  $\sigma_{ER}$ is highly reduced due to re-separation of the nuclei by the quasi-fission in the same entrance channel and by a considerable probability of fission of the CN called fusion-fission \cite{hofmann2016remarks}. The $\sigma_{ER}$ depends  on specific entrance channel parameters $z$, $\chi_{m}$, $\eta_A$ and $\alpha_Z$ as mentioned above, which are defined in earlier papers (for example \cite{giardina2000effect}) and has also been provided in Supplemental Material \cite{SM}.  A slight change in the entrance channel parameters makes a large difference in the $\sigma_{ER}$ \cite{gehlot2019evaporation}. Such variations can be observed with the experimental  $\sigma_{ER}$ versus any particular entrance channel parameter. We have studied the roles of the entrance channel in different classes of reactions that are made on the basis of deformation in the projectile and target nuclei. They include the spherical-spherical, deformed-spherical, deformed-deformed and spherical-deformed. Fig. \ref{Spe-DefandDer-Spe} displays the nature of experimental $\sigma_{ER}$ vs $z$ and $\chi_m$ in different deformation classes in semi-log plots. Similarly, trends of other entrance channel parameters on the $\sigma_{ER}$ have been shown in Fig.\ref{meanfissility}, Fig.\ref{chargeasymmetry} and Fig.\ref{massasymmetry}.\\
%%%%%%%%%%%%%%%%%%%%%%%%%%%%%%%%%%%%%%%%%%%%%%%%%%%%%%%%%%%%%%%%%%%%%%%%%%%%%%%%
\indent A strange characteristic has been noticed in the plots of Fig. \ref{Spe-DefandDer-Spe}-\ref{massasymmetry}. Though the ER cross-section data versus $z$  and $\chi_m$ can be presented on semi-log plot for spherical-spherical, deformed-spherical and deformed-deformed cases, but spherical-deformed case has to be differentiated with the range of compound nuclear atomic numbers, viz., preactinides ($z=77-88$), actinides ($Z=89-103$) and transactinides ($Z=104-118$) as shown in Fig.\ref{Spe-DefandDer-Spe}-\ref{meanfissility}. This strategy fails when we consider the parameter $\alpha_Z$ and $\eta_A$; there deformed-deformed case also has to be divided as the spherical-deformed case as shown in Fig.\ref{chargeasymmetry}-\ref{massasymmetry}.\\    
%\begin{figure}
   % \centering
  %  \centering
 %   \includegraphics[width=\linewidth]{cold ER even.pdf}
 %   \caption{Evaporation residue cross-sections for cold fusion reactions $^{50}_{22}Ti+^{208}_{82}Pb$, $^{50}_{22}Ti+^{208}_{82}Pb$, $^{54}_{24}Cr$+$^{208}_{82}Pb$, $^{54}_{24}Cr+^{208}_{82}Pb$, $^{58}_{26}Fe+^{208}_{82}Pb$, $^{58}_{26}Fe+^{208}_{82}Pb$, $^{64}_{28}Ni+^{208}_{82}Pb$, $^{62}_{28}Ni+^{208}_{82}Pb$, $^{70}_{30}Zn+^{208}_{82}Pb$, $^{70}_{30}Zn+^{208}_{82}Pb$ to synthesize the even superheavy elements as a function of entrance channel parameters (a) Coulomb interaction $z$, (b) mean fissility $\chi_m$, (c) mass asymmetry $\eta$, and (d) charge asymmetry $\alpha$.}
  %  \label{coldevpeven}
%\end{figure}
%%%%%%%%%%%%%%%%%%%%%%%%%%%%%%%%%%%%%%%%%%%%%%%%%%%%%%%%%%%%%%%%%%%%%%%%
%%%%%%%%%%%%%%%%%%%%%%%%%%%%%%%%%%%%%%%%%%%%%%%%%%%%%%%%%%%%%%%%%%%%%%%%
%%%%%%%%%%%%%%%%%%%%%%%%%%%%%%%%%%%%%%%%%%%%%%%%%%%%%%%%%%%%%%%%%%%%%%%%
%%%%%%%%%%%%%%%%%%%%%%%%%%%%%%%%%%%%%%%%%%%%%%%%%%%%%%%%%%%%%%%%%%%%%%%%
\indent Having observed the nature of the reactions in Fig.\ref{Spe-DefandDer-Spe}-\ref{massasymmetry}, we may infer certain general facts as follows: (i) the fusion of two spherical nuclei yields the largest $\sigma_{ER}$, (ii) the second largest $\sigma_{ER}$ is obtained when both the projectile and target are deformed, (iii) fusion of either projectile or target nucleus is spherical or deformed yields the lowest $\sigma_{ER}$, (iv) the highest $\sigma_{ER}$ that can obtained is dictated by the properties of the entrance channel parameters such as Coulomb interaction parameter, mean fissility parameter, charge asymmetry parameter, and mass asymmetry parameter. Furthermore, to get systematic trends of the curves seen in Fig. \ref{Spe-DefandDer-Spe}-\ref{massasymmetry}, every curve has been fitted with a suitable mathematical expression as given in Table I. These empirical equations can be useful while planning an experiment for synthesizing any new superheavy elemental isotopes. Before doing so we have first examined the reactions employed in the synthesis of SHEs $Z=119-120$ and thus we have provided the expected $\sigma_{ER}$ in Table II. This exercise results in a gloomy picture because of the large uncertainties. The uncertainties are larger than the empirically predicted values. When the uncertainty with - sign is larger than the value itself, it implies that the net value is closed to zero and not at all negative. The similar scenario is observed in Table III when we examine the reactions planned for future in different laboratories. To shed useful light on this issue we made use of theoretical tools as discussed below.\\
%%%%%%%%%%%%%%%%%%%%%%%%%%%%%%%%%%%%%%%%%%%%%%%%%%%%%%%%
\begin{table}%\tiny
\caption{The equations fitted with the data shown for $\ln\sigma_{ER}~(x=z,\chi_m, \eta_A, \alpha_Z)$ as a function of the entrance channel parameter $x=z/\chi_m/\eta_A/\alpha_Z)$ in Fig.1-4. Here, the first column stands for the type of reaction (TOR) with respect to the deformation parameter of the projectile and target nuclei. The last column gives the uncertainty in evaporation residue cross-section ($\sigma_{ER}$) as obtained from the least square fittings.}
\resizebox{\linewidth}{!}{
\begin{tabular}{|c|l|c|c|c|}
\hline
TOR&$\ln\sigma_{ER}~(x=z)$&\begin{tabular}[c]{@{}c@{}} Range of CN \\atomic no. $Z$  \end{tabular} & \begin{tabular}[c]{@{}c@{}}Range of entrance\\ ch. parameters \end{tabular}&uncertainty\\ \hline
S-S & \begin{tabular}[c]{@{}c@{}}15.62-4.53$\times10^{-2}$x-6.46$\times10^{-4}$x$^2$\\+1.65$\times10^{-6}$x$^3$ \end{tabular} &75$\le Z\le$105&21$\le z\le$189 &$\pm$ 3.88$\;\sigma_{ER}$\\ \hline
\multirow{3}{*}{S-D} & 12.78+2.81$\times10^{-2}$x-4.07$\times10^{-4}$x$^2$&77$\le Z\le$88&21$\le z\le$159 &$\pm$1$\;\sigma_{ER}$\\ \cline{2-5} 
 & 22.11-0.63x+5.27$\times10^{-3}$x$^2$-1.36$\times10^{-5}$x$^3$&89$\le Z\le$102&24$\le z\le$211 &$\pm$16.5$\;\sigma_{ER}$\\ \cline{2-5} 
 & 19.85-0.45x+2.28$\times10^{-3}$x$^2$-4.55$\times10^{-6}$x$^3$&104$\le Z\le$118 &87$\le z\le$197&$\pm$1.46$\;\sigma_{ER}$\\ \hline
D-S & 3160.51-42.23x+0.19x$^2$-2.79$\times10^{-4}$x$^3$&106$\le Z\le$113 &94$\le z\le$247 &$\pm$0.88$\;\sigma_{ER}$\\ \hline
D-D& 152.53-2.78x+1.52$\times10^{-2}$x$^2$-2.79$\times10^{-5}$x$^3$&82$\le Z\le$108 &79$\le z\le$124&$\pm$10.7$\;\sigma_{ER}$\\\hline
\multicolumn{4}{|c|}{$\ln\sigma_{ER}$~(x$=\chi_m)$} &\\ \hline
S-S & -23.71+239.47x-457.51x$^2$+233.52x$^3$&75$\le Z\le$105& 0.32$\le \chi_m\le$0.76 &$\pm$14.8$\;\sigma_{ER}$\\ \hline
\multirow{3}{*}{S-D} & 49.36-244.53x+553.81x$^2$-424.24x$^3$&77$\le Z\le$88&0.29$\le \chi_m\le$0.67 & $\pm$0.7$\;\sigma_{ER}$\\ \cline{2-5} & 143.29-723.71x+1213.24x$^2$-678.15x$^3$&89$\le Z\le$102&0.35$\le \chi_m\le$0.77 & $\pm$4.27$\;\sigma_{ER}$\\ \cline{2-5} 
 & 113.05-414.53x+479.53x$^2$-199.25x$^3$&104$\le Z\le$118 &0.56$\le \chi_m\le$0.79& $\pm$1.39$\;\sigma_{ER}$\\ \hline
D-S & -19.01+92.99x-99.85x$^2$&106$\le Z\le$113 &0.5$\le \chi_m\le$0.65&$\pm$1.63$\;\sigma_{ER}$ \\ \hline
D-D & 455.15-1700.57x+2052.54x$^2$-827.91x$^3$&82$\le Z\le$118&0.55$\le \chi_m\le$0.87& $\pm$4.59$\;\sigma_{ER}$\\ \hline
\multicolumn{4}{|c|}{$\ln\sigma_{ER}$~(x$=\alpha_Z)$} & \\ \hline
S-S & -139.25+424.78x-413.93x$^2$+143.98x$^3$&75$\le Z\le$105 &0.57$\le \alpha_Z\le$0.95 & $\pm$16$\;\sigma_{ER}$\\ \hline
\multirow{3}{*}{S-D} & -16.78+68.88x-39.33x$^2$&77$\le Z\le$88&0.48$\le \alpha_Z\le$0.95 &$\pm$2.41$\;\sigma_{ER}$\\ \cline{2-5} 
 & 0.75-20.83x+28.55x$^2$&89$\le Z\le$102&0.11$\le \alpha_Z\le$0.95 &$\pm$29.6$\;\sigma_{ER}$ \\ \cline{2-5} 
 & -72.84+200.82x-254.62x$^2$+133.55x$^3$&104$\le Z\le$118 &0.64$\le \alpha_Z\le$0.84&$\pm$1.81$\;\sigma_{ER}$ \\ \hline
D-S & -158.98+500.11x-411.46x$^2$&106$\le Z\le$113&0.46$\le \alpha_Z\le$0.55 &$\pm$1.6$\;\sigma_{ER}$ \\ \hline
\multirow{2}{*}{D-D} & -2560.13+6503.67x-4110.66x$^2$&82$\le Z\le$88&0.76$\le \alpha_Z\le$0.78&$\pm$0.31$\;\sigma_{ER}$ \\ \cline{2-5} 
 & -386.23+827.31x-444.78x$^2$&105$\le Z\le$108&0.77$\le \alpha_Z\le$0.82 &$\pm$4.32$\;\sigma_{ER}$\\ \hline
\multicolumn{4}{|c|}{$\ln\sigma_{ER}$~(x$=\eta_A)$} & \\ \hline
S-S & -44.57+51.54x+55.93x$^2$-47.76x$^3$&752$\le Z\le$105&0.61$\le \eta_A\le$0.96 & $\pm$10.8$\;\sigma_{ER}$\\ \hline
\multirow{3}{*}{S-D} & -14.56+59.26x-31.68x$^2$&77$\le Z\le$88&0.52$\le \eta_A\le$0.94 & $\pm$2.2$\;\sigma_{ER}$\\ \cline{2-5} 
 & 1.68-24.26x+30.64x$^2$&89$\le Z\le$102&0.12$\le \eta_A\le$0.96 &$\pm$32.4$\;\sigma_{ER}$ \\ \cline{2-5} 
 & -122.77+434.35x-606.79x$^2$+302.61x$^3$&104$\le Z\le$118&0.66$\le \eta_A\le$0.88 &$\pm$2.1$\;\sigma_{ER}$ \\ \hline
D-S & -137.06+389.91x-292.17x$^2$&106$\le Z\le$113&0.49$\le \eta_A\le$0.58 &$\pm$1.42$\;\sigma_{ER}$ \\ \hline
\multirow{2}{*}{D-D} & -1325.02+3233.12x-1953.96x$^2$&82$\le Z\le$88&0.79$\le \eta_A\le$0.81 &$\pm$0.32$\;\sigma_{ER}$\\ \cline{2-5} 
 & -1117+2539.41x-1450.19x$^2$&105$\le Z\le$108&0.81$\le \eta_A\le$0.85 & $\pm$5.82$\;\sigma_{ER}$\\ \hline
\end{tabular}}
\label{entrance_channel_parameters}
\end{table}
%%%%%%%%%%%%%%%%%%%%%%%%%%%%%%%%%%%%%%%%%%%%%%%%%%%%%%%%
%\resizebox{\linewidth}{!}{
\begin{table*}
\caption{Details of unsuccessful experiments in synthesis of SHEs Z=119-120. Type of reaction (TOR) is mentioned in 10th column in commensurate with Table I. Symbol $\langle\sigma_{ER}\rangle(pb)$ represents an average of $\sigma_{ER}(z)$, $\sigma_{ER}(\chi_m)$, $\sigma_{ER}(\eta_A)$, and $\sigma_{ER}(\alpha_z)$. $E^*$ denotes the excitation energy and $\beta_2$ the deformation parameter. Note that when the uncertainty with - sign is larger than the value itself, it implies that the net value is closed to zero and not at all negative.}
\begin{tabular}{|l|c|c|c|c|c|c|c|c|c|c|c|c|c|c|c|}
\hline
\multicolumn{1}{|c|}{\multirow{2}{*}{Reaction}} & \multirow{2}{*}{\begin{tabular}[c]{@{}c@{}}E$^*$\\ (MeV)\end{tabular}} & \multirow{2}{*}{$z$} & \multirow{2}{*}{$\chi_m$} & \multirow{2}{*}{$\eta_A$} & \multirow{2}{*}{$\alpha_Z$} & \multicolumn{3}{c|}{$\beta_2$} & TOR& \multirow{2}{*}{$\langle\sigma_{ER}\rangle(pb)$}  \\ \cline{7-9}
\multicolumn{1}{|c|}{} &  &  &&&  & Proj. & Targ. & Comp. &&$_{(z,\chi_m,\eta_A,\alpha_Z)}$  \\ \hline
$^{50}_{22}Ti+^{249}_{97}Bk\rightarrow^{299}_{119}Uue   $\cite{khuyagbaatar2013superheavy} & 32.4 & 213.9 & 0.84 & 0.67 & 0.63 & 0 & 0.24 & -0.020 &  S-D&0.8$\pm$1.5   \\ \hline
$^{50}_{22}Ti+^{249}_{98}Cf\rightarrow^{299}_{120}Ubn $  \cite{khuyagbaatar2013superheavy}  & 42.8 & 216.1 & 0.85 & 0.67 & 0.63 & 0 & 0.235 & -0.035 &  S-D&0.8$\pm$1.5   \\ \hline
$^{51}_{23}V+^{248}_{96}Cm\rightarrow^{299}_{119}Uue $  \cite{khuyagbaatar2013superheavy}  & 36.8 & 221.0 & 0.86 & 0.66 & 0.61 & 0 & 0.24 & -0.020 & S-D&0.5$\pm$1.1  \\ \hline
$^{51}_{23}V+^{249}_{97}Bk\rightarrow^{300}_{120}Ubn $  \cite{khuyagbaatar2013superheavy}  & 35.9 & 223.1 & 0.79 & 0.62 & 0.57 & 0 & 0.24 & 0 &  S-D&0.6$\pm$1.1  \\ \hline
$^{54}_{24}Cr+^{243}_{95}Am\rightarrow^{297}_{119}Uue$   \cite{khuyagbaatar2013superheavy}  & 31.5 & 227.5 & 0.86 & 0.64 & 0.6 & 0.18 & 0.22 & -0.080 &  D-D&0.8$\pm$6.3   \\ \hline
$^{58}_{26}Fe+^{237}_{93}Np\rightarrow^{295}_{119}Uue $  \cite{khuyagbaatar2013superheavy} & 29.9 & 240.4 & 0.89 & 0.61 & 0.56 & 0.199 & 0.215 & 0.072 & D-D&0.2$\pm$1.2  \\ \hline
$^{59}_{27}Co+^{238}_{92}U\rightarrow^{297}_{119}Uue $  \cite{khuyagbaatar2013superheavy}  & 36.7 & 246.2 & 0.9 & 0.6 & 0.55 & 0.14 & 0.22 & -0.080 & D-D&0.1$\pm$0.9  \\ \hline
$^{54}_{24}Cr+^{248}_{96}Cm\rightarrow^{302}_{120}Ubn $  \cite{khuyagbaatar2013superheavy}  & 42 & 229.0 & 0.76 & 0.67 & 0.63 & 0.18 & 0.24 & -0.080 &  D-D&0.4$\pm$2.9    \\ \hline
$^{55}_{25}Mn+^{243}_{95}Am\rightarrow^{298}_{120}Ubn$   \cite{khuyagbaatar2013superheavy}  & 34.5 & 236.5 & 0.8 & 0.58 & 0.53 & 0.19 & 0.22 & -0.080 &  D-D&0.3$\pm$2.1 \\ \hline
$^{59}_{27}Co+^{237}_{93}Np\rightarrow^{296}_{120}Ubn$   \cite{khuyagbaatar2013superheavy}   & 32.9 & 249.1 & 0.77 & 0.66 & 0.62 & 0.14 & 0.22 & -0.090 &  D-D&0.1$\pm$0.9    \\ \hline
$^{58}_{26}Fe+^{244}_{94}Pu\rightarrow^{302}_{120}Ubn$\cite{khuyagbaatar2013superheavy,ts2009attempt}  & 29.4 & 241.5 & 0.77 & 0.66 & 0.62 & 0.19 & 0.22 & 0 &  D-D&0.7$\pm$5.7  \\ \hline
$^{64}_{28}Ni+^{238}_{92}U\rightarrow^{302}120 $  \cite{ts2009attempt,Hofmann2008report}  & 32.2 & 252.6 & 0.65 & 0.64 & 0.6 & -0.08 & 0.22 & 0 & D-D&0.1$\pm$0.3   \\ \hline
$^{54}_{24}Cr+^{248}_{96}Cm\rightarrow^{302}_{120}Ubn$   \cite{Morita2015report}  & 33.0 & 229.0 & 0.8 & 0.6 & 0.55 & 0.18 & 0.26 & 0 & D-D&0.4$\pm$2.9    \\ \hline
$^{58}_{26}Fe+^{244}_{94}Pu\rightarrow^{302}_{120}Ubn  $ \cite{khuyagbaatar2013superheavy} & 33.9 & 241.5 & 0.76 & 0.67 & 0.63 & 0.19 & 0.22 & 0 & D-D&0.2$\pm$1.4 \\ \hline
$^{55}_{25}Mn+^{244}_{94}Pu\rightarrow^{299}_{119}Uue  $ \cite{khuyagbaatar2013superheavy} & 37.7 & 233.8 & 0.88 & 0.63 & 0.58 & 0.19 & 0.22 & -0.020 &  D-D&0.3$\pm$2.1   \\ \hline
\end{tabular}
\label{Failuretable}
\end{table*}
%%%%%%%%%%%%%%%%%%%%%%%%%%%%%%%%%%%%%%%%%%%%%%%%%%%%%%%%%%%%%%%%%%%%%%
\begin{table*}
\caption{Prediction on the possible outcome of the fusion reactions planned to synthesize the SHEs $Z=119 ~\&~120$ in various labs. The first column gives the laboratory name and the 10th column type of reaction (TOR). Symbol $\langle\sigma_{ER}\rangle$ in the 11th column and $\beta_2$ carry the same meaning as in Table \ref{Failuretable}. Hence, when the uncertainty with - sign is larger than the value itself, it implies that the net value is closed to zero and not at all negative. The energy $E_{cm}^{opt}$ indicates the optimal beam energy in MeV suggested for obtaining the maximum cross-section $\sigma_{ER}^{max}$ in $pb$ using a particular model calculation.}
\resizebox{\textwidth}{!}{
\begin{tabular}{|c|c|c|c|c|c|c|c|c|c|c|c|c|c|c|c|c|c|c|c|c|c|c|}
\hline
\multirow{2}{*}{\begin{tabular}[c]{@{}c@{}}Name \\ of lab \end{tabular}}&\multirow{2}{*}{Reaction} &
   \multirow{2}{*}{z} &
  \multirow{2}{*}{$\chi_{m}$} &
  \multirow{2}{*}{$\eta_A$} &
  \multirow{2}{*}{$\alpha_Z$} &
   \multicolumn{3}{c|}{$\beta_2$}& \multirow{2}{*}{TOR}&\multirow{2}{*}{$\langle \sigma_{ER}\rangle (pb)$} &\multicolumn{2}{c|}{DNS}&\multicolumn{2}{c|}{ASM} 
   \\ \cline{7-9}\cline{12-15}
   &     && &&& Proj. &  Targ. &  Comp.  &&$_{(z,\chi_m,\eta_A,\alpha_Z)}$  &$E_{cm}^{opt}$ &$\sigma^{max}_{ER}(xn)$&E$^{opt}_{cm}$&$\sigma^{max}_{ER}(xn)$    \\ \hline
LBNL&$^{86}_{36}Kr+^{209}_{83}Bi\rightarrow^{295}_{119}Uue$\cite{khuyagbaatar2013superheavy}  &  288.7 & 0.95 & 0.42 & 0.39 &0& 0& 0.072 &S-S&34.5$\pm$392.5 &313 &261(1n)&319&547.3(1n)  \\\hline
LBNL&$^{87}_{37}Rb+^{208}_{82}Pb\rightarrow^{295}_{119}Uue$  \cite{hoffman2000transuranium} &293.0 &  0.96 & 0.41 & 0.38 &0& 0 & 0.072&S-S&38.7$\pm$440.3 &317 &268(1n)&324&536.4(1n) \\ \hline
%LBNL&$^{50}_{22}Ti+^{249}_{97}Bk\rightarrow^{299}_{119}Uue$ \cite{dmitriev2016status}&  213.9 & 0.84 & 0.67 & 0.63  &0& 0.235 & -0.018& S-D&0.8$\pm$1.5&220&32(3n)&214&3.9(4n) \\ \hline
JINR&$^{45}_{21}Sc+^{249}_{98}Cf\rightarrow^{294}_{119}Uue$ \cite{khuyagbaatar2013superheavy} &    209.0 & 0.84 &  0.69 &  0.65 &0 &  0.24 &  0.08&S-D&1.6$\pm$3.1   &217&18.8(3n) &208&3.8(3n)\\\hline 
%JINR&$^{50}_{22}Ti+^{249}_{98}Cf\rightarrow^{299}_{120}Ubn$ \cite{khuyagbaatar2013superheavy} &216.1 & 0.85 & 0.67 & 0.63  &0 & 0.235 & -0.035&S-D&0.8$\pm$1.5 &213&169(3n)&219&2.5(3n) \\ \hline
RIKEN&$^{54}_{24}Cr+^{248}_{96}Cm\rightarrow^{302}_{120}Ubn$\cite{dmitriev2016status}  &229.0 & 0.87 & 0.64 & 0.60  &0.36 & 0.235 & 0&D-D&0.2$\pm$1.4&236 &0.36nb(3n) &228&6.93(3n) \\ \hline
\end{tabular}}
\label{listtable}
\end{table*}
%%%%%%%%%%%%%%%%%%%%%%%%%%%%%%%%%%%%%%%%%%%%%%Discussion on entrance channel parameter effects continue..
\begin{figure}
\centering
    \includegraphics[width=\linewidth]{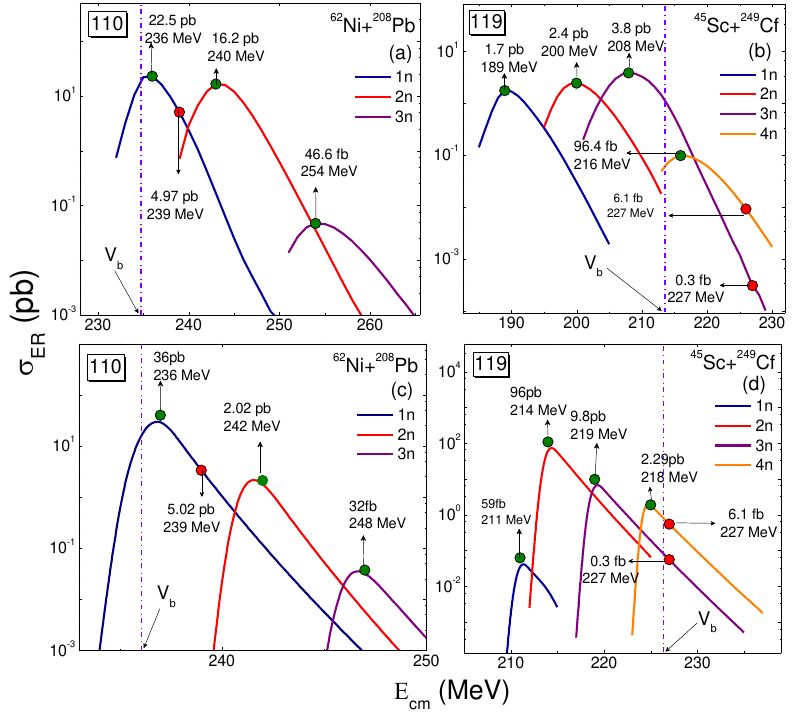}
    \caption{Variation of evaporation residue cross-section ($\sigma_{ER}$) with center of mass energy (E$_{cm}$) for different neutron evaporation channels using the advanced statistical model for the reaction $^{62}Ni+^{208}Pb\rightarrow^{269}Ds$ (a) and the reaction $^{45}_{21}Sc+^{249}_{98}Cf \rightarrow ^{294}_{119}Uue$ (b). While (c) and (d) represent the similar results as obtained with the dinuclear system model. The maximum $\sigma_{ER}$ for every case is marked with green sphere and the $\sigma_{ER}$ at the energy used in experiment is marked with the pink sphere. The projectile energy ($E_{cm}$) are taken from Ref. \cite{hofmann2016remarks} for (a) and (c), while \cite{khuyagbaatar2013superheavy} for (b) and (d). Dotted vertical lines represent fusion barrier energy ($V_B$) \cite{nandi2021search}.}
    \label{Sc+Cf and Ni+Pb ERgraph}
\end{figure}
%%%%%%%%%%%%%%%%%%%%%%%%%%%%%%%%%%%%%%%%%%%%%%%%%%%%%%%%%%%%%%%%%%%%%%%%%%%%%%%%%%%%%%%%%%
%\indent Deformation of projectile and target plays an important role in the heavy-ion reactions, especially those meant for synthesis of superheavy nuclei \cite{itkis2015fusion} as shown in Fig. 1 in Supplemental Material \cite{SM}. Deformation driven nuclear shapes can be represented by the deformation parameters, in particular, the $\beta_2$ for the nuclei \cite{deformation} which are not exactly spherical nor prolate or oblate are displayed in Table III. A careful analysis on the successful reactions leads us to constitute the following rules:  (i) the fusion of two spherical nuclei yields a large $\sigma_{ER}$ of the order of nb and (ii) fusion of one partner either projectile or target nucleus is spherical and other being deformed yields $\sigma_{ER}$ of the order of pb. If we extend the analysis on the unsuccessful reactions, we can set two more rules: (iii) fusion of two deformed nuclei cannot produce cross-section of the order of pb and (iv) deformed CN as a SHN have a better survival probability than that of the spherical SHN. However, these rules violate the reaction $^{45}_{21}Sc+^{249}_{98}Cf \rightarrow ^{294}_{119}Uue$ (Table \ref{Failuretable}). Let us check whether this null result has any relation with the excitation energy used for the experiment.\\
%%%%%%%%%%%%%%%%%%%%%%%%%%%%%%%%%%%%% Consideration of excitation energy begins: average l-value
\indent Main goal of the present work is to find a way to synthesize the SHE, which is a system having a large number of nucleons having intrinsically a high density of excited states and thus a suitable statistical model can deal with its properties \cite{rauscher1997nuclear}. Excitation energy dependent evaporation residue cross-section with subsequent emission of $x$ neutrons can be well evaluated by the ASM as
given by \cite{sridhar2018search,sridhar2019studies,manjunatha2018investigations}
\begin{multline}
    \sigma_{ER-ASM}^{xn}=\frac{\pi}{k^2}\sum_{\ell=0}^\infty(2\ell+1)T(E,\ell)\\P_{CN}(E_{cm},\ell)P_{sur}^{xn}(E^*,\ell) \label{sigmaEr}
    \end{multline}
\noindent Where $k$ is the wave number, average angular momentum $\ell=\langle\ell\rangle$, 
$P^{xn}_{sur}$ is the survival probability of the compound nucleus to ground state by emitting neutrons or lighter particles, $P_{CN}$ is the compound nucleus formation probability, the excitation energy $E^*=E_{cm}+Q$, the reaction $Q$-value is calculated \cite{moller1993nuclear} and $T_\ell$ is the barrier penetration factor \cite{sridhar2018search,sridhar2019studies,manjunatha2018investigations}, which is a function of E$_{cm}$ and $\ell$. The value of $x$ depends on the $E^*$ used for the reaction. Note that $\ell$ takes a vital role and its  value can be deduced as done in \citet{capurro1997average}. While the $\sigma_{ER}$ by DNS model \cite{giardina2000effect} is stated as 
\begin{align}
    \sigma_{ER-DNS}^{Z,A}=\sum_{\ell=0}^{\infty} (2\ell+1)\sigma_{\ell}^{fus}(E_{cm},\ell)P_{sur}^{xn}(E^*,\ell) \label{2}.
    \end{align}
Where $\sigma_{\ell}^{fus}$ is the partial capture cross-section which represents the transition of the colliding nuclei over the
Coulomb barrier and the formation of the initial DNS. Note that the deformation takes significant role in calculation of the fusion cross section. Ideally the dynamical deformation is required. However, static deformation is only used in the ASM and DNS models. Whereas the shell effect plays its role in the calculation of survival probability for both the models. Details of both the ASM and DNS models for excitation function study have been provided in Supplemental Material \cite{SM}.\\
%%%%%%%%%%%%%%%%%%%%%%%%%%%%%%%%%%%%%%%% Discussion of the \sigma_{ER}^{xn} vs beam energy
\indent The $\sigma_{ER}^{xn}$ can be calculated as a function of $E_{cm}$ using Eqn. (1) or (2) to find the optimum $E_{cm}$ in obtaining the maximum cross-section for $xn$ channel ($\sigma_{ER}^{max}$) \cite{Manjunath2021PLB}. We have first tested such possibility using the ASM through a successful reaction $^{62}Ni+^{208}Pb\rightarrow^{269}Ds$ for which the experimental $\sigma_{ER}$ at $E_{cm}=239$ MeV is known and thus marked in Fig. \ref{Sc+Cf and Ni+Pb ERgraph}. 
%\begin{figure}[h!]
  %  \centering
%    \includegraphics[width=\linewidth]{Er comp with exp.pdf}
%    \caption{Variation of evaporation residue cross-section ($\sigma_{ER}$) with center of mass energy (E$_{cm}$) for $3$ neutron evaporation channels in case of the reaction $^{62}Ni+^{208}Pb\rightarrow^{269}Ds$ (a) and $4$ neutron evaporation channels in case of the reaction $^{45}_{21}Sc+^{249}_{98}Cf \rightarrow ^{294}_{119}Uue$ (b). The highest $\sigma_{ER}$ value for every reaction channel is marked with pink sphere and the $(\sigma_{ER})$ at the energy experiment was performed is marked with the violet sphere. The projectile energy ($E_{cm}$) are taken from Ref. \cite{hofmann2016remarks} (a) and \cite{khuyagbaatar2013superheavy} (b). Dotted lines are to guide the eye.}
%    \label{Sc+Cf and Ni+Pb ERgraph}
%\end{figure}
%%%%%%%%%%%%%%%%%%%%%%%%%%%%%%%%%%%%%%
At this energy, $x=1$ is the major and $x=2$ the minor channel yielding cross-section of 4.97 and 0.7 pb, respectively. Hence, total theoretical cross-section $5.67 pb$ is in agreement with the measured cross-section $3.5^{+2.7}_{-1.8}pb$ \cite{hofmann1995production}. Many such comparisons have been compiled in a recent article \cite{Manjunath2021PLB}. With this confidence level, we have calculated the $\sigma_{ER}^{xn}$ for the reaction $^{45}_{21}Sc+^{249}_{98}Cf \rightarrow ^{294}_{119}Uue$ and the results shown in Fig. \ref{Sc+Cf and Ni+Pb ERgraph} (b) gives us a notion that the cross-section at the beam energy at which the experiment was carried out provides $\sigma_{ER}^{xn}$ = 6.4 fb only and thereby, beyond the reach of the experimental sensitivity at that point of time. On the other hand, if experiment be done at 208 MeV, it can lead to $\sigma_{ER}^{xn}=3.97$ pb, which is well within the present detection limit. This $\sigma_{ER}^{xn}$ versus $E_{cm}$ study let us constitute the most vital input for an experiment, we call it the optimal energy criterion: the optimum beam energy for an allowed reaction with respect to the first four criteria must be determined by a suitable calculation to check whether the predicted largest $\sigma_{ER}^{xn}$ is within the detection limit of a certain setup. This fact can be represented well by the DNS model also as shown in Fig. \ref{Sc+Cf and Ni+Pb ERgraph} (c) and (d) and also in Table  III and thus a model independent way. Significance of optimal energy selection for SHE synthesis has been experimentally demonstrated very recently by \citet{tanaka2020study}. \\%effects of $\langle\ell\rangle$ and excitation energy ($E^*$) of the synthesis of the SHEs and has been discussed in Supplemental Material \cite{SM}.\\
%%%%%%%%%%%%%%%%%%%%%%%%%%%%%%%%%%%%%%%%%%%%%%%%%%%%%%%%%%%%%
\indent The above mentioned optimal energy criterion has been implemented on the planned reactions made by several labs  \cite{Khuyagbaatar2013report,hoffman2000transuranium,dmitriev2016status} for synthesizing SHEs $Z=119,120$ using the heavy-ion reactions and shown in Table \ref{listtable}. Here, it is evident that all these reactions can be utilised to measure the $\sigma_{ER}^{max}$ at the optimal energy listed there. A remarkable point is that the first two reactions in Table III lead to a very large $\sigma_{ER}^{max}$ and thus should be attempted at the earliest possible. Furthermore, the optimal energy predicted by the ASM and DNS models differ by 1-9 MeV and the $\sigma_{ER}^{max}$ is the largest at the corresponding optimal energy, they agree within a factor of 8. Hence, we recommend conducting experiments at both the optimal energies to test the power of each calculation.\\
%rule (iv) suggests that synthesizing a magic superheavy nuclei will not be possible. This inference is in commensurate with prediction using relativistic mean-field model that a broad valley of shell stabilization around Z = 120 and N = 172–184 is more likely than a magic superheavy nuclei \cite{bender2001shell}. This prediction has been further supported by several later works \cite{cwiok2005shape, koura2014estimating,heenen2015shapes,nazarewicz2018limits}. On this ground, the last reaction of Table \ref{listtable} will not yield any measurable production cross-section.\\ %Hence, one can expect that day is not far off when we discover new members in the periodic table.\\
%%%%%%%%%%%%%%%%%%%%%%%%%%%%%%%%%%%%%%%%%%%%%%%%%%%%%%%%%%%%%%%%%%%%%%%%%%%%%%%%%%%%%
%%%%%%%%%%%%%%%%%%%%%%%%%%%%%%%%%%%%%%%%%%%%%%%%%%%%%%%%%% Conclusion
\indent To summarize, a careful analysis of the experimental facts with the heavy ion reactions provides us certain criteria in the light of deformation parameters of the projectile and target systems (Fig. 1-4) that can be of great value to meet the challenges for synthesizing the new elements in the eighth period of the periodic table as mentioned above. The analysis based on these criteria does yield much less than 1 $pb$ evaporation residue cross-section for the unsuccessful experiments listed in Table \ref{Failuretable}. However, these cross-sections suffer from a large uncertainty. Such a scenario is noticed as well for the planned experiments shown on the 11th column of Table \ref{listtable}. Root cause of such unusually large errors is nothing but the arbitrary choice of the beam energies used for the experiments. The optimal beam energy for obtaining the maximum evaporation residue cross-section can be chosen from the theoretical excitation function curves from the ASM as well as DNS model (Fig. 5). Furthermore, such curves can be used to find the expected cross-section at the specific beam energies used for the experiments. We saw the optimal energies and the evaporation residue cross-sections from these two models agree well within 9 MeV and a factor of 8, respectively. \\
\indent Let us consider a very recent paper \cite{khuyagbaatar2020search} that  describes a despondent fact of not detecting any events from the $^{50}Ti + ^{249}Bk \rightarrow ^{295}Uue+4n$ and $^{50}Ti + ^{249}Cf \rightarrow ^{296}Ubn+3n$ reactions at cross-section sensitivity levels of 65 and 200 $fb$, respectively, at a beam energy of $E_{lab}=281.5$ MeV. We provide theoretical cross-sections using ASM and DNS models for these reactions at this energy as well as at the optimal energies by each model in Table \ref{PRC2020}. Both the models predict a few tens of $fb$ cross-section at the beam energy used. Thus, the cause of debacle is well explained. Interestingly, the cross-sections shoot up to at least a few $pb$ at the optimal energies. Similar analysis of the proposed reactions in different labs shown in Table \ref{listtable} gives us a pretty green signal. Especially, first two reactions in the Table producing $Z=119$ can have a few hundred $pb$ cross-sections. Hence, we may see shortly that new members of the periodic table are being discovered with the presently available facility in the laboratories.\\
%%%%%%%%%%%%%%%%%%%%%%%%%%%%%%%%%%%%%%%%%%%%%%%%%%%%%%%%%%%%%%
\begin{table}
\caption{Examining the outcome of the two reactions $^{50}Ti + ^{249}Bk$ and $^{50}Ti + ^{249}Cf$ reported recently \cite{khuyagbaatar2020search}. $\sigma_{ER}^{EE}$ stands for $\sigma_{ER}$ at experimental energy ($E_{lab}$=281.5 MeV or $E_{cm}$=234.43 MeV) and other symbols are same as used for Table \ref{listtable}. Cross-sections are given in $pb$ and energies in MeV.}
\resizebox{\linewidth}{!}{
\begin{tabular}{|l|c|c|c|c|c|c|c|c|c|}
\hline
\multicolumn{1}{|c|}{\multirow{2}{*}{Reaction}} &  \multicolumn{3}{c|}{ASM} & \multicolumn{3}{c|}{DNS} \\ \cline{2-7} 
\multicolumn{1}{|c|}{}   & $\sigma_{ER}^{EE}(xn)$ & E$_{cm}^{opt}$ & $\sigma_{ER}^{max}(xn)$ & $\sigma_{ER}^{EE}(xn)$& E$_{cm}^{opt}$ & $\sigma_{ER}^{max}(xn)$ \\ \hline
$^{50}Ti+^{249}Bk$ & 0.015 (4n) & 214 & 3.98 (4n) & 0.011 (4n) & 220 & 6.3 (4n) \\ \hline
$^{50}Ti+^{249}Cf$ & 0.011 (3n) & 219 & 2.46 (3n) & 0.039 (3n) & 213 & 16.9 (3n) \\ \hline
\end{tabular}}
\label{PRC2020}
\end{table}
%%%%%%%%%%%%%%%%%%%%%%%%%%%%%%%%%%%%%%%%%%%%%%%%%%%%%%%%%%%
\indent We thank Tilak Ghosh for useful comments.
\bibliography{apssamp}% Produces the bibliography via Bibtex

\begin{thebibliography}{103}%
\makeatletter
\providecommand \@ifxundefined [1]{%
 \@ifx{#1\undefined}
}%
\providecommand \@ifnum [1]{%
 \ifnum #1\expandafter \@firstoftwo
 \else \expandafter \@secondoftwo
 \fi
}%
\providecommand \@ifx [1]{%
 \ifx #1\expandafter \@firstoftwo
 \else \expandafter \@secondoftwo
 \fi
}%
\providecommand \natexlab [1]{#1}%
\providecommand \enquote  [1]{``#1''}%
\providecommand \bibnamefont  [1]{#1}%
\providecommand \bibfnamefont [1]{#1}%
\providecommand \citenamefont [1]{#1}%
\providecommand \href@noop [0]{\@secondoftwo}%
\providecommand \href [0]{\begingroup \@sanitize@url \@href}%
\providecommand \@href[1]{\@@startlink{#1}\@@href}%
\providecommand \@@href[1]{\endgroup#1\@@endlink}%
\providecommand \@sanitize@url [0]{\catcode `\\12\catcode `\$12\catcode
  `\&12\catcode `\#12\catcode `\^12\catcode `\_12\catcode `\%12\relax}%
\providecommand \@@startlink[1]{}%
\providecommand \@@endlink[0]{}%
\providecommand \url  [0]{\begingroup\@sanitize@url \@url }%
\providecommand \@url [1]{\endgroup\@href {#1}{\urlprefix }}%
\providecommand \urlprefix  [0]{URL }%
\providecommand \Eprint [0]{\href }%
\providecommand \doibase [0]{https://doi.org/}%
\providecommand \selectlanguage [0]{\@gobble}%
\providecommand \bibinfo  [0]{\@secondoftwo}%
\providecommand \bibfield  [0]{\@secondoftwo}%
\providecommand \translation [1]{[#1]}%
\providecommand \BibitemOpen [0]{}%
\providecommand \bibitemStop [0]{}%
\providecommand \bibitemNoStop [0]{.\EOS\space}%
\providecommand \EOS [0]{\spacefactor3000\relax}%
\providecommand \BibitemShut  [1]{\csname bibitem#1\endcsname}%
\let\auto@bib@innerbib\@empty
%</preamble>
\bibitem [{\citenamefont {Hofmann}\ and\ \citenamefont
  {M{\"u}nzenberg}(2000)}]{hofmann2000discovery}%
  \BibitemOpen
  \bibfield  {author} {\bibinfo {author} {\bibfnamefont {S.}~\bibnamefont
  {Hofmann}}\ and\ \bibinfo {author} {\bibfnamefont {G.}~\bibnamefont
  {M{\"u}nzenberg}},\ }\href@noop {} {\bibfield  {journal} {\bibinfo  {journal}
  {Reviews of Modern Physics}\ }\textbf {\bibinfo {volume} {72}},\ \bibinfo
  {pages} {733} (\bibinfo {year} {2000})}\BibitemShut {NoStop}%
\bibitem [{\citenamefont {Oganessian}(2007)}]{oganessian2007heaviest}%
  \BibitemOpen
  \bibfield  {author} {\bibinfo {author} {\bibfnamefont {Y.}~\bibnamefont
  {Oganessian}},\ }\href@noop {} {\bibfield  {journal} {\bibinfo  {journal}
  {Journal of Physics G: Nuclear and Particle Physics}\ }\textbf {\bibinfo
  {volume} {34}},\ \bibinfo {pages} {R165} (\bibinfo {year}
  {2007})}\BibitemShut {NoStop}%
\bibitem [{\citenamefont {Oganessian}\ and\ \citenamefont
  {Utyonkov}(2015)}]{oganessian2015super}%
  \BibitemOpen
  \bibfield  {author} {\bibinfo {author} {\bibfnamefont {Y.~T.}\ \bibnamefont
  {Oganessian}}\ and\ \bibinfo {author} {\bibfnamefont {V.}~\bibnamefont
  {Utyonkov}},\ }\href@noop {} {\bibfield  {journal} {\bibinfo  {journal}
  {Reports on Progress in Physics}\ }\textbf {\bibinfo {volume} {78}},\
  \bibinfo {pages} {036301} (\bibinfo {year} {2015})}\BibitemShut {NoStop}%
\bibitem [{\citenamefont {Lee}(2017)}]{lee2017elemental}%
  \BibitemOpen
  \bibfield  {author} {\bibinfo {author} {\bibfnamefont {M.~S.}\ \bibnamefont
  {Lee}},\ }\href@noop {} {\bibfield  {journal} {\bibinfo  {journal} {Science}\
  }\textbf {\bibinfo {volume} {357}},\ \bibinfo {pages} {461} (\bibinfo {year}
  {2017})}\BibitemShut {NoStop}%
\bibitem [{\citenamefont {Hofmann}\ \emph
  {et~al.}(2016{\natexlab{a}})\citenamefont {Hofmann}, \citenamefont {Heinz},
  \citenamefont {Mann}, \citenamefont {Maurer}, \citenamefont {M{\"u}nzenberg},
  \citenamefont {Antalic}, \citenamefont {Barth}, \citenamefont {Burkhard},
  \citenamefont {Dahl}, \citenamefont {Eberhardt} \emph
  {et~al.}}]{hofmann2016review}%
  \BibitemOpen
  \bibfield  {author} {\bibinfo {author} {\bibfnamefont {S.}~\bibnamefont
  {Hofmann}}, \bibinfo {author} {\bibfnamefont {S.}~\bibnamefont {Heinz}},
  \bibinfo {author} {\bibfnamefont {R.}~\bibnamefont {Mann}}, \bibinfo {author}
  {\bibfnamefont {J.}~\bibnamefont {Maurer}}, \bibinfo {author} {\bibfnamefont
  {G.}~\bibnamefont {M{\"u}nzenberg}}, \bibinfo {author} {\bibfnamefont
  {S.}~\bibnamefont {Antalic}}, \bibinfo {author} {\bibfnamefont
  {W.}~\bibnamefont {Barth}}, \bibinfo {author} {\bibfnamefont
  {H.}~\bibnamefont {Burkhard}}, \bibinfo {author} {\bibfnamefont
  {L.}~\bibnamefont {Dahl}}, \bibinfo {author} {\bibfnamefont {K.}~\bibnamefont
  {Eberhardt}}, \emph {et~al.},\ }\href@noop {} {\bibfield  {journal} {\bibinfo
   {journal} {The European Physical Journal A}\ }\textbf {\bibinfo {volume}
  {52}},\ \bibinfo {pages} {180} (\bibinfo {year}
  {2016}{\natexlab{a}})}\BibitemShut {NoStop}%
\bibitem [{\citenamefont {Giuliani}\ \emph {et~al.}(2019)\citenamefont
  {Giuliani}, \citenamefont {Matheson}, \citenamefont {Nazarewicz},
  \citenamefont {Olsen}, \citenamefont {Reinhard}, \citenamefont {Sadhukhan},
  \citenamefont {Schuetrumpf}, \citenamefont {Schunck},\ and\ \citenamefont
  {Schwerdtfeger}}]{giuliani2019colloquium}%
  \BibitemOpen
  \bibfield  {author} {\bibinfo {author} {\bibfnamefont {S.}~\bibnamefont
  {Giuliani}}, \bibinfo {author} {\bibfnamefont {Z.}~\bibnamefont {Matheson}},
  \bibinfo {author} {\bibfnamefont {W.}~\bibnamefont {Nazarewicz}}, \bibinfo
  {author} {\bibfnamefont {E.}~\bibnamefont {Olsen}}, \bibinfo {author}
  {\bibfnamefont {P.-G.}\ \bibnamefont {Reinhard}}, \bibinfo {author}
  {\bibfnamefont {J.}~\bibnamefont {Sadhukhan}}, \bibinfo {author}
  {\bibfnamefont {B.}~\bibnamefont {Schuetrumpf}}, \bibinfo {author}
  {\bibfnamefont {N.}~\bibnamefont {Schunck}},\ and\ \bibinfo {author}
  {\bibfnamefont {P.}~\bibnamefont {Schwerdtfeger}},\ }\href@noop {} {\bibfield
   {journal} {\bibinfo  {journal} {Reviews of Modern Physics}\ }\textbf
  {\bibinfo {volume} {91}},\ \bibinfo {pages} {011001} (\bibinfo {year}
  {2019})}\BibitemShut {NoStop}%
\bibitem [{\citenamefont {Haba}(2019)}]{haba2019new}%
  \BibitemOpen
  \bibfield  {author} {\bibinfo {author} {\bibfnamefont {H.}~\bibnamefont
  {Haba}},\ }\href@noop {} {\bibfield  {journal} {\bibinfo  {journal} {Nature
  Chemistry}\ }\textbf {\bibinfo {volume} {11}},\ \bibinfo {pages} {10}
  (\bibinfo {year} {2019})}\BibitemShut {NoStop}%
\bibitem [{\citenamefont {Novikov}\ \emph {et~al.}(2020)\citenamefont
  {Novikov}, \citenamefont {Kozulin}, \citenamefont {Knyazheva}, \citenamefont
  {Itkis}, \citenamefont {Karpov}, \citenamefont {Itkis}, \citenamefont
  {Diatlov}, \citenamefont {Cheralu}, \citenamefont {Gall}, \citenamefont
  {Asfari} \emph {et~al.}}]{novikov2020formation}%
  \BibitemOpen
  \bibfield  {author} {\bibinfo {author} {\bibfnamefont {K.}~\bibnamefont
  {Novikov}}, \bibinfo {author} {\bibfnamefont {E.}~\bibnamefont {Kozulin}},
  \bibinfo {author} {\bibfnamefont {G.}~\bibnamefont {Knyazheva}}, \bibinfo
  {author} {\bibfnamefont {I.}~\bibnamefont {Itkis}}, \bibinfo {author}
  {\bibfnamefont {A.}~\bibnamefont {Karpov}}, \bibinfo {author} {\bibfnamefont
  {M.}~\bibnamefont {Itkis}}, \bibinfo {author} {\bibfnamefont
  {I.}~\bibnamefont {Diatlov}}, \bibinfo {author} {\bibfnamefont
  {M.}~\bibnamefont {Cheralu}}, \bibinfo {author} {\bibfnamefont
  {B.}~\bibnamefont {Gall}}, \bibinfo {author} {\bibfnamefont {Z.}~\bibnamefont
  {Asfari}}, \emph {et~al.},\ }\href@noop {} {\bibfield  {journal} {\bibinfo
  {journal} {Bulletin of the Russian Academy of Sciences: Physics}\ }\textbf
  {\bibinfo {volume} {84}},\ \bibinfo {pages} {495} (\bibinfo {year}
  {2020})}\BibitemShut {NoStop}%
\bibitem [{\citenamefont {Hamilton}\ \emph {et~al.}(2013)\citenamefont
  {Hamilton}, \citenamefont {Hofmann},\ and\ \citenamefont
  {Oganessian}}]{hamilton2013search}%
  \BibitemOpen
  \bibfield  {author} {\bibinfo {author} {\bibfnamefont {J.~H.}\ \bibnamefont
  {Hamilton}}, \bibinfo {author} {\bibfnamefont {S.}~\bibnamefont {Hofmann}},\
  and\ \bibinfo {author} {\bibfnamefont {Y.~T.}\ \bibnamefont {Oganessian}},\
  }\href@noop {} {\bibfield  {journal} {\bibinfo  {journal} {Annual Review of
  Nuclear and Particle Science}\ }\textbf {\bibinfo {volume} {63}},\ \bibinfo
  {pages} {383} (\bibinfo {year} {2013})}\BibitemShut {NoStop}%
\bibitem [{\citenamefont {Lesko}\ \emph {et~al.}(1983)\citenamefont {Lesko},
  \citenamefont {Gil}, \citenamefont {Lazzarini}, \citenamefont {Metag},
  \citenamefont {Seamster},\ and\ \citenamefont
  {Vandenbosch}}]{lesko1983properties}%
  \BibitemOpen
  \bibfield  {author} {\bibinfo {author} {\bibfnamefont {K.}~\bibnamefont
  {Lesko}}, \bibinfo {author} {\bibfnamefont {S.}~\bibnamefont {Gil}}, \bibinfo
  {author} {\bibfnamefont {A.}~\bibnamefont {Lazzarini}}, \bibinfo {author}
  {\bibfnamefont {V.}~\bibnamefont {Metag}}, \bibinfo {author} {\bibfnamefont
  {A.}~\bibnamefont {Seamster}},\ and\ \bibinfo {author} {\bibfnamefont
  {R.}~\bibnamefont {Vandenbosch}},\ }\href@noop {} {\bibfield  {journal}
  {\bibinfo  {journal} {Physical Review C}\ }\textbf {\bibinfo {volume} {27}},\
  \bibinfo {pages} {2999} (\bibinfo {year} {1983})}\BibitemShut {NoStop}%
\bibitem [{\citenamefont {Fazio}\ \emph {et~al.}(2004)\citenamefont {Fazio},
  \citenamefont {Giardina}, \citenamefont {Lamberto}, \citenamefont {Ruggeri},
  \citenamefont {Sacc{\'a}}, \citenamefont {Palamara}, \citenamefont {Muminov},
  \citenamefont {Nasirov}, \citenamefont {Yakhshiev}, \citenamefont {Hanappe}
  \emph {et~al.}}]{fazio2004formation}%
  \BibitemOpen
  \bibfield  {author} {\bibinfo {author} {\bibfnamefont {G.}~\bibnamefont
  {Fazio}}, \bibinfo {author} {\bibfnamefont {G.}~\bibnamefont {Giardina}},
  \bibinfo {author} {\bibfnamefont {A.}~\bibnamefont {Lamberto}}, \bibinfo
  {author} {\bibfnamefont {R.}~\bibnamefont {Ruggeri}}, \bibinfo {author}
  {\bibfnamefont {C.}~\bibnamefont {Sacc{\'a}}}, \bibinfo {author}
  {\bibfnamefont {R.}~\bibnamefont {Palamara}}, \bibinfo {author}
  {\bibfnamefont {A.}~\bibnamefont {Muminov}}, \bibinfo {author} {\bibfnamefont
  {A.}~\bibnamefont {Nasirov}}, \bibinfo {author} {\bibfnamefont
  {U.}~\bibnamefont {Yakhshiev}}, \bibinfo {author} {\bibfnamefont
  {F.}~\bibnamefont {Hanappe}}, \emph {et~al.},\ }\href@noop {} {\bibfield
  {journal} {\bibinfo  {journal} {The European Physical Journal A-Hadrons and
  Nuclei}\ }\textbf {\bibinfo {volume} {19}},\ \bibinfo {pages} {89} (\bibinfo
  {year} {2004})}\BibitemShut {NoStop}%
\bibitem [{\citenamefont {Mandaglio}\ \emph {et~al.}(2018)\citenamefont
  {Mandaglio}, \citenamefont {Anastasi}, \citenamefont {Curciarello},
  \citenamefont {Fazio}, \citenamefont {Giardina},\ and\ \citenamefont
  {Nasirov}}]{mandaglio2018effects}%
  \BibitemOpen
  \bibfield  {author} {\bibinfo {author} {\bibfnamefont {G.}~\bibnamefont
  {Mandaglio}}, \bibinfo {author} {\bibfnamefont {A.}~\bibnamefont {Anastasi}},
  \bibinfo {author} {\bibfnamefont {F.}~\bibnamefont {Curciarello}}, \bibinfo
  {author} {\bibfnamefont {G.}~\bibnamefont {Fazio}}, \bibinfo {author}
  {\bibfnamefont {G.}~\bibnamefont {Giardina}},\ and\ \bibinfo {author}
  {\bibfnamefont {A.}~\bibnamefont {Nasirov}},\ }\href@noop {} {\bibfield
  {journal} {\bibinfo  {journal} {Physical Review C}\ }\textbf {\bibinfo
  {volume} {98}},\ \bibinfo {pages} {044616} (\bibinfo {year}
  {2018})}\BibitemShut {NoStop}%
\bibitem [{\citenamefont {Ramamurthy}\ and\ \citenamefont
  {Kapoor}(1985)}]{ramamurthy1985interpretation}%
  \BibitemOpen
  \bibfield  {author} {\bibinfo {author} {\bibfnamefont {V.}~\bibnamefont
  {Ramamurthy}}\ and\ \bibinfo {author} {\bibfnamefont {S.}~\bibnamefont
  {Kapoor}},\ }\href@noop {} {\bibfield  {journal} {\bibinfo  {journal}
  {Physical Review Letters}\ }\textbf {\bibinfo {volume} {54}},\ \bibinfo
  {pages} {178} (\bibinfo {year} {1985})}\BibitemShut {NoStop}%
\bibitem [{\citenamefont {Soheyli}\ and\ \citenamefont
  {Khalili}(2012)}]{soheyli2012non}%
  \BibitemOpen
  \bibfield  {author} {\bibinfo {author} {\bibfnamefont {S.}~\bibnamefont
  {Soheyli}}\ and\ \bibinfo {author} {\bibfnamefont {M.}~\bibnamefont
  {Khalili}},\ }\href@noop {} {\bibfield  {journal} {\bibinfo  {journal}
  {Physical Review C}\ }\textbf {\bibinfo {volume} {85}},\ \bibinfo {pages}
  {034610} (\bibinfo {year} {2012})}\BibitemShut {NoStop}%
\bibitem [{\citenamefont {Nhan~Hao}\ \emph {et~al.}(2019)\citenamefont
  {Nhan~Hao}, \citenamefont {Duy}, \citenamefont {Chae}, \citenamefont
  {Quang~Hung},\ and\ \citenamefont {Nhu~Le}}]{nhan2019investigation}%
  \BibitemOpen
  \bibfield  {author} {\bibinfo {author} {\bibfnamefont {T.}~\bibnamefont
  {Nhan~Hao}}, \bibinfo {author} {\bibfnamefont {N.}~\bibnamefont {Duy}},
  \bibinfo {author} {\bibfnamefont {K.}~\bibnamefont {Chae}}, \bibinfo {author}
  {\bibfnamefont {N.}~\bibnamefont {Quang~Hung}},\ and\ \bibinfo {author}
  {\bibfnamefont {N.}~\bibnamefont {Nhu~Le}},\ }\href@noop {} {\bibfield
  {journal} {\bibinfo  {journal} {International Journal of Modern Physics E}\
  }\textbf {\bibinfo {volume} {28}},\ \bibinfo {pages} {1950056} (\bibinfo
  {year} {2019})}\BibitemShut {NoStop}%
\bibitem [{\citenamefont {Giardina}\ \emph {et~al.}(2000)\citenamefont
  {Giardina}, \citenamefont {Hofmann}, \citenamefont {Muminov},\ and\
  \citenamefont {Nasirov}}]{giardina2000effect}%
  \BibitemOpen
  \bibfield  {author} {\bibinfo {author} {\bibfnamefont {G.}~\bibnamefont
  {Giardina}}, \bibinfo {author} {\bibfnamefont {S.}~\bibnamefont {Hofmann}},
  \bibinfo {author} {\bibfnamefont {A.}~\bibnamefont {Muminov}},\ and\ \bibinfo
  {author} {\bibfnamefont {A.}~\bibnamefont {Nasirov}},\ }\href@noop {}
  {\bibfield  {journal} {\bibinfo  {journal} {The European Physical Journal A}\
  }\textbf {\bibinfo {volume} {8}},\ \bibinfo {pages} {205} (\bibinfo {year}
  {2000})}\BibitemShut {NoStop}%
\bibitem [{\citenamefont {Manjunatha}\ \emph {et~al.}(2020)\citenamefont
  {Manjunatha}, \citenamefont {Sowmya}, \citenamefont {Manjunatha},
  \citenamefont {Damodara~Gupta}, \citenamefont {Seenappa}, \citenamefont
  {Sridhar}, \citenamefont {Ganesh},\ and\ \citenamefont
  {Nandi}}]{manjunath2020entrance}%
  \BibitemOpen
  \bibfield  {author} {\bibinfo {author} {\bibfnamefont {H.~C.}\ \bibnamefont
  {Manjunatha}}, \bibinfo {author} {\bibfnamefont {N.}~\bibnamefont {Sowmya}},
  \bibinfo {author} {\bibfnamefont {N.}~\bibnamefont {Manjunatha}}, \bibinfo
  {author} {\bibfnamefont {P.~S.}\ \bibnamefont {Damodara~Gupta}}, \bibinfo
  {author} {\bibfnamefont {L.}~\bibnamefont {Seenappa}}, \bibinfo {author}
  {\bibfnamefont {K.~N.}\ \bibnamefont {Sridhar}}, \bibinfo {author}
  {\bibfnamefont {T.}~\bibnamefont {Ganesh}},\ and\ \bibinfo {author}
  {\bibfnamefont {T.}~\bibnamefont {Nandi}},\ }\href@noop {} {\bibfield
  {journal} {\bibinfo  {journal} {Physical Review C}\ }\textbf {\bibinfo
  {volume} {102}},\ \bibinfo {pages} {064605} (\bibinfo {year}
  {2020})}\BibitemShut {NoStop}%
\bibitem [{\citenamefont {Itkis}\ \emph {et~al.}(2015)\citenamefont {Itkis},
  \citenamefont {Vardaci}, \citenamefont {Itkis}, \citenamefont {Knyazheva},\
  and\ \citenamefont {Kozulin}}]{itkis2015fusion}%
  \BibitemOpen
  \bibfield  {author} {\bibinfo {author} {\bibfnamefont {M.}~\bibnamefont
  {Itkis}}, \bibinfo {author} {\bibfnamefont {E.}~\bibnamefont {Vardaci}},
  \bibinfo {author} {\bibfnamefont {I.}~\bibnamefont {Itkis}}, \bibinfo
  {author} {\bibfnamefont {G.}~\bibnamefont {Knyazheva}},\ and\ \bibinfo
  {author} {\bibfnamefont {E.}~\bibnamefont {Kozulin}},\ }\href@noop {}
  {\bibfield  {journal} {\bibinfo  {journal} {Nuclear Physics A}\ }\textbf
  {\bibinfo {volume} {944}},\ \bibinfo {pages} {204} (\bibinfo {year}
  {2015})}\BibitemShut {NoStop}%
\bibitem [{\citenamefont {Itkis}\ \emph {et~al.}(2008)\citenamefont {Itkis},
  \citenamefont {Bogachev}, \citenamefont {Itkis}, \citenamefont {Kliman},
  \citenamefont {Knyazheva}, \citenamefont {Kondratiev}, \citenamefont
  {Kozulin}, \citenamefont {Krupa}, \citenamefont {Ts.~OGANESSIAN},
  \citenamefont {Pokrovsky} \emph {et~al.}}]{itkis2008processes}%
  \BibitemOpen
  \bibfield  {author} {\bibinfo {author} {\bibfnamefont {M.~G.}\ \bibnamefont
  {Itkis}}, \bibinfo {author} {\bibfnamefont {A.~A.}\ \bibnamefont {Bogachev}},
  \bibinfo {author} {\bibfnamefont {I.~M.}\ \bibnamefont {Itkis}}, \bibinfo
  {author} {\bibfnamefont {J.}~\bibnamefont {Kliman}}, \bibinfo {author}
  {\bibfnamefont {G.~N.}\ \bibnamefont {Knyazheva}}, \bibinfo {author}
  {\bibfnamefont {N.~A.}\ \bibnamefont {Kondratiev}}, \bibinfo {author}
  {\bibfnamefont {E.}~\bibnamefont {Kozulin}}, \bibinfo {author} {\bibfnamefont
  {L.}~\bibnamefont {Krupa}}, \bibinfo {author} {\bibfnamefont
  {Y.}~\bibnamefont {Ts.~OGANESSIAN}}, \bibinfo {author} {\bibfnamefont
  {I.~V.}\ \bibnamefont {Pokrovsky}}, \emph {et~al.},\ }\href@noop {}
  {\bibfield  {journal} {\bibinfo  {journal} {Dynamical Aspects Of Nuclear
  Fission}\ }\textbf {\bibinfo {volume} {$10.1142/9789812837530_0004$}},\
  \bibinfo {pages} {36} (\bibinfo {year} {2008})}\BibitemShut {NoStop}%
\bibitem [{\citenamefont {Schmitt}\ \emph {et~al.}(2019)\citenamefont
  {Schmitt}, \citenamefont {Mazurek},\ and\ \citenamefont
  {Nadtochy}}]{schmitt2019new}%
  \BibitemOpen
  \bibfield  {author} {\bibinfo {author} {\bibfnamefont {C.}~\bibnamefont
  {Schmitt}}, \bibinfo {author} {\bibfnamefont {K.}~\bibnamefont {Mazurek}},\
  and\ \bibinfo {author} {\bibfnamefont {P.~N.}\ \bibnamefont {Nadtochy}},\
  }\href@noop {} {\bibfield  {journal} {\bibinfo  {journal} {Physical Review
  C}\ }\textbf {\bibinfo {volume} {100}},\ \bibinfo {pages} {064606} (\bibinfo
  {year} {2019})}\BibitemShut {NoStop}%
\bibitem [{\citenamefont {Sekizawa}\ and\ \citenamefont
  {Hagino}(2019)}]{sekizawa2019time}%
  \BibitemOpen
  \bibfield  {author} {\bibinfo {author} {\bibfnamefont {K.}~\bibnamefont
  {Sekizawa}}\ and\ \bibinfo {author} {\bibfnamefont {K.}~\bibnamefont
  {Hagino}},\ }\href@noop {} {\bibfield  {journal} {\bibinfo  {journal}
  {Physical Review C}\ }\textbf {\bibinfo {volume} {99}},\ \bibinfo {pages}
  {051602} (\bibinfo {year} {2019})}\BibitemShut {NoStop}%
\bibitem [{\citenamefont {Sridhar}\ \emph {et~al.}(2018)\citenamefont
  {Sridhar}, \citenamefont {Manjunatha},\ and\ \citenamefont
  {Ramalingam}}]{sridhar2018search}%
  \BibitemOpen
  \bibfield  {author} {\bibinfo {author} {\bibfnamefont {K.~N.}\ \bibnamefont
  {Sridhar}}, \bibinfo {author} {\bibfnamefont {H.~C.}\ \bibnamefont
  {Manjunatha}},\ and\ \bibinfo {author} {\bibfnamefont {H.}~\bibnamefont
  {Ramalingam}},\ }\href@noop {} {\bibfield  {journal} {\bibinfo  {journal}
  {Physical Review C}\ }\textbf {\bibinfo {volume} {98}},\ \bibinfo {pages}
  {064605} (\bibinfo {year} {2018})}\BibitemShut {NoStop}%
\bibitem [{\citenamefont {Sridhar}\ \emph {et~al.}(2019)\citenamefont
  {Sridhar}, \citenamefont {Manjunatha},\ and\ \citenamefont
  {Ramalingam}}]{sridhar2019studies}%
  \BibitemOpen
  \bibfield  {author} {\bibinfo {author} {\bibfnamefont {K.~N.}\ \bibnamefont
  {Sridhar}}, \bibinfo {author} {\bibfnamefont {H.~C.}\ \bibnamefont
  {Manjunatha}},\ and\ \bibinfo {author} {\bibfnamefont {H.~B.}\ \bibnamefont
  {Ramalingam}},\ }\href@noop {} {\bibfield  {journal} {\bibinfo  {journal}
  {Indian Journal of Physics}\ ,\ \bibinfo {pages} {1}} (\bibinfo {year}
  {2019})}\BibitemShut {NoStop}%
\bibitem [{\citenamefont {Manjunatha}\ \emph {et~al.}(2018)\citenamefont
  {Manjunatha}, \citenamefont {Sridhar},\ and\ \citenamefont
  {Sowmya}}]{manjunatha2018investigations}%
  \BibitemOpen
  \bibfield  {author} {\bibinfo {author} {\bibfnamefont {H.~C.}\ \bibnamefont
  {Manjunatha}}, \bibinfo {author} {\bibfnamefont {K.~N.}\ \bibnamefont
  {Sridhar}},\ and\ \bibinfo {author} {\bibfnamefont {N.}~\bibnamefont
  {Sowmya}},\ }\href@noop {} {\bibfield  {journal} {\bibinfo  {journal}
  {Physical Review C}\ }\textbf {\bibinfo {volume} {98}},\ \bibinfo {pages}
  {024308} (\bibinfo {year} {2018})}\BibitemShut {NoStop}%
\bibitem [{\citenamefont {Hofmann}\ \emph
  {et~al.}(2016{\natexlab{b}})\citenamefont {Hofmann}, \citenamefont {Heinz},
  \citenamefont {Mann}, \citenamefont {Maurer}, \citenamefont {M{\"u}nzenberg},
  \citenamefont {Antalic}, \citenamefont {Barth}, \citenamefont {Dahl},
  \citenamefont {Eberhardt}, \citenamefont {Grzywacz} \emph
  {et~al.}}]{hofmann2016remarks}%
  \BibitemOpen
  \bibfield  {author} {\bibinfo {author} {\bibfnamefont {S.}~\bibnamefont
  {Hofmann}}, \bibinfo {author} {\bibfnamefont {S.}~\bibnamefont {Heinz}},
  \bibinfo {author} {\bibfnamefont {R.}~\bibnamefont {Mann}}, \bibinfo {author}
  {\bibfnamefont {J.}~\bibnamefont {Maurer}}, \bibinfo {author} {\bibfnamefont
  {G.}~\bibnamefont {M{\"u}nzenberg}}, \bibinfo {author} {\bibfnamefont
  {S.}~\bibnamefont {Antalic}}, \bibinfo {author} {\bibfnamefont
  {W.}~\bibnamefont {Barth}}, \bibinfo {author} {\bibfnamefont
  {L.}~\bibnamefont {Dahl}}, \bibinfo {author} {\bibfnamefont {K.}~\bibnamefont
  {Eberhardt}}, \bibinfo {author} {\bibfnamefont {R.}~\bibnamefont {Grzywacz}},
  \emph {et~al.},\ }\href@noop {} {\bibfield  {journal} {\bibinfo  {journal}
  {The European Physical Journal A}\ }\textbf {\bibinfo {volume} {52}},\
  \bibinfo {pages} {116} (\bibinfo {year} {2016}{\natexlab{b}})}\BibitemShut
  {NoStop}%
\bibitem [{SM()}]{SM}%
  \BibitemOpen
  \href@noop {} {\bibinfo  {journal} {See Supplemental Material for (a)
  Definitions of entrance channel parameters, (b) Advance Statistical Model and
  (c) Dinuclear System Model at http://link.aps.org/supplemental/...........}\
  }\BibitemShut {NoStop}%
\bibitem [{\citenamefont {Gehlot}\ \emph {et~al.}(2019)\citenamefont {Gehlot},
  \citenamefont {Vinodkumar}, \citenamefont {Madhavan}, \citenamefont {Nath},
  \citenamefont {Jhingan}, \citenamefont {Varughese}, \citenamefont {Banerjee},
  \citenamefont {Shamlath}, \citenamefont {Laveen}, \citenamefont {Shareef}
  \emph {et~al.}}]{gehlot2019evaporation}%
  \BibitemOpen
\bibfield  {journal} {  }\bibfield  {author} {\bibinfo {author} {\bibfnamefont
  {J.}~\bibnamefont {Gehlot}}, \bibinfo {author} {\bibfnamefont
  {A.}~\bibnamefont {Vinodkumar}}, \bibinfo {author} {\bibfnamefont
  {N.}~\bibnamefont {Madhavan}}, \bibinfo {author} {\bibfnamefont
  {S.}~\bibnamefont {Nath}}, \bibinfo {author} {\bibfnamefont {A.}~\bibnamefont
  {Jhingan}}, \bibinfo {author} {\bibfnamefont {T.}~\bibnamefont {Varughese}},
  \bibinfo {author} {\bibfnamefont {T.}~\bibnamefont {Banerjee}}, \bibinfo
  {author} {\bibfnamefont {A.}~\bibnamefont {Shamlath}}, \bibinfo {author}
  {\bibfnamefont {P.}~\bibnamefont {Laveen}}, \bibinfo {author} {\bibfnamefont
  {M.}~\bibnamefont {Shareef}}, \emph {et~al.},\ }\href@noop {} {\bibfield
  {journal} {\bibinfo  {journal} {Physical Review C}\ }\textbf {\bibinfo
  {volume} {99}},\ \bibinfo {pages} {034615} (\bibinfo {year}
  {2019})}\BibitemShut {NoStop}%
\bibitem [{\citenamefont {Lukyanov}\ \emph {et~al.}(2009)\citenamefont
  {Lukyanov}, \citenamefont {Penionzhkevich}, \citenamefont {Astabatian},
  \citenamefont {Demekhina}, \citenamefont {Dlouhy}, \citenamefont {Ivanov},
  \citenamefont {Kalpakchieva}, \citenamefont {Kulko}, \citenamefont
  {Markarian}, \citenamefont {Maslov} \emph {et~al.}}]{lukyanov2009study}%
  \BibitemOpen
  \bibfield  {author} {\bibinfo {author} {\bibfnamefont {S.}~\bibnamefont
  {Lukyanov}}, \bibinfo {author} {\bibfnamefont {Y.~E.}\ \bibnamefont
  {Penionzhkevich}}, \bibinfo {author} {\bibfnamefont {R.}~\bibnamefont
  {Astabatian}}, \bibinfo {author} {\bibfnamefont {N.}~\bibnamefont
  {Demekhina}}, \bibinfo {author} {\bibfnamefont {Z.}~\bibnamefont {Dlouhy}},
  \bibinfo {author} {\bibfnamefont {M.}~\bibnamefont {Ivanov}}, \bibinfo
  {author} {\bibfnamefont {R.}~\bibnamefont {Kalpakchieva}}, \bibinfo {author}
  {\bibfnamefont {A.}~\bibnamefont {Kulko}}, \bibinfo {author} {\bibfnamefont
  {E.}~\bibnamefont {Markarian}}, \bibinfo {author} {\bibfnamefont
  {V.}~\bibnamefont {Maslov}}, \emph {et~al.},\ }\href@noop {} {\bibfield
  {journal} {\bibinfo  {journal} {Physics Letters B}\ }\textbf {\bibinfo
  {volume} {670}},\ \bibinfo {pages} {321} (\bibinfo {year}
  {2009})}\BibitemShut {NoStop}%
\bibitem [{\citenamefont {Ramler}\ \emph {et~al.}(1959)\citenamefont {Ramler},
  \citenamefont {Wing}, \citenamefont {Henderson},\ and\ \citenamefont
  {Huizenga}}]{ramler1959excitation}%
  \BibitemOpen
  \bibfield  {author} {\bibinfo {author} {\bibfnamefont {W.}~\bibnamefont
  {Ramler}}, \bibinfo {author} {\bibfnamefont {J.}~\bibnamefont {Wing}},
  \bibinfo {author} {\bibfnamefont {D.}~\bibnamefont {Henderson}},\ and\
  \bibinfo {author} {\bibfnamefont {J.}~\bibnamefont {Huizenga}},\ }\href@noop
  {} {\bibfield  {journal} {\bibinfo  {journal} {Physical Review}\ }\textbf
  {\bibinfo {volume} {114}},\ \bibinfo {pages} {154} (\bibinfo {year}
  {1959})}\BibitemShut {NoStop}%
\bibitem [{\citenamefont {Vinodkumar}\ \emph {et~al.}(2009)\citenamefont
  {Vinodkumar}, \citenamefont {Loveland}, \citenamefont {Sprunger},
  \citenamefont {Prisbrey}, \citenamefont {Trinczek}, \citenamefont {Dombsky},
  \citenamefont {Machule}, \citenamefont {Kolata},\ and\ \citenamefont
  {Roberts}}]{vinodkumar2009fusion}%
  \BibitemOpen
  \bibfield  {author} {\bibinfo {author} {\bibfnamefont {A.}~\bibnamefont
  {Vinodkumar}}, \bibinfo {author} {\bibfnamefont {W.}~\bibnamefont
  {Loveland}}, \bibinfo {author} {\bibfnamefont {P.}~\bibnamefont {Sprunger}},
  \bibinfo {author} {\bibfnamefont {L.}~\bibnamefont {Prisbrey}}, \bibinfo
  {author} {\bibfnamefont {M.}~\bibnamefont {Trinczek}}, \bibinfo {author}
  {\bibfnamefont {M.}~\bibnamefont {Dombsky}}, \bibinfo {author} {\bibfnamefont
  {P.}~\bibnamefont {Machule}}, \bibinfo {author} {\bibfnamefont
  {J.}~\bibnamefont {Kolata}},\ and\ \bibinfo {author} {\bibfnamefont
  {A.}~\bibnamefont {Roberts}},\ }\href@noop {} {\bibfield  {journal} {\bibinfo
   {journal} {Physical Review-Section C-Nuclear Physics}\ }\textbf {\bibinfo
  {volume} {22}},\ \bibinfo {pages} {54609} (\bibinfo {year}
  {2009})}\BibitemShut {NoStop}%
\bibitem [{\citenamefont {Penionzhkevich}\ \emph {et~al.}(2008)\citenamefont
  {Penionzhkevich}, \citenamefont {Lukyanov}, \citenamefont {Astabatyan},
  \citenamefont {Demekhina}, \citenamefont {Ivanov}, \citenamefont
  {Kalpakchieva}, \citenamefont {Kulko}, \citenamefont {Markaryan},
  \citenamefont {Maslov}, \citenamefont {Muzychka} \emph
  {et~al.}}]{penionzhkevich2008complete}%
  \BibitemOpen
  \bibfield  {author} {\bibinfo {author} {\bibfnamefont {Y.~E.}\ \bibnamefont
  {Penionzhkevich}}, \bibinfo {author} {\bibfnamefont {S.}~\bibnamefont
  {Lukyanov}}, \bibinfo {author} {\bibfnamefont {R.}~\bibnamefont
  {Astabatyan}}, \bibinfo {author} {\bibfnamefont {N.}~\bibnamefont
  {Demekhina}}, \bibinfo {author} {\bibfnamefont {M.}~\bibnamefont {Ivanov}},
  \bibinfo {author} {\bibfnamefont {R.}~\bibnamefont {Kalpakchieva}}, \bibinfo
  {author} {\bibfnamefont {A.}~\bibnamefont {Kulko}}, \bibinfo {author}
  {\bibfnamefont {E.}~\bibnamefont {Markaryan}}, \bibinfo {author}
  {\bibfnamefont {V.}~\bibnamefont {Maslov}}, \bibinfo {author} {\bibfnamefont
  {Y.~A.}\ \bibnamefont {Muzychka}}, \emph {et~al.},\ }\href@noop {} {\bibfield
   {journal} {\bibinfo  {journal} {Journal of Physics G: Nuclear and Particle
  Physics}\ }\textbf {\bibinfo {volume} {36}},\ \bibinfo {pages} {025104}
  (\bibinfo {year} {2008})}\BibitemShut {NoStop}%
\bibitem [{\citenamefont {Dasgupta}\ \emph {et~al.}(2010)\citenamefont
  {Dasgupta}, \citenamefont {Hinde}, \citenamefont {Sheehy},\ and\
  \citenamefont {Bouriquet}}]{dasgupta2010suppression}%
  \BibitemOpen
  \bibfield  {author} {\bibinfo {author} {\bibfnamefont {M.}~\bibnamefont
  {Dasgupta}}, \bibinfo {author} {\bibfnamefont {D.}~\bibnamefont {Hinde}},
  \bibinfo {author} {\bibfnamefont {S.}~\bibnamefont {Sheehy}},\ and\ \bibinfo
  {author} {\bibfnamefont {B.}~\bibnamefont {Bouriquet}},\ }\href@noop {}
  {\bibfield  {journal} {\bibinfo  {journal} {Physical Review C}\ }\textbf
  {\bibinfo {volume} {81}},\ \bibinfo {pages} {024608} (\bibinfo {year}
  {2010})}\BibitemShut {NoStop}%
\bibitem [{\citenamefont {Signorini}\ \emph {et~al.}(2004)\citenamefont
  {Signorini}, \citenamefont {Yoshida}, \citenamefont {Watanabe}, \citenamefont
  {Pierroutsakou}, \citenamefont {Stroe}, \citenamefont {Fukuda}, \citenamefont
  {Mazzocco}, \citenamefont {Fukuda}, \citenamefont {Mizoi}, \citenamefont
  {Ishihara} \emph {et~al.}}]{signorini2004subbarrier}%
  \BibitemOpen
  \bibfield  {author} {\bibinfo {author} {\bibfnamefont {C.}~\bibnamefont
  {Signorini}}, \bibinfo {author} {\bibfnamefont {A.}~\bibnamefont {Yoshida}},
  \bibinfo {author} {\bibfnamefont {Y.}~\bibnamefont {Watanabe}}, \bibinfo
  {author} {\bibfnamefont {D.}~\bibnamefont {Pierroutsakou}}, \bibinfo {author}
  {\bibfnamefont {L.}~\bibnamefont {Stroe}}, \bibinfo {author} {\bibfnamefont
  {T.}~\bibnamefont {Fukuda}}, \bibinfo {author} {\bibfnamefont
  {M.}~\bibnamefont {Mazzocco}}, \bibinfo {author} {\bibfnamefont
  {N.}~\bibnamefont {Fukuda}}, \bibinfo {author} {\bibfnamefont
  {Y.}~\bibnamefont {Mizoi}}, \bibinfo {author} {\bibfnamefont
  {M.}~\bibnamefont {Ishihara}}, \emph {et~al.},\ }\href@noop {} {\bibfield
  {journal} {\bibinfo  {journal} {Nuclear Physics A}\ }\textbf {\bibinfo
  {volume} {735}},\ \bibinfo {pages} {329} (\bibinfo {year}
  {2004})}\BibitemShut {NoStop}%
\bibitem [{\citenamefont {Gasques}\ \emph {et~al.}(2009)\citenamefont
  {Gasques}, \citenamefont {Hinde}, \citenamefont {Dasgupta}, \citenamefont
  {Mukherjee},\ and\ \citenamefont {Thomas}}]{gasques2009suppression}%
  \BibitemOpen
  \bibfield  {author} {\bibinfo {author} {\bibfnamefont {L.}~\bibnamefont
  {Gasques}}, \bibinfo {author} {\bibfnamefont {D.}~\bibnamefont {Hinde}},
  \bibinfo {author} {\bibfnamefont {M.}~\bibnamefont {Dasgupta}}, \bibinfo
  {author} {\bibfnamefont {A.}~\bibnamefont {Mukherjee}},\ and\ \bibinfo
  {author} {\bibfnamefont {R.}~\bibnamefont {Thomas}},\ }\href@noop {}
  {\bibfield  {journal} {\bibinfo  {journal} {Physical Review C}\ }\textbf
  {\bibinfo {volume} {79}},\ \bibinfo {pages} {034605} (\bibinfo {year}
  {2009})}\BibitemShut {NoStop}%
\bibitem [{\citenamefont {{Le Beyec}}\ \emph {et~al.}(1972)\citenamefont {{Le
  Beyec}}, \citenamefont {Lefort},\ and\ \citenamefont
  {Sarda}}]{LEBEYEC1972405}%
  \BibitemOpen
  \bibfield  {author} {\bibinfo {author} {\bibfnamefont {Y.}~\bibnamefont {{Le
  Beyec}}}, \bibinfo {author} {\bibfnamefont {M.}~\bibnamefont {Lefort}},\ and\
  \bibinfo {author} {\bibfnamefont {M.}~\bibnamefont {Sarda}},\ }\href
  {https://doi.org/https://doi.org/10.1016/0375-9474(72)90268-0} {\bibfield
  {journal} {\bibinfo  {journal} {Nuclear Physics A}\ }\textbf {\bibinfo
  {volume} {192}},\ \bibinfo {pages} {405} (\bibinfo {year}
  {1972})}\BibitemShut {NoStop}%
\bibitem [{\citenamefont {Morton}\ \emph {et~al.}(1995)\citenamefont {Morton},
  \citenamefont {Hinde}, \citenamefont {Leigh}, \citenamefont {Lestone},
  \citenamefont {Dasgupta}, \citenamefont {Mein}, \citenamefont {Newton},\ and\
  \citenamefont {Timmers}}]{PhysRevC.52.243}%
  \BibitemOpen
  \bibfield  {author} {\bibinfo {author} {\bibfnamefont {C.~R.}\ \bibnamefont
  {Morton}}, \bibinfo {author} {\bibfnamefont {D.~J.}\ \bibnamefont {Hinde}},
  \bibinfo {author} {\bibfnamefont {J.~R.}\ \bibnamefont {Leigh}}, \bibinfo
  {author} {\bibfnamefont {J.~P.}\ \bibnamefont {Lestone}}, \bibinfo {author}
  {\bibfnamefont {M.}~\bibnamefont {Dasgupta}}, \bibinfo {author}
  {\bibfnamefont {J.~C.}\ \bibnamefont {Mein}}, \bibinfo {author}
  {\bibfnamefont {J.~O.}\ \bibnamefont {Newton}},\ and\ \bibinfo {author}
  {\bibfnamefont {H.}~\bibnamefont {Timmers}},\ }\href
  {https://doi.org/10.1103/PhysRevC.52.243} {\bibfield  {journal} {\bibinfo
  {journal} {Phys. Rev. C}\ }\textbf {\bibinfo {volume} {52}},\ \bibinfo
  {pages} {243} (\bibinfo {year} {1995})}\BibitemShut {NoStop}%
\bibitem [{\citenamefont {Oganessian}\ \emph
  {et~al.}(2001{\natexlab{a}})\citenamefont {Oganessian}, \citenamefont
  {Utyonkov}, \citenamefont {Lobanov}, \citenamefont {Abdullin}, \citenamefont
  {Polyakov}, \citenamefont {Shirokovsky}, \citenamefont {Tsyganov},
  \citenamefont {Mezentsev}, \citenamefont {Iliev}, \citenamefont {Subbotin},
  \citenamefont {Sukhov}, \citenamefont {Subotic}, \citenamefont {Ivanov},
  \citenamefont {Voinov}, \citenamefont {Zagrebaev}, \citenamefont {Moody},
  \citenamefont {Wild}, \citenamefont {Stoyer}, \citenamefont {Stoyer},\ and\
  \citenamefont {Lougheed}}]{PhysRevC.64.054606}%
  \BibitemOpen
  \bibfield  {author} {\bibinfo {author} {\bibfnamefont {Y.~T.}\ \bibnamefont
  {Oganessian}}, \bibinfo {author} {\bibfnamefont {V.~K.}\ \bibnamefont
  {Utyonkov}}, \bibinfo {author} {\bibfnamefont {Y.~V.}\ \bibnamefont
  {Lobanov}}, \bibinfo {author} {\bibfnamefont {F.~S.}\ \bibnamefont
  {Abdullin}}, \bibinfo {author} {\bibfnamefont {A.~N.}\ \bibnamefont
  {Polyakov}}, \bibinfo {author} {\bibfnamefont {I.~V.}\ \bibnamefont
  {Shirokovsky}}, \bibinfo {author} {\bibfnamefont {Y.~S.}\ \bibnamefont
  {Tsyganov}}, \bibinfo {author} {\bibfnamefont {A.~N.}\ \bibnamefont
  {Mezentsev}}, \bibinfo {author} {\bibfnamefont {S.}~\bibnamefont {Iliev}},
  \bibinfo {author} {\bibfnamefont {V.~G.}\ \bibnamefont {Subbotin}}, \bibinfo
  {author} {\bibfnamefont {A.~M.}\ \bibnamefont {Sukhov}}, \bibinfo {author}
  {\bibfnamefont {K.}~\bibnamefont {Subotic}}, \bibinfo {author} {\bibfnamefont
  {O.~V.}\ \bibnamefont {Ivanov}}, \bibinfo {author} {\bibfnamefont {A.~N.}\
  \bibnamefont {Voinov}}, \bibinfo {author} {\bibfnamefont {V.~I.}\
  \bibnamefont {Zagrebaev}}, \bibinfo {author} {\bibfnamefont {K.~J.}\
  \bibnamefont {Moody}}, \bibinfo {author} {\bibfnamefont {J.~F.}\ \bibnamefont
  {Wild}}, \bibinfo {author} {\bibfnamefont {N.~J.}\ \bibnamefont {Stoyer}},
  \bibinfo {author} {\bibfnamefont {M.~A.}\ \bibnamefont {Stoyer}},\ and\
  \bibinfo {author} {\bibfnamefont {R.~W.}\ \bibnamefont {Lougheed}},\ }\href
  {https://doi.org/10.1103/PhysRevC.64.054606} {\bibfield  {journal} {\bibinfo
  {journal} {Phys. Rev. C}\ }\textbf {\bibinfo {volume} {64}},\ \bibinfo
  {pages} {054606} (\bibinfo {year} {2001}{\natexlab{a}})}\BibitemShut
  {NoStop}%
\bibitem [{\citenamefont {Khuyagbaatar}\ \emph {et~al.}(2010)\citenamefont
  {Khuyagbaatar}, \citenamefont {He{\ss}berger}, \citenamefont {Hofmann},
  \citenamefont {Ackermann}, \citenamefont {Comas}, \citenamefont {Heinz},
  \citenamefont {Heredia}, \citenamefont {Kindler}, \citenamefont {Kojouharov},
  \citenamefont {Lommel} \emph {et~al.}}]{khuyagbaatar2010new}%
  \BibitemOpen
  \bibfield  {author} {\bibinfo {author} {\bibfnamefont {J.}~\bibnamefont
  {Khuyagbaatar}}, \bibinfo {author} {\bibfnamefont {F.}~\bibnamefont
  {He{\ss}berger}}, \bibinfo {author} {\bibfnamefont {S.}~\bibnamefont
  {Hofmann}}, \bibinfo {author} {\bibfnamefont {D.}~\bibnamefont {Ackermann}},
  \bibinfo {author} {\bibfnamefont {V.}~\bibnamefont {Comas}}, \bibinfo
  {author} {\bibfnamefont {S.}~\bibnamefont {Heinz}}, \bibinfo {author}
  {\bibfnamefont {J.}~\bibnamefont {Heredia}}, \bibinfo {author} {\bibfnamefont
  {B.}~\bibnamefont {Kindler}}, \bibinfo {author} {\bibfnamefont
  {I.}~\bibnamefont {Kojouharov}}, \bibinfo {author} {\bibfnamefont
  {B.}~\bibnamefont {Lommel}}, \emph {et~al.},\ }\href@noop {} {\bibfield
  {journal} {\bibinfo  {journal} {The European Physical Journal A}\ }\textbf
  {\bibinfo {volume} {46}},\ \bibinfo {pages} {59} (\bibinfo {year}
  {2010})}\BibitemShut {NoStop}%
\bibitem [{\citenamefont {G{\"a}ggeler}\ \emph {et~al.}(1989)\citenamefont
  {G{\"a}ggeler}, \citenamefont {Jost}, \citenamefont {T{\"u}rler},
  \citenamefont {Armbruster}, \citenamefont {Br{\"u}chle}, \citenamefont
  {Folger}, \citenamefont {He{\ss}berger}, \citenamefont {Hofmann},
  \citenamefont {M{\"u}nzenberg}, \citenamefont {Ninov} \emph
  {et~al.}}]{gaggeler1989cold}%
  \BibitemOpen
  \bibfield  {author} {\bibinfo {author} {\bibfnamefont {H.}~\bibnamefont
  {G{\"a}ggeler}}, \bibinfo {author} {\bibfnamefont {D.}~\bibnamefont {Jost}},
  \bibinfo {author} {\bibfnamefont {A.}~\bibnamefont {T{\"u}rler}}, \bibinfo
  {author} {\bibfnamefont {P.}~\bibnamefont {Armbruster}}, \bibinfo {author}
  {\bibfnamefont {W.}~\bibnamefont {Br{\"u}chle}}, \bibinfo {author}
  {\bibfnamefont {H.}~\bibnamefont {Folger}}, \bibinfo {author} {\bibfnamefont
  {F.}~\bibnamefont {He{\ss}berger}}, \bibinfo {author} {\bibfnamefont
  {S.}~\bibnamefont {Hofmann}}, \bibinfo {author} {\bibfnamefont
  {G.}~\bibnamefont {M{\"u}nzenberg}}, \bibinfo {author} {\bibfnamefont
  {V.}~\bibnamefont {Ninov}}, \emph {et~al.},\ }\href@noop {} {\bibfield
  {journal} {\bibinfo  {journal} {Nuclear Physics A}\ }\textbf {\bibinfo
  {volume} {502}},\ \bibinfo {pages} {561} (\bibinfo {year}
  {1989})}\BibitemShut {NoStop}%
\bibitem [{\citenamefont {Belozerov}\ \emph {et~al.}(2003)\citenamefont
  {Belozerov}, \citenamefont {Chelnokov}, \citenamefont {Chepigin},
  \citenamefont {Drobina}, \citenamefont {Gorshkov}, \citenamefont
  {Kabachenko}, \citenamefont {Malyshev}, \citenamefont {Merkin}, \citenamefont
  {Oganessian}, \citenamefont {Popeko} \emph
  {et~al.}}]{belozerov2003spontaneous}%
  \BibitemOpen
  \bibfield  {author} {\bibinfo {author} {\bibfnamefont {A.}~\bibnamefont
  {Belozerov}}, \bibinfo {author} {\bibfnamefont {M.}~\bibnamefont
  {Chelnokov}}, \bibinfo {author} {\bibfnamefont {V.}~\bibnamefont {Chepigin}},
  \bibinfo {author} {\bibfnamefont {T.}~\bibnamefont {Drobina}}, \bibinfo
  {author} {\bibfnamefont {V.}~\bibnamefont {Gorshkov}}, \bibinfo {author}
  {\bibfnamefont {A.}~\bibnamefont {Kabachenko}}, \bibinfo {author}
  {\bibfnamefont {O.}~\bibnamefont {Malyshev}}, \bibinfo {author}
  {\bibfnamefont {I.}~\bibnamefont {Merkin}}, \bibinfo {author} {\bibfnamefont
  {Y.~T.}\ \bibnamefont {Oganessian}}, \bibinfo {author} {\bibfnamefont
  {A.}~\bibnamefont {Popeko}}, \emph {et~al.},\ }\href@noop {} {\bibfield
  {journal} {\bibinfo  {journal} {The European Physical Journal A-Hadrons and
  Nuclei}\ }\textbf {\bibinfo {volume} {16}},\ \bibinfo {pages} {447} (\bibinfo
  {year} {2003})}\BibitemShut {NoStop}%
\bibitem [{\citenamefont {Oganessian}\ \emph
  {et~al.}(2001{\natexlab{b}})\citenamefont {Oganessian}, \citenamefont
  {Utyonkov}, \citenamefont {Lobanov}, \citenamefont {Abdullin}, \citenamefont
  {Polyakov}, \citenamefont {Shirokovsky}, \citenamefont {Tsyganov},
  \citenamefont {Mezentsev}, \citenamefont {Iliev}, \citenamefont {Subbotin}
  \emph {et~al.}}]{oganessian2001measurements}%
  \BibitemOpen
  \bibfield  {author} {\bibinfo {author} {\bibfnamefont {Y.~T.}\ \bibnamefont
  {Oganessian}}, \bibinfo {author} {\bibfnamefont {V.}~\bibnamefont
  {Utyonkov}}, \bibinfo {author} {\bibfnamefont {Y.~V.}\ \bibnamefont
  {Lobanov}}, \bibinfo {author} {\bibfnamefont {F.~S.}\ \bibnamefont
  {Abdullin}}, \bibinfo {author} {\bibfnamefont {A.}~\bibnamefont {Polyakov}},
  \bibinfo {author} {\bibfnamefont {I.}~\bibnamefont {Shirokovsky}}, \bibinfo
  {author} {\bibfnamefont {Y.~S.}\ \bibnamefont {Tsyganov}}, \bibinfo {author}
  {\bibfnamefont {A.}~\bibnamefont {Mezentsev}}, \bibinfo {author}
  {\bibfnamefont {S.}~\bibnamefont {Iliev}}, \bibinfo {author} {\bibfnamefont
  {V.}~\bibnamefont {Subbotin}}, \emph {et~al.},\ }\href@noop {} {\bibfield
  {journal} {\bibinfo  {journal} {Physical Review C}\ }\textbf {\bibinfo
  {volume} {64}},\ \bibinfo {pages} {054606} (\bibinfo {year}
  {2001}{\natexlab{b}})}\BibitemShut {NoStop}%
\bibitem [{\citenamefont {He{\ss}berger}\ \emph {et~al.}(2001)\citenamefont
  {He{\ss}berger}, \citenamefont {Hofmann}, \citenamefont {Ackermann},
  \citenamefont {Ninov}, \citenamefont {Leino}, \citenamefont {M{\"u}nzenberg},
  \citenamefont {Saro}, \citenamefont {Lavrentev}, \citenamefont {Popeko},
  \citenamefont {Yeremin} \emph {et~al.}}]{hessberger2001decay}%
  \BibitemOpen
  \bibfield  {author} {\bibinfo {author} {\bibfnamefont {F.}~\bibnamefont
  {He{\ss}berger}}, \bibinfo {author} {\bibfnamefont {S.}~\bibnamefont
  {Hofmann}}, \bibinfo {author} {\bibfnamefont {D.}~\bibnamefont {Ackermann}},
  \bibinfo {author} {\bibfnamefont {V.}~\bibnamefont {Ninov}}, \bibinfo
  {author} {\bibfnamefont {M.}~\bibnamefont {Leino}}, \bibinfo {author}
  {\bibfnamefont {G.}~\bibnamefont {M{\"u}nzenberg}}, \bibinfo {author}
  {\bibfnamefont {S.}~\bibnamefont {Saro}}, \bibinfo {author} {\bibfnamefont
  {A.}~\bibnamefont {Lavrentev}}, \bibinfo {author} {\bibfnamefont
  {A.}~\bibnamefont {Popeko}}, \bibinfo {author} {\bibfnamefont
  {A.}~\bibnamefont {Yeremin}}, \emph {et~al.},\ }\href@noop {} {\bibfield
  {journal} {\bibinfo  {journal} {The European Physical Journal A-Hadrons and
  Nuclei}\ }\textbf {\bibinfo {volume} {12}},\ \bibinfo {pages} {57} (\bibinfo
  {year} {2001})}\BibitemShut {NoStop}%
\bibitem [{\citenamefont {Andreev}\ \emph {et~al.}(1989)\citenamefont
  {Andreev}, \citenamefont {Bogdanov}, \citenamefont {Eremin}, \citenamefont
  {Kabachenko}, \citenamefont {Orlova}, \citenamefont {Ter-Akop'yan},\ and\
  \citenamefont {Chepigin}}]{andreev1989measurement}%
  \BibitemOpen
  \bibfield  {author} {\bibinfo {author} {\bibfnamefont {A.~N.}\ \bibnamefont
  {Andreev}}, \bibinfo {author} {\bibfnamefont {D.~D.}\ \bibnamefont
  {Bogdanov}}, \bibinfo {author} {\bibfnamefont {A.~V.}\ \bibnamefont
  {Eremin}}, \bibinfo {author} {\bibfnamefont {A.~P.}\ \bibnamefont
  {Kabachenko}}, \bibinfo {author} {\bibfnamefont {O.~A.}\ \bibnamefont
  {Orlova}}, \bibinfo {author} {\bibfnamefont {G.~M.}\ \bibnamefont
  {Ter-Akop'yan}},\ and\ \bibinfo {author} {\bibfnamefont {V.~I.}\ \bibnamefont
  {Chepigin}},\ }\href@noop {} {\bibfield  {journal} {\bibinfo  {journal}
  {Soviet Journal of Nuclear Physics (English Translation)}\ }\textbf {\bibinfo
  {volume} {50}},\ \bibinfo {pages} {381} (\bibinfo {year} {1989})}\BibitemShut
  {NoStop}%
\bibitem [{\citenamefont {Hofmann}(2004)}]{hofmann2004properties}%
  \BibitemOpen
  \bibfield  {author} {\bibinfo {author} {\bibfnamefont {S.}~\bibnamefont
  {Hofmann}},\ }\href@noop {} {\bibfield  {journal} {\bibinfo  {journal} {AIP
  Conference Proceedings}\ }\textbf {\bibinfo {volume} {704}},\ \bibinfo
  {pages} {21} (\bibinfo {year} {2004})}\BibitemShut {NoStop}%
\bibitem [{\citenamefont {M{\"u}nzenberg}\ \emph {et~al.}(1981)\citenamefont
  {M{\"u}nzenberg}, \citenamefont {Hofmann}, \citenamefont {He{\ss}berger},
  \citenamefont {Reisdorf}, \citenamefont {Schmidt}, \citenamefont {Schneider},
  \citenamefont {Armbruster}, \citenamefont {Sahm},\ and\ \citenamefont
  {Thuma}}]{munzenberg1981identification}%
  \BibitemOpen
  \bibfield  {author} {\bibinfo {author} {\bibfnamefont {G.}~\bibnamefont
  {M{\"u}nzenberg}}, \bibinfo {author} {\bibfnamefont {S.}~\bibnamefont
  {Hofmann}}, \bibinfo {author} {\bibfnamefont {F.}~\bibnamefont
  {He{\ss}berger}}, \bibinfo {author} {\bibfnamefont {W.}~\bibnamefont
  {Reisdorf}}, \bibinfo {author} {\bibfnamefont {K.}~\bibnamefont {Schmidt}},
  \bibinfo {author} {\bibfnamefont {J.}~\bibnamefont {Schneider}}, \bibinfo
  {author} {\bibfnamefont {P.}~\bibnamefont {Armbruster}}, \bibinfo {author}
  {\bibfnamefont {C.}~\bibnamefont {Sahm}},\ and\ \bibinfo {author}
  {\bibfnamefont {B.}~\bibnamefont {Thuma}},\ }\href@noop {} {\bibfield
  {journal} {\bibinfo  {journal} {Zeitschrift f{\"u}r Physik A Atoms and
  Nuclei}\ }\textbf {\bibinfo {volume} {300}},\ \bibinfo {pages} {107}
  (\bibinfo {year} {1981})}\BibitemShut {NoStop}%
\bibitem [{\citenamefont {M{\"u}nzenberg}\ \emph {et~al.}(1984)\citenamefont
  {M{\"u}nzenberg}, \citenamefont {Reisdorf}, \citenamefont {Hofmann},
  \citenamefont {Agarwal}, \citenamefont {He{\ss}berger}, \citenamefont
  {Poppensieker}, \citenamefont {Schneider}, \citenamefont {Schneider},
  \citenamefont {Schmidt}, \citenamefont {Sch{\"o}tt} \emph
  {et~al.}}]{munzenberg1984evidence}%
  \BibitemOpen
  \bibfield  {author} {\bibinfo {author} {\bibfnamefont {G.}~\bibnamefont
  {M{\"u}nzenberg}}, \bibinfo {author} {\bibfnamefont {W.}~\bibnamefont
  {Reisdorf}}, \bibinfo {author} {\bibfnamefont {S.}~\bibnamefont {Hofmann}},
  \bibinfo {author} {\bibfnamefont {Y.}~\bibnamefont {Agarwal}}, \bibinfo
  {author} {\bibfnamefont {F.}~\bibnamefont {He{\ss}berger}}, \bibinfo {author}
  {\bibfnamefont {K.}~\bibnamefont {Poppensieker}}, \bibinfo {author}
  {\bibfnamefont {J.}~\bibnamefont {Schneider}}, \bibinfo {author}
  {\bibfnamefont {W.}~\bibnamefont {Schneider}}, \bibinfo {author}
  {\bibfnamefont {K.-H.}\ \bibnamefont {Schmidt}}, \bibinfo {author}
  {\bibfnamefont {H.-J.}\ \bibnamefont {Sch{\"o}tt}}, \emph {et~al.},\
  }\href@noop {} {\bibfield  {journal} {\bibinfo  {journal} {Zeitschrift
  f{\"u}r Physik A Atoms and Nuclei}\ }\textbf {\bibinfo {volume} {315}},\
  \bibinfo {pages} {145} (\bibinfo {year} {1984})}\BibitemShut {NoStop}%
\bibitem [{\citenamefont {Morita}\ \emph
  {et~al.}(2004{\natexlab{a}})\citenamefont {Morita}, \citenamefont {Morimoto},
  \citenamefont {Kaji}, \citenamefont {Goto}, \citenamefont {Haba},
  \citenamefont {Ideguchi}, \citenamefont {Kanungo}, \citenamefont {Katori},
  \citenamefont {Koura}, \citenamefont {Kudo} \emph
  {et~al.}}]{morita2004status}%
  \BibitemOpen
  \bibfield  {author} {\bibinfo {author} {\bibfnamefont {K.}~\bibnamefont
  {Morita}}, \bibinfo {author} {\bibfnamefont {K.}~\bibnamefont {Morimoto}},
  \bibinfo {author} {\bibfnamefont {D.}~\bibnamefont {Kaji}}, \bibinfo {author}
  {\bibfnamefont {S.}~\bibnamefont {Goto}}, \bibinfo {author} {\bibfnamefont
  {H.}~\bibnamefont {Haba}}, \bibinfo {author} {\bibfnamefont {E.}~\bibnamefont
  {Ideguchi}}, \bibinfo {author} {\bibfnamefont {R.}~\bibnamefont {Kanungo}},
  \bibinfo {author} {\bibfnamefont {K.}~\bibnamefont {Katori}}, \bibinfo
  {author} {\bibfnamefont {H.}~\bibnamefont {Koura}}, \bibinfo {author}
  {\bibfnamefont {H.}~\bibnamefont {Kudo}}, \emph {et~al.},\ }\href@noop {}
  {\bibfield  {journal} {\bibinfo  {journal} {Nuclear Physics A}\ }\textbf
  {\bibinfo {volume} {734}},\ \bibinfo {pages} {101} (\bibinfo {year}
  {2004}{\natexlab{a}})}\BibitemShut {NoStop}%
\bibitem [{\citenamefont {Hofmann}\ \emph
  {et~al.}(1995{\natexlab{a}})\citenamefont {Hofmann}, \citenamefont {Ninov},
  \citenamefont {He{\ss}berger}, \citenamefont {Armbruster}, \citenamefont
  {Folger}, \citenamefont {M{\"u}nzenberg}, \citenamefont {Sch{\"o}tt},
  \citenamefont {Popeko}, \citenamefont {Yeremin}, \citenamefont {Andreyev}
  \emph {et~al.}}]{hofmann1995production}%
  \BibitemOpen
  \bibfield  {author} {\bibinfo {author} {\bibfnamefont {S.}~\bibnamefont
  {Hofmann}}, \bibinfo {author} {\bibfnamefont {V.}~\bibnamefont {Ninov}},
  \bibinfo {author} {\bibfnamefont {F.}~\bibnamefont {He{\ss}berger}}, \bibinfo
  {author} {\bibfnamefont {P.}~\bibnamefont {Armbruster}}, \bibinfo {author}
  {\bibfnamefont {H.}~\bibnamefont {Folger}}, \bibinfo {author} {\bibfnamefont
  {G.}~\bibnamefont {M{\"u}nzenberg}}, \bibinfo {author} {\bibfnamefont
  {H.}~\bibnamefont {Sch{\"o}tt}}, \bibinfo {author} {\bibfnamefont
  {A.}~\bibnamefont {Popeko}}, \bibinfo {author} {\bibfnamefont
  {A.}~\bibnamefont {Yeremin}}, \bibinfo {author} {\bibfnamefont
  {A.}~\bibnamefont {Andreyev}}, \emph {et~al.},\ }\href@noop {} {\bibfield
  {journal} {\bibinfo  {journal} {Zeitschrift f{\"u}r Physik A Hadrons and
  Nuclei}\ }\textbf {\bibinfo {volume} {350}},\ \bibinfo {pages} {277}
  (\bibinfo {year} {1995}{\natexlab{a}})}\BibitemShut {NoStop}%
\bibitem [{\citenamefont {Hofmann}\ \emph
  {et~al.}(1995{\natexlab{b}})\citenamefont {Hofmann}, \citenamefont {Ninov},
  \citenamefont {He{\ss}berger}, \citenamefont {Armbruster}, \citenamefont
  {Folger}, \citenamefont {M{\"u}nzenberg}, \citenamefont {Sch{\"o}tt},
  \citenamefont {Popeko}, \citenamefont {Yeremin}, \citenamefont {Andreyev}
  \emph {et~al.}}]{hofmann1995new}%
  \BibitemOpen
  \bibfield  {author} {\bibinfo {author} {\bibfnamefont {S.}~\bibnamefont
  {Hofmann}}, \bibinfo {author} {\bibfnamefont {V.}~\bibnamefont {Ninov}},
  \bibinfo {author} {\bibfnamefont {F.}~\bibnamefont {He{\ss}berger}}, \bibinfo
  {author} {\bibfnamefont {P.}~\bibnamefont {Armbruster}}, \bibinfo {author}
  {\bibfnamefont {H.}~\bibnamefont {Folger}}, \bibinfo {author} {\bibfnamefont
  {G.}~\bibnamefont {M{\"u}nzenberg}}, \bibinfo {author} {\bibfnamefont
  {H.}~\bibnamefont {Sch{\"o}tt}}, \bibinfo {author} {\bibfnamefont
  {A.}~\bibnamefont {Popeko}}, \bibinfo {author} {\bibfnamefont
  {A.}~\bibnamefont {Yeremin}}, \bibinfo {author} {\bibfnamefont
  {A.}~\bibnamefont {Andreyev}}, \emph {et~al.},\ }\href@noop {} {\bibfield
  {journal} {\bibinfo  {journal} {Zeitschrift f{\"u}r Physik A Hadrons and
  Nuclei}\ }\textbf {\bibinfo {volume} {350}},\ \bibinfo {pages} {281}
  (\bibinfo {year} {1995}{\natexlab{b}})}\BibitemShut {NoStop}%
\bibitem [{\citenamefont {Hofmann}\ \emph {et~al.}(1996)\citenamefont
  {Hofmann}, \citenamefont {Ninov}, \citenamefont {Hessberger}, \citenamefont
  {Armbruster}, \citenamefont {Folger}, \citenamefont {M{\"u}nzenberg},
  \citenamefont {Sch{\"o}tt}, \citenamefont {Popeko}, \citenamefont {Yeremin},
  \citenamefont {Saro} \emph {et~al.}}]{hofmann1996new}%
  \BibitemOpen
  \bibfield  {author} {\bibinfo {author} {\bibfnamefont {S.}~\bibnamefont
  {Hofmann}}, \bibinfo {author} {\bibfnamefont {V.}~\bibnamefont {Ninov}},
  \bibinfo {author} {\bibfnamefont {F.}~\bibnamefont {Hessberger}}, \bibinfo
  {author} {\bibfnamefont {P.}~\bibnamefont {Armbruster}}, \bibinfo {author}
  {\bibfnamefont {H.}~\bibnamefont {Folger}}, \bibinfo {author} {\bibfnamefont
  {G.}~\bibnamefont {M{\"u}nzenberg}}, \bibinfo {author} {\bibfnamefont
  {H.}~\bibnamefont {Sch{\"o}tt}}, \bibinfo {author} {\bibfnamefont
  {A.}~\bibnamefont {Popeko}}, \bibinfo {author} {\bibfnamefont
  {A.}~\bibnamefont {Yeremin}}, \bibinfo {author} {\bibfnamefont
  {S.}~\bibnamefont {Saro}}, \emph {et~al.},\ }\href@noop {} {\bibfield
  {journal} {\bibinfo  {journal} {Zeitschrift f{\"u}r Physik A Hadrons and
  Nuclei}\ }\textbf {\bibinfo {volume} {354}},\ \bibinfo {pages} {229}
  (\bibinfo {year} {1996})}\BibitemShut {NoStop}%
\bibitem [{\citenamefont {Morita}\ \emph
  {et~al.}(2004{\natexlab{b}})\citenamefont {Morita}, \citenamefont {Morimoto},
  \citenamefont {Kaji}, \citenamefont {Akiyama}, \citenamefont {Goto},
  \citenamefont {Haba}, \citenamefont {Ideguchi}, \citenamefont {Kanungo},
  \citenamefont {Katori}, \citenamefont {Koura} \emph
  {et~al.}}]{morita2004experiment}%
  \BibitemOpen
  \bibfield  {author} {\bibinfo {author} {\bibfnamefont {K.}~\bibnamefont
  {Morita}}, \bibinfo {author} {\bibfnamefont {K.}~\bibnamefont {Morimoto}},
  \bibinfo {author} {\bibfnamefont {D.}~\bibnamefont {Kaji}}, \bibinfo {author}
  {\bibfnamefont {T.}~\bibnamefont {Akiyama}}, \bibinfo {author} {\bibfnamefont
  {S.-i.}\ \bibnamefont {Goto}}, \bibinfo {author} {\bibfnamefont
  {H.}~\bibnamefont {Haba}}, \bibinfo {author} {\bibfnamefont {E.}~\bibnamefont
  {Ideguchi}}, \bibinfo {author} {\bibfnamefont {R.}~\bibnamefont {Kanungo}},
  \bibinfo {author} {\bibfnamefont {K.}~\bibnamefont {Katori}}, \bibinfo
  {author} {\bibfnamefont {H.}~\bibnamefont {Koura}}, \emph {et~al.},\
  }\href@noop {} {\bibfield  {journal} {\bibinfo  {journal} {Journal of the
  physical Society of Japan}\ }\textbf {\bibinfo {volume} {73}},\ \bibinfo
  {pages} {2593} (\bibinfo {year} {2004}{\natexlab{b}})}\BibitemShut {NoStop}%
\bibitem [{\citenamefont {Hinde}\ \emph {et~al.}(1982)\citenamefont {Hinde},
  \citenamefont {Leigh}, \citenamefont {Newton}, \citenamefont {Galster},\ and\
  \citenamefont {Sie}}]{hinde1982fission}%
  \BibitemOpen
  \bibfield  {author} {\bibinfo {author} {\bibfnamefont {D.}~\bibnamefont
  {Hinde}}, \bibinfo {author} {\bibfnamefont {J.}~\bibnamefont {Leigh}},
  \bibinfo {author} {\bibfnamefont {J.}~\bibnamefont {Newton}}, \bibinfo
  {author} {\bibfnamefont {W.}~\bibnamefont {Galster}},\ and\ \bibinfo {author}
  {\bibfnamefont {S.}~\bibnamefont {Sie}},\ }\href@noop {} {\bibfield
  {journal} {\bibinfo  {journal} {Nuclear Physics A}\ }\textbf {\bibinfo
  {volume} {385}},\ \bibinfo {pages} {109} (\bibinfo {year}
  {1982})}\BibitemShut {NoStop}%
\bibitem [{\citenamefont {Mahata}\ \emph {et~al.}(2003)\citenamefont {Mahata},
  \citenamefont {Kailas}, \citenamefont {Shrivastava}, \citenamefont
  {Chatterjee}, \citenamefont {Navin}, \citenamefont {Singh}, \citenamefont
  {Santra},\ and\ \citenamefont {Tomar}}]{mahata2003fusion}%
  \BibitemOpen
  \bibfield  {author} {\bibinfo {author} {\bibfnamefont {K.}~\bibnamefont
  {Mahata}}, \bibinfo {author} {\bibfnamefont {S.}~\bibnamefont {Kailas}},
  \bibinfo {author} {\bibfnamefont {A.}~\bibnamefont {Shrivastava}}, \bibinfo
  {author} {\bibfnamefont {A.}~\bibnamefont {Chatterjee}}, \bibinfo {author}
  {\bibfnamefont {A.}~\bibnamefont {Navin}}, \bibinfo {author} {\bibfnamefont
  {P.}~\bibnamefont {Singh}}, \bibinfo {author} {\bibfnamefont
  {S.}~\bibnamefont {Santra}},\ and\ \bibinfo {author} {\bibfnamefont
  {B.}~\bibnamefont {Tomar}},\ }\href@noop {} {\bibfield  {journal} {\bibinfo
  {journal} {Nuclear Physics A}\ }\textbf {\bibinfo {volume} {720}},\ \bibinfo
  {pages} {209} (\bibinfo {year} {2003})}\BibitemShut {NoStop}%
\bibitem [{\citenamefont {Andreyev}\ \emph
  {et~al.}(1997{\natexlab{a}})\citenamefont {Andreyev}, \citenamefont
  {Bogdanov}, \citenamefont {Chepigin}, \citenamefont {Kabachenko},
  \citenamefont {Malyshev}, \citenamefont {Oganessian}, \citenamefont {Popeko},
  \citenamefont {Roh{\'a}c}, \citenamefont {Sagaidak}, \citenamefont
  {Ter-Akopian} \emph {et~al.}}]{andreyev1997statistical}%
  \BibitemOpen
  \bibfield  {author} {\bibinfo {author} {\bibfnamefont {A.}~\bibnamefont
  {Andreyev}}, \bibinfo {author} {\bibfnamefont {D.}~\bibnamefont {Bogdanov}},
  \bibinfo {author} {\bibfnamefont {V.}~\bibnamefont {Chepigin}}, \bibinfo
  {author} {\bibfnamefont {A.}~\bibnamefont {Kabachenko}}, \bibinfo {author}
  {\bibfnamefont {O.}~\bibnamefont {Malyshev}}, \bibinfo {author}
  {\bibfnamefont {Y.~T.}\ \bibnamefont {Oganessian}}, \bibinfo {author}
  {\bibfnamefont {A.}~\bibnamefont {Popeko}}, \bibinfo {author} {\bibfnamefont
  {J.}~\bibnamefont {Roh{\'a}c}}, \bibinfo {author} {\bibfnamefont
  {R.}~\bibnamefont {Sagaidak}}, \bibinfo {author} {\bibfnamefont
  {G.}~\bibnamefont {Ter-Akopian}}, \emph {et~al.},\ }\href@noop {} {\bibfield
  {journal} {\bibinfo  {journal} {Nuclear Physics A}\ }\textbf {\bibinfo
  {volume} {626}},\ \bibinfo {pages} {857} (\bibinfo {year}
  {1997}{\natexlab{a}})}\BibitemShut {NoStop}%
\bibitem [{\citenamefont {Andreyev}\ \emph
  {et~al.}(1997{\natexlab{b}})\citenamefont {Andreyev}, \citenamefont
  {Bogdanov}, \citenamefont {Chepigin}, \citenamefont {Kabachenko},
  \citenamefont {Malyshev}, \citenamefont {Muzichka}, \citenamefont
  {Oganessian}, \citenamefont {Popeko}, \citenamefont {Pustylnik},
  \citenamefont {Sagaidak} \emph {et~al.}}]{andreyev1997decay}%
  \BibitemOpen
  \bibfield  {author} {\bibinfo {author} {\bibfnamefont {A.}~\bibnamefont
  {Andreyev}}, \bibinfo {author} {\bibfnamefont {D.}~\bibnamefont {Bogdanov}},
  \bibinfo {author} {\bibfnamefont {V.}~\bibnamefont {Chepigin}}, \bibinfo
  {author} {\bibfnamefont {A.}~\bibnamefont {Kabachenko}}, \bibinfo {author}
  {\bibfnamefont {O.}~\bibnamefont {Malyshev}}, \bibinfo {author}
  {\bibfnamefont {Y.~A.}\ \bibnamefont {Muzichka}}, \bibinfo {author}
  {\bibfnamefont {Y.~T.}\ \bibnamefont {Oganessian}}, \bibinfo {author}
  {\bibfnamefont {A.}~\bibnamefont {Popeko}}, \bibinfo {author} {\bibfnamefont
  {B.}~\bibnamefont {Pustylnik}}, \bibinfo {author} {\bibfnamefont
  {R.}~\bibnamefont {Sagaidak}}, \emph {et~al.},\ }\href@noop {} {\bibfield
  {journal} {\bibinfo  {journal} {Nuclear Physics A}\ }\textbf {\bibinfo
  {volume} {620}},\ \bibinfo {pages} {229} (\bibinfo {year}
  {1997}{\natexlab{b}})}\BibitemShut {NoStop}%
\bibitem [{\citenamefont {Haba}\ \emph {et~al.}(2014)\citenamefont {Haba},
  \citenamefont {Huang}, \citenamefont {Kaji}, \citenamefont {Kanaya},
  \citenamefont {Kudou}, \citenamefont {Morimoto}, \citenamefont {Morita},
  \citenamefont {Murakami}, \citenamefont {Ozeki}, \citenamefont {Sakai} \emph
  {et~al.}}]{haba2014production}%
  \BibitemOpen
  \bibfield  {author} {\bibinfo {author} {\bibfnamefont {H.}~\bibnamefont
  {Haba}}, \bibinfo {author} {\bibfnamefont {M.}~\bibnamefont {Huang}},
  \bibinfo {author} {\bibfnamefont {D.}~\bibnamefont {Kaji}}, \bibinfo {author}
  {\bibfnamefont {J.}~\bibnamefont {Kanaya}}, \bibinfo {author} {\bibfnamefont
  {Y.}~\bibnamefont {Kudou}}, \bibinfo {author} {\bibfnamefont
  {K.}~\bibnamefont {Morimoto}}, \bibinfo {author} {\bibfnamefont
  {K.}~\bibnamefont {Morita}}, \bibinfo {author} {\bibfnamefont
  {M.}~\bibnamefont {Murakami}}, \bibinfo {author} {\bibfnamefont
  {K.}~\bibnamefont {Ozeki}}, \bibinfo {author} {\bibfnamefont
  {R.}~\bibnamefont {Sakai}}, \emph {et~al.},\ }\href@noop {} {\bibfield
  {journal} {\bibinfo  {journal} {Physical Review C}\ }\textbf {\bibinfo
  {volume} {89}},\ \bibinfo {pages} {024618} (\bibinfo {year}
  {2014})}\BibitemShut {NoStop}%
\bibitem [{\citenamefont {Haba}\ \emph {et~al.}(2012)\citenamefont {Haba},
  \citenamefont {Kaji}, \citenamefont {Kudou}, \citenamefont {Morimoto},
  \citenamefont {Morita}, \citenamefont {Ozeki}, \citenamefont {Sakai},
  \citenamefont {Sumita}, \citenamefont {Yoneda}, \citenamefont {Kasamatsu}
  \emph {et~al.}}]{haba2012production}%
  \BibitemOpen
  \bibfield  {author} {\bibinfo {author} {\bibfnamefont {H.}~\bibnamefont
  {Haba}}, \bibinfo {author} {\bibfnamefont {D.}~\bibnamefont {Kaji}}, \bibinfo
  {author} {\bibfnamefont {Y.}~\bibnamefont {Kudou}}, \bibinfo {author}
  {\bibfnamefont {K.}~\bibnamefont {Morimoto}}, \bibinfo {author}
  {\bibfnamefont {K.}~\bibnamefont {Morita}}, \bibinfo {author} {\bibfnamefont
  {K.}~\bibnamefont {Ozeki}}, \bibinfo {author} {\bibfnamefont
  {R.}~\bibnamefont {Sakai}}, \bibinfo {author} {\bibfnamefont
  {T.}~\bibnamefont {Sumita}}, \bibinfo {author} {\bibfnamefont
  {A.}~\bibnamefont {Yoneda}}, \bibinfo {author} {\bibfnamefont
  {Y.}~\bibnamefont {Kasamatsu}}, \emph {et~al.},\ }\href@noop {} {\bibfield
  {journal} {\bibinfo  {journal} {Physical Review C}\ }\textbf {\bibinfo
  {volume} {85}},\ \bibinfo {pages} {024611} (\bibinfo {year}
  {2012})}\BibitemShut {NoStop}%
\bibitem [{\citenamefont {Dvorak}\ \emph {et~al.}(2008)\citenamefont {Dvorak},
  \citenamefont {Br{\"u}chle}, \citenamefont {Chelnokov}, \citenamefont
  {D{\"u}llmann}, \citenamefont {Dvorakova}, \citenamefont {Eberhardt},
  \citenamefont {J{\"a}ger}, \citenamefont {Kr{\"u}cken}, \citenamefont
  {Kuznetsov}, \citenamefont {Nagame} \emph {et~al.}}]{dvorak2008observation}%
  \BibitemOpen
  \bibfield  {author} {\bibinfo {author} {\bibfnamefont {J.}~\bibnamefont
  {Dvorak}}, \bibinfo {author} {\bibfnamefont {W.}~\bibnamefont {Br{\"u}chle}},
  \bibinfo {author} {\bibfnamefont {M.}~\bibnamefont {Chelnokov}}, \bibinfo
  {author} {\bibfnamefont {C.~E.}\ \bibnamefont {D{\"u}llmann}}, \bibinfo
  {author} {\bibfnamefont {Z.}~\bibnamefont {Dvorakova}}, \bibinfo {author}
  {\bibfnamefont {K.}~\bibnamefont {Eberhardt}}, \bibinfo {author}
  {\bibfnamefont {E.}~\bibnamefont {J{\"a}ger}}, \bibinfo {author}
  {\bibfnamefont {R.}~\bibnamefont {Kr{\"u}cken}}, \bibinfo {author}
  {\bibfnamefont {A.}~\bibnamefont {Kuznetsov}}, \bibinfo {author}
  {\bibfnamefont {Y.}~\bibnamefont {Nagame}}, \emph {et~al.},\ }\href@noop {}
  {\bibfield  {journal} {\bibinfo  {journal} {Physical review letters}\
  }\textbf {\bibinfo {volume} {100}},\ \bibinfo {pages} {132503} (\bibinfo
  {year} {2008})}\BibitemShut {NoStop}%
\bibitem [{\citenamefont {Chakrabarty}\ \emph {et~al.}(2000)\citenamefont
  {Chakrabarty}, \citenamefont {Tomar}, \citenamefont {Goswami}, \citenamefont
  {Gubbi}, \citenamefont {Manohar}, \citenamefont {Sharma}, \citenamefont
  {Bindukumar},\ and\ \citenamefont {Mukherjee}}]{chakrabarty2000complete}%
  \BibitemOpen
  \bibfield  {author} {\bibinfo {author} {\bibfnamefont {S.}~\bibnamefont
  {Chakrabarty}}, \bibinfo {author} {\bibfnamefont {B.}~\bibnamefont {Tomar}},
  \bibinfo {author} {\bibfnamefont {A.}~\bibnamefont {Goswami}}, \bibinfo
  {author} {\bibfnamefont {G.}~\bibnamefont {Gubbi}}, \bibinfo {author}
  {\bibfnamefont {S.}~\bibnamefont {Manohar}}, \bibinfo {author} {\bibfnamefont
  {A.}~\bibnamefont {Sharma}}, \bibinfo {author} {\bibfnamefont
  {B.}~\bibnamefont {Bindukumar}},\ and\ \bibinfo {author} {\bibfnamefont
  {S.}~\bibnamefont {Mukherjee}},\ }\href@noop {} {\bibfield  {journal}
  {\bibinfo  {journal} {Nuclear Physics A}\ }\textbf {\bibinfo {volume}
  {678}},\ \bibinfo {pages} {355} (\bibinfo {year} {2000})}\BibitemShut
  {NoStop}%
\bibitem [{\citenamefont {Kumar}\ \emph {et~al.}(2013)\citenamefont {Kumar},
  \citenamefont {Ahmad}, \citenamefont {Ali}, \citenamefont {Rizvi},
  \citenamefont {Agarwal}, \citenamefont {Kumar}, \citenamefont {Golda},\ and\
  \citenamefont {Chaubey}}]{kumar2013low}%
  \BibitemOpen
  \bibfield  {author} {\bibinfo {author} {\bibfnamefont {K.}~\bibnamefont
  {Kumar}}, \bibinfo {author} {\bibfnamefont {T.}~\bibnamefont {Ahmad}},
  \bibinfo {author} {\bibfnamefont {S.}~\bibnamefont {Ali}}, \bibinfo {author}
  {\bibfnamefont {I.}~\bibnamefont {Rizvi}}, \bibinfo {author} {\bibfnamefont
  {A.}~\bibnamefont {Agarwal}}, \bibinfo {author} {\bibfnamefont
  {R.}~\bibnamefont {Kumar}}, \bibinfo {author} {\bibfnamefont
  {K.}~\bibnamefont {Golda}},\ and\ \bibinfo {author} {\bibfnamefont
  {A.}~\bibnamefont {Chaubey}},\ }\href@noop {} {\bibfield  {journal} {\bibinfo
   {journal} {Physical Review C}\ }\textbf {\bibinfo {volume} {87}},\ \bibinfo
  {pages} {044608} (\bibinfo {year} {2013})}\BibitemShut {NoStop}%
\bibitem [{\citenamefont {Zhang}\ \emph {et~al.}(2014)\citenamefont {Zhang},
  \citenamefont {Fang}, \citenamefont {Gomes}, \citenamefont {Lubian},
  \citenamefont {Liu}, \citenamefont {Zhou}, \citenamefont {Li}, \citenamefont
  {Wang}, \citenamefont {Guo}, \citenamefont {Qiang} \emph
  {et~al.}}]{zhang2014complete}%
  \BibitemOpen
  \bibfield  {author} {\bibinfo {author} {\bibfnamefont {N.}~\bibnamefont
  {Zhang}}, \bibinfo {author} {\bibfnamefont {Y.}~\bibnamefont {Fang}},
  \bibinfo {author} {\bibfnamefont {P.}~\bibnamefont {Gomes}}, \bibinfo
  {author} {\bibfnamefont {J.}~\bibnamefont {Lubian}}, \bibinfo {author}
  {\bibfnamefont {M.}~\bibnamefont {Liu}}, \bibinfo {author} {\bibfnamefont
  {X.}~\bibnamefont {Zhou}}, \bibinfo {author} {\bibfnamefont {G.}~\bibnamefont
  {Li}}, \bibinfo {author} {\bibfnamefont {J.}~\bibnamefont {Wang}}, \bibinfo
  {author} {\bibfnamefont {S.}~\bibnamefont {Guo}}, \bibinfo {author}
  {\bibfnamefont {Y.}~\bibnamefont {Qiang}}, \emph {et~al.},\ }\href@noop {}
  {\bibfield  {journal} {\bibinfo  {journal} {Physical Review C}\ }\textbf
  {\bibinfo {volume} {90}},\ \bibinfo {pages} {024621} (\bibinfo {year}
  {2014})}\BibitemShut {NoStop}%
\bibitem [{\citenamefont {Sharma}\ \emph {et~al.}(2015)\citenamefont {Sharma},
  \citenamefont {Singh}, \citenamefont {Singh}, \citenamefont {Yadav},
  \citenamefont {Sharma}, \citenamefont {Bala}, \citenamefont {Kumar},
  \citenamefont {Singh}, \citenamefont {Prasad} \emph
  {et~al.}}]{sharma2015systematic}%
  \BibitemOpen
  \bibfield  {author} {\bibinfo {author} {\bibfnamefont {M.~K.}\ \bibnamefont
  {Sharma}}, \bibinfo {author} {\bibfnamefont {P.~P.}\ \bibnamefont {Singh}},
  \bibinfo {author} {\bibfnamefont {D.~P.}\ \bibnamefont {Singh}}, \bibinfo
  {author} {\bibfnamefont {A.}~\bibnamefont {Yadav}}, \bibinfo {author}
  {\bibfnamefont {V.~R.}\ \bibnamefont {Sharma}}, \bibinfo {author}
  {\bibfnamefont {I.}~\bibnamefont {Bala}}, \bibinfo {author} {\bibfnamefont
  {R.}~\bibnamefont {Kumar}}, \bibinfo {author} {\bibfnamefont
  {B.}~\bibnamefont {Singh}}, \bibinfo {author} {\bibfnamefont
  {R.}~\bibnamefont {Prasad}}, \emph {et~al.},\ }\href@noop {} {\bibfield
  {journal} {\bibinfo  {journal} {Physical Review C}\ }\textbf {\bibinfo
  {volume} {91}},\ \bibinfo {pages} {014603} (\bibinfo {year}
  {2015})}\BibitemShut {NoStop}%
\bibitem [{\citenamefont {Scholz}\ \emph {et~al.}(2014)\citenamefont {Scholz},
  \citenamefont {Endres}, \citenamefont {Hennig}, \citenamefont {Netterdon},
  \citenamefont {Becker}, \citenamefont {Endres}, \citenamefont {Mayer},
  \citenamefont {Giesen}, \citenamefont {Rogalla}, \citenamefont {Schl{\"u}ter}
  \emph {et~al.}}]{scholz2014measurement}%
  \BibitemOpen
  \bibfield  {author} {\bibinfo {author} {\bibfnamefont {P.}~\bibnamefont
  {Scholz}}, \bibinfo {author} {\bibfnamefont {A.}~\bibnamefont {Endres}},
  \bibinfo {author} {\bibfnamefont {A.}~\bibnamefont {Hennig}}, \bibinfo
  {author} {\bibfnamefont {L.}~\bibnamefont {Netterdon}}, \bibinfo {author}
  {\bibfnamefont {H.}~\bibnamefont {Becker}}, \bibinfo {author} {\bibfnamefont
  {J.}~\bibnamefont {Endres}}, \bibinfo {author} {\bibfnamefont
  {J.}~\bibnamefont {Mayer}}, \bibinfo {author} {\bibfnamefont
  {U.}~\bibnamefont {Giesen}}, \bibinfo {author} {\bibfnamefont
  {D.}~\bibnamefont {Rogalla}}, \bibinfo {author} {\bibfnamefont
  {F.}~\bibnamefont {Schl{\"u}ter}}, \emph {et~al.},\ }\href@noop {} {\bibfield
   {journal} {\bibinfo  {journal} {Physical Review C}\ }\textbf {\bibinfo
  {volume} {90}},\ \bibinfo {pages} {065807} (\bibinfo {year}
  {2014})}\BibitemShut {NoStop}%
\bibitem [{\citenamefont {Fang}\ \emph {et~al.}(2013)\citenamefont {Fang},
  \citenamefont {Gomes}, \citenamefont {Lubian}, \citenamefont {Zhou},
  \citenamefont {Zhang}, \citenamefont {Han}, \citenamefont {Liu},
  \citenamefont {Zheng}, \citenamefont {Guo}, \citenamefont {Wang} \emph
  {et~al.}}]{fang2013fusion}%
  \BibitemOpen
  \bibfield  {author} {\bibinfo {author} {\bibfnamefont {Y.}~\bibnamefont
  {Fang}}, \bibinfo {author} {\bibfnamefont {P.}~\bibnamefont {Gomes}},
  \bibinfo {author} {\bibfnamefont {J.}~\bibnamefont {Lubian}}, \bibinfo
  {author} {\bibfnamefont {X.}~\bibnamefont {Zhou}}, \bibinfo {author}
  {\bibfnamefont {Y.}~\bibnamefont {Zhang}}, \bibinfo {author} {\bibfnamefont
  {J.}~\bibnamefont {Han}}, \bibinfo {author} {\bibfnamefont {M.}~\bibnamefont
  {Liu}}, \bibinfo {author} {\bibfnamefont {Y.}~\bibnamefont {Zheng}}, \bibinfo
  {author} {\bibfnamefont {S.}~\bibnamefont {Guo}}, \bibinfo {author}
  {\bibfnamefont {J.}~\bibnamefont {Wang}}, \emph {et~al.},\ }\href@noop {}
  {\bibfield  {journal} {\bibinfo  {journal} {Physical Review C}\ }\textbf
  {\bibinfo {volume} {87}},\ \bibinfo {pages} {024604} (\bibinfo {year}
  {2013})}\BibitemShut {NoStop}%
\bibitem [{\citenamefont {Fang}\ \emph {et~al.}(2015)\citenamefont {Fang},
  \citenamefont {Gomes}, \citenamefont {Lubian}, \citenamefont {Liu},
  \citenamefont {Zhou}, \citenamefont {Junior}, \citenamefont {Zhang},
  \citenamefont {Zhang}, \citenamefont {Li}, \citenamefont {Wang} \emph
  {et~al.}}]{fang2015complete}%
  \BibitemOpen
  \bibfield  {author} {\bibinfo {author} {\bibfnamefont {Y.}~\bibnamefont
  {Fang}}, \bibinfo {author} {\bibfnamefont {P.}~\bibnamefont {Gomes}},
  \bibinfo {author} {\bibfnamefont {J.}~\bibnamefont {Lubian}}, \bibinfo
  {author} {\bibfnamefont {M.}~\bibnamefont {Liu}}, \bibinfo {author}
  {\bibfnamefont {X.}~\bibnamefont {Zhou}}, \bibinfo {author} {\bibfnamefont
  {D.~M.}\ \bibnamefont {Junior}}, \bibinfo {author} {\bibfnamefont
  {N.}~\bibnamefont {Zhang}}, \bibinfo {author} {\bibfnamefont
  {Y.}~\bibnamefont {Zhang}}, \bibinfo {author} {\bibfnamefont
  {G.}~\bibnamefont {Li}}, \bibinfo {author} {\bibfnamefont {J.}~\bibnamefont
  {Wang}}, \emph {et~al.},\ }\href@noop {} {\bibfield  {journal} {\bibinfo
  {journal} {Physical Review C}\ }\textbf {\bibinfo {volume} {91}},\ \bibinfo
  {pages} {014608} (\bibinfo {year} {2015})}\BibitemShut {NoStop}%
\bibitem [{\citenamefont {Penionzhkevich}\ \emph {et~al.}(2007)\citenamefont
  {Penionzhkevich}, \citenamefont {Astabatyan}, \citenamefont {Demekhina},
  \citenamefont {Gulbekian}, \citenamefont {Kalpakchieva}, \citenamefont
  {Kulko}, \citenamefont {Lukyanov}, \citenamefont {Markaryan}, \citenamefont
  {Maslov}, \citenamefont {Muzychka} \emph
  {et~al.}}]{penionzhkevich2007excitation}%
  \BibitemOpen
  \bibfield  {author} {\bibinfo {author} {\bibfnamefont {Y.~E.}\ \bibnamefont
  {Penionzhkevich}}, \bibinfo {author} {\bibfnamefont {R.}~\bibnamefont
  {Astabatyan}}, \bibinfo {author} {\bibfnamefont {N.}~\bibnamefont
  {Demekhina}}, \bibinfo {author} {\bibfnamefont {G.}~\bibnamefont
  {Gulbekian}}, \bibinfo {author} {\bibfnamefont {R.}~\bibnamefont
  {Kalpakchieva}}, \bibinfo {author} {\bibfnamefont {A.~A.}\ \bibnamefont
  {Kulko}}, \bibinfo {author} {\bibfnamefont {S.~M.}\ \bibnamefont {Lukyanov}},
  \bibinfo {author} {\bibfnamefont {E.}~\bibnamefont {Markaryan}}, \bibinfo
  {author} {\bibfnamefont {V.}~\bibnamefont {Maslov}}, \bibinfo {author}
  {\bibfnamefont {Y.~A.}\ \bibnamefont {Muzychka}}, \emph {et~al.},\
  }\href@noop {} {\bibfield  {journal} {\bibinfo  {journal} {The European
  Physical Journal A}\ }\textbf {\bibinfo {volume} {31}},\ \bibinfo {pages}
  {185} (\bibinfo {year} {2007})}\BibitemShut {NoStop}%
\bibitem [{\citenamefont {Singh}\ \emph {et~al.}(2009)\citenamefont {Singh},
  \citenamefont {Singh}, \citenamefont {Yadav}, \citenamefont {Sharma},
  \citenamefont {Singh}, \citenamefont {Golda}, \citenamefont {Kumar},
  \citenamefont {Sinha}, \citenamefont {Prasad} \emph
  {et~al.}}]{singh2009investigation}%
  \BibitemOpen
  \bibfield  {author} {\bibinfo {author} {\bibfnamefont {D.~P.}\ \bibnamefont
  {Singh}}, \bibinfo {author} {\bibfnamefont {P.~P.}\ \bibnamefont {Singh}},
  \bibinfo {author} {\bibfnamefont {A.}~\bibnamefont {Yadav}}, \bibinfo
  {author} {\bibfnamefont {M.~K.}\ \bibnamefont {Sharma}}, \bibinfo {author}
  {\bibfnamefont {B.}~\bibnamefont {Singh}}, \bibinfo {author} {\bibfnamefont
  {K.}~\bibnamefont {Golda}}, \bibinfo {author} {\bibfnamefont
  {R.}~\bibnamefont {Kumar}}, \bibinfo {author} {\bibfnamefont
  {A.}~\bibnamefont {Sinha}}, \bibinfo {author} {\bibfnamefont
  {R.}~\bibnamefont {Prasad}}, \emph {et~al.},\ }\href@noop {} {\bibfield
  {journal} {\bibinfo  {journal} {Physical Review C}\ }\textbf {\bibinfo
  {volume} {80}},\ \bibinfo {pages} {014601} (\bibinfo {year}
  {2009})}\BibitemShut {NoStop}%
\bibitem [{\citenamefont {Shrivastava}\ \emph {et~al.}(1999)\citenamefont
  {Shrivastava}, \citenamefont {Kailas}, \citenamefont {Chatterjee},
  \citenamefont {Samant}, \citenamefont {Navin}, \citenamefont {Singh},\ and\
  \citenamefont {Tomar}}]{shrivastava1999shell}%
  \BibitemOpen
  \bibfield  {author} {\bibinfo {author} {\bibfnamefont {A.}~\bibnamefont
  {Shrivastava}}, \bibinfo {author} {\bibfnamefont {S.}~\bibnamefont {Kailas}},
  \bibinfo {author} {\bibfnamefont {A.}~\bibnamefont {Chatterjee}}, \bibinfo
  {author} {\bibfnamefont {A.}~\bibnamefont {Samant}}, \bibinfo {author}
  {\bibfnamefont {A.}~\bibnamefont {Navin}}, \bibinfo {author} {\bibfnamefont
  {P.}~\bibnamefont {Singh}},\ and\ \bibinfo {author} {\bibfnamefont
  {B.}~\bibnamefont {Tomar}},\ }\href@noop {} {\bibfield  {journal} {\bibinfo
  {journal} {Physical review letters}\ }\textbf {\bibinfo {volume} {82}},\
  \bibinfo {pages} {699} (\bibinfo {year} {1999})}\BibitemShut {NoStop}%
\bibitem [{\citenamefont {Sikkeland}\ \emph {et~al.}(1970)\citenamefont
  {Sikkeland}, \citenamefont {Silva}, \citenamefont {Ghiorso},\ and\
  \citenamefont {Nurmia}}]{sikkeland1970study}%
  \BibitemOpen
  \bibfield  {author} {\bibinfo {author} {\bibfnamefont {T.}~\bibnamefont
  {Sikkeland}}, \bibinfo {author} {\bibfnamefont {R.}~\bibnamefont {Silva}},
  \bibinfo {author} {\bibfnamefont {A.}~\bibnamefont {Ghiorso}},\ and\ \bibinfo
  {author} {\bibfnamefont {M.}~\bibnamefont {Nurmia}},\ }\href@noop {}
  {\bibfield  {journal} {\bibinfo  {journal} {Physical Review C}\ }\textbf
  {\bibinfo {volume} {1}},\ \bibinfo {pages} {1564} (\bibinfo {year}
  {1970})}\BibitemShut {NoStop}%
\bibitem [{\citenamefont {Mayorov}\ \emph {et~al.}(2014)\citenamefont
  {Mayorov}, \citenamefont {Werke}, \citenamefont {Alfonso}, \citenamefont
  {Bennett},\ and\ \citenamefont {Folden~III}}]{mayorov2014production}%
  \BibitemOpen
  \bibfield  {author} {\bibinfo {author} {\bibfnamefont {D.}~\bibnamefont
  {Mayorov}}, \bibinfo {author} {\bibfnamefont {T.}~\bibnamefont {Werke}},
  \bibinfo {author} {\bibfnamefont {M.}~\bibnamefont {Alfonso}}, \bibinfo
  {author} {\bibfnamefont {M.}~\bibnamefont {Bennett}},\ and\ \bibinfo {author}
  {\bibfnamefont {C.}~\bibnamefont {Folden~III}},\ }\href@noop {} {\bibfield
  {journal} {\bibinfo  {journal} {Physical Review C}\ }\textbf {\bibinfo
  {volume} {90}},\ \bibinfo {pages} {024602} (\bibinfo {year}
  {2014})}\BibitemShut {NoStop}%
\bibitem [{\citenamefont {Baba}\ \emph {et~al.}(1988)\citenamefont {Baba},
  \citenamefont {Hata}, \citenamefont {Ichikawa}, \citenamefont {Sekine},
  \citenamefont {Nagame}, \citenamefont {Yokoyama}, \citenamefont {Shoji},
  \citenamefont {Saito}, \citenamefont {Takahashi}, \citenamefont {Baba} \emph
  {et~al.}}]{baba1988evaporation}%
  \BibitemOpen
  \bibfield  {author} {\bibinfo {author} {\bibfnamefont {S.}~\bibnamefont
  {Baba}}, \bibinfo {author} {\bibfnamefont {K.}~\bibnamefont {Hata}}, \bibinfo
  {author} {\bibfnamefont {S.}~\bibnamefont {Ichikawa}}, \bibinfo {author}
  {\bibfnamefont {T.}~\bibnamefont {Sekine}}, \bibinfo {author} {\bibfnamefont
  {Y.}~\bibnamefont {Nagame}}, \bibinfo {author} {\bibfnamefont
  {A.}~\bibnamefont {Yokoyama}}, \bibinfo {author} {\bibfnamefont
  {M.}~\bibnamefont {Shoji}}, \bibinfo {author} {\bibfnamefont
  {T.}~\bibnamefont {Saito}}, \bibinfo {author} {\bibfnamefont
  {N.}~\bibnamefont {Takahashi}}, \bibinfo {author} {\bibfnamefont
  {H.}~\bibnamefont {Baba}}, \emph {et~al.},\ }\href@noop {} {\bibfield
  {journal} {\bibinfo  {journal} {Zeitschrift f{\"u}r Physik A Atomic Nuclei}\
  }\textbf {\bibinfo {volume} {331}},\ \bibinfo {pages} {53} (\bibinfo {year}
  {1988})}\BibitemShut {NoStop}%
\bibitem [{\citenamefont {Maiti}\ and\ \citenamefont
  {Lahiri}(2011)}]{maiti2011production}%
  \BibitemOpen
  \bibfield  {author} {\bibinfo {author} {\bibfnamefont {M.}~\bibnamefont
  {Maiti}}\ and\ \bibinfo {author} {\bibfnamefont {S.}~\bibnamefont {Lahiri}},\
  }\href@noop {} {\bibfield  {journal} {\bibinfo  {journal} {Physical Review
  C}\ }\textbf {\bibinfo {volume} {84}},\ \bibinfo {pages} {067601} (\bibinfo
  {year} {2011})}\BibitemShut {NoStop}%
\bibitem [{\citenamefont {Vermeulen}\ \emph {et~al.}(1984)\citenamefont
  {Vermeulen}, \citenamefont {Clerc}, \citenamefont {Sahm}, \citenamefont
  {Schmidt}, \citenamefont {Keller}, \citenamefont {M{\"u}nzenberg},\ and\
  \citenamefont {Reisdorf}}]{vermeulen1984cross}%
  \BibitemOpen
  \bibfield  {author} {\bibinfo {author} {\bibfnamefont {D.}~\bibnamefont
  {Vermeulen}}, \bibinfo {author} {\bibfnamefont {H.-G.}\ \bibnamefont
  {Clerc}}, \bibinfo {author} {\bibfnamefont {C.-C.}\ \bibnamefont {Sahm}},
  \bibinfo {author} {\bibfnamefont {K.-H.}\ \bibnamefont {Schmidt}}, \bibinfo
  {author} {\bibfnamefont {J.}~\bibnamefont {Keller}}, \bibinfo {author}
  {\bibfnamefont {G.}~\bibnamefont {M{\"u}nzenberg}},\ and\ \bibinfo {author}
  {\bibfnamefont {W.}~\bibnamefont {Reisdorf}},\ }\href@noop {} {\bibfield
  {journal} {\bibinfo  {journal} {Zeitschrift f{\"u}r Physik A Atoms and
  Nuclei}\ }\textbf {\bibinfo {volume} {318}},\ \bibinfo {pages} {157}
  (\bibinfo {year} {1984})}\BibitemShut {NoStop}%
\bibitem [{\citenamefont {Mayorov}\ \emph {et~al.}(2015)\citenamefont
  {Mayorov}, \citenamefont {Werke}, \citenamefont {Alfonso}, \citenamefont
  {Tereshatov}, \citenamefont {Bennett}, \citenamefont {Frey},\ and\
  \citenamefont {Folden~III}}]{mayorov2015evaporation}%
  \BibitemOpen
  \bibfield  {author} {\bibinfo {author} {\bibfnamefont {D.}~\bibnamefont
  {Mayorov}}, \bibinfo {author} {\bibfnamefont {T.}~\bibnamefont {Werke}},
  \bibinfo {author} {\bibfnamefont {M.}~\bibnamefont {Alfonso}}, \bibinfo
  {author} {\bibfnamefont {E.}~\bibnamefont {Tereshatov}}, \bibinfo {author}
  {\bibfnamefont {M.}~\bibnamefont {Bennett}}, \bibinfo {author} {\bibfnamefont
  {M.}~\bibnamefont {Frey}},\ and\ \bibinfo {author} {\bibfnamefont
  {C.}~\bibnamefont {Folden~III}},\ }\href@noop {} {\bibfield  {journal}
  {\bibinfo  {journal} {Physical Review C}\ }\textbf {\bibinfo {volume} {92}},\
  \bibinfo {pages} {054601} (\bibinfo {year} {2015})}\BibitemShut {NoStop}%
\bibitem [{\citenamefont {Corradi}\ \emph {et~al.}(2005)\citenamefont
  {Corradi}, \citenamefont {Behera}, \citenamefont {Fioretto}, \citenamefont
  {Gadea}, \citenamefont {Latina}, \citenamefont {Stefanini}, \citenamefont
  {Szilner}, \citenamefont {Trotta}, \citenamefont {Wu}, \citenamefont
  {Beghini} \emph {et~al.}}]{corradi2005excitation}%
  \BibitemOpen
  \bibfield  {author} {\bibinfo {author} {\bibfnamefont {L.}~\bibnamefont
  {Corradi}}, \bibinfo {author} {\bibfnamefont {B.}~\bibnamefont {Behera}},
  \bibinfo {author} {\bibfnamefont {E.}~\bibnamefont {Fioretto}}, \bibinfo
  {author} {\bibfnamefont {A.}~\bibnamefont {Gadea}}, \bibinfo {author}
  {\bibfnamefont {A.}~\bibnamefont {Latina}}, \bibinfo {author} {\bibfnamefont
  {A.}~\bibnamefont {Stefanini}}, \bibinfo {author} {\bibfnamefont
  {S.}~\bibnamefont {Szilner}}, \bibinfo {author} {\bibfnamefont
  {M.}~\bibnamefont {Trotta}}, \bibinfo {author} {\bibfnamefont
  {Y.}~\bibnamefont {Wu}}, \bibinfo {author} {\bibfnamefont {S.}~\bibnamefont
  {Beghini}}, \emph {et~al.},\ }\href@noop {} {\bibfield  {journal} {\bibinfo
  {journal} {Physical Review C}\ }\textbf {\bibinfo {volume} {71}},\ \bibinfo
  {pages} {014609} (\bibinfo {year} {2005})}\BibitemShut {NoStop}%
\bibitem [{\citenamefont {Sahm}\ \emph
  {et~al.}(1985{\natexlab{a}})\citenamefont {Sahm}, \citenamefont {Clerc},
  \citenamefont {Schmidt}, \citenamefont {Reisdorf}, \citenamefont
  {Armbruster}, \citenamefont {Hessberger}, \citenamefont {Keller},
  \citenamefont {M{\"u}nzenberg},\ and\ \citenamefont
  {Vermeulen}}]{sahm1985fusion}%
  \BibitemOpen
  \bibfield  {author} {\bibinfo {author} {\bibfnamefont {C.}~\bibnamefont
  {Sahm}}, \bibinfo {author} {\bibfnamefont {H.}~\bibnamefont {Clerc}},
  \bibinfo {author} {\bibfnamefont {K.-H.}\ \bibnamefont {Schmidt}}, \bibinfo
  {author} {\bibfnamefont {W.}~\bibnamefont {Reisdorf}}, \bibinfo {author}
  {\bibfnamefont {P.}~\bibnamefont {Armbruster}}, \bibinfo {author}
  {\bibfnamefont {F.}~\bibnamefont {Hessberger}}, \bibinfo {author}
  {\bibfnamefont {J.}~\bibnamefont {Keller}}, \bibinfo {author} {\bibfnamefont
  {G.}~\bibnamefont {M{\"u}nzenberg}},\ and\ \bibinfo {author} {\bibfnamefont
  {D.}~\bibnamefont {Vermeulen}},\ }\href@noop {} {\bibfield  {journal}
  {\bibinfo  {journal} {Nuclear Physics A}\ }\textbf {\bibinfo {volume}
  {441}},\ \bibinfo {pages} {316} (\bibinfo {year}
  {1985}{\natexlab{a}})}\BibitemShut {NoStop}%
\bibitem [{\citenamefont {Sahm}\ \emph
  {et~al.}(1985{\natexlab{b}})\citenamefont {Sahm}, \citenamefont {Clerc},
  \citenamefont {Schmidt}, \citenamefont {Reisdorf}, \citenamefont
  {Armbruster}, \citenamefont {Hessberger}, \citenamefont {Keller},
  \citenamefont {Münzenberg},\ and\ \citenamefont {Vermeulen}}]{SAHM1985316}%
  \BibitemOpen
  \bibfield  {author} {\bibinfo {author} {\bibfnamefont {C.}~\bibnamefont
  {Sahm}}, \bibinfo {author} {\bibfnamefont {H.}~\bibnamefont {Clerc}},
  \bibinfo {author} {\bibfnamefont {K.-H.}\ \bibnamefont {Schmidt}}, \bibinfo
  {author} {\bibfnamefont {W.}~\bibnamefont {Reisdorf}}, \bibinfo {author}
  {\bibfnamefont {P.}~\bibnamefont {Armbruster}}, \bibinfo {author}
  {\bibfnamefont {F.}~\bibnamefont {Hessberger}}, \bibinfo {author}
  {\bibfnamefont {J.}~\bibnamefont {Keller}}, \bibinfo {author} {\bibfnamefont
  {G.}~\bibnamefont {Münzenberg}},\ and\ \bibinfo {author} {\bibfnamefont
  {D.}~\bibnamefont {Vermeulen}},\ }\href
  {https://doi.org/https://doi.org/10.1016/0375-9474(85)90036-3} {\bibfield
  {journal} {\bibinfo  {journal} {Nuclear Physics A}\ }\textbf {\bibinfo
  {volume} {441}},\ \bibinfo {pages} {316 } (\bibinfo {year}
  {1985}{\natexlab{b}})}\BibitemShut {NoStop}%
\bibitem [{\citenamefont {Schmidt}\ \emph {et~al.}(1981)\citenamefont
  {Schmidt}, \citenamefont {Armbruster}, \citenamefont {Hessberger},
  \citenamefont {M{\"u}nzenberg}, \citenamefont {Reisdorf}, \citenamefont
  {Sahm}, \citenamefont {Vermeulen}, \citenamefont {Clerc}, \citenamefont
  {Keller},\ and\ \citenamefont {Schulte}}]{schmidt1981barrier}%
  \BibitemOpen
  \bibfield  {author} {\bibinfo {author} {\bibfnamefont {K.-H.}\ \bibnamefont
  {Schmidt}}, \bibinfo {author} {\bibfnamefont {P.}~\bibnamefont {Armbruster}},
  \bibinfo {author} {\bibfnamefont {F.}~\bibnamefont {Hessberger}}, \bibinfo
  {author} {\bibfnamefont {G.}~\bibnamefont {M{\"u}nzenberg}}, \bibinfo
  {author} {\bibfnamefont {W.}~\bibnamefont {Reisdorf}}, \bibinfo {author}
  {\bibfnamefont {C.-C.}\ \bibnamefont {Sahm}}, \bibinfo {author}
  {\bibfnamefont {D.}~\bibnamefont {Vermeulen}}, \bibinfo {author}
  {\bibfnamefont {H.-G.}\ \bibnamefont {Clerc}}, \bibinfo {author}
  {\bibfnamefont {J.}~\bibnamefont {Keller}},\ and\ \bibinfo {author}
  {\bibfnamefont {H.}~\bibnamefont {Schulte}},\ }\href@noop {} {\bibfield
  {journal} {\bibinfo  {journal} {Zeitschrift f{\"u}r Physik A Atoms and
  Nuclei}\ }\textbf {\bibinfo {volume} {301}},\ \bibinfo {pages} {21} (\bibinfo
  {year} {1981})}\BibitemShut {NoStop}%
\bibitem [{\citenamefont {Mitsuoka}\ \emph {et~al.}(2000)\citenamefont
  {Mitsuoka}, \citenamefont {Ikezoe}, \citenamefont {Nishio},\ and\
  \citenamefont {Lu}}]{PhysRevC.62.054603}%
  \BibitemOpen
  \bibfield  {author} {\bibinfo {author} {\bibfnamefont {S.}~\bibnamefont
  {Mitsuoka}}, \bibinfo {author} {\bibfnamefont {H.}~\bibnamefont {Ikezoe}},
  \bibinfo {author} {\bibfnamefont {K.}~\bibnamefont {Nishio}},\ and\ \bibinfo
  {author} {\bibfnamefont {J.}~\bibnamefont {Lu}},\ }\href
  {https://doi.org/10.1103/PhysRevC.62.054603} {\bibfield  {journal} {\bibinfo
  {journal} {Phys. Rev. C}\ }\textbf {\bibinfo {volume} {62}},\ \bibinfo
  {pages} {054603} (\bibinfo {year} {2000})}\BibitemShut {NoStop}%
\bibitem [{\citenamefont {Vandenbosch}\ \emph {et~al.}(1958)\citenamefont
  {Vandenbosch}, \citenamefont {Thomas}, \citenamefont {Vandenbosch},
  \citenamefont {Glass},\ and\ \citenamefont {Seaborg}}]{PhysRev.111.1358}%
  \BibitemOpen
  \bibfield  {author} {\bibinfo {author} {\bibfnamefont {R.}~\bibnamefont
  {Vandenbosch}}, \bibinfo {author} {\bibfnamefont {T.~D.}\ \bibnamefont
  {Thomas}}, \bibinfo {author} {\bibfnamefont {S.~E.}\ \bibnamefont
  {Vandenbosch}}, \bibinfo {author} {\bibfnamefont {R.~A.}\ \bibnamefont
  {Glass}},\ and\ \bibinfo {author} {\bibfnamefont {G.~T.}\ \bibnamefont
  {Seaborg}},\ }\href {https://doi.org/10.1103/PhysRev.111.1358} {\bibfield
  {journal} {\bibinfo  {journal} {Phys. Rev.}\ }\textbf {\bibinfo {volume}
  {111}},\ \bibinfo {pages} {1358} (\bibinfo {year} {1958})}\BibitemShut
  {NoStop}%
\bibitem [{\citenamefont {Delagrange}\ \emph {et~al.}(1978)\citenamefont
  {Delagrange}, \citenamefont {Fleury},\ and\ \citenamefont
  {Alexander}}]{PhysRevC.17.1706}%
  \BibitemOpen
  \bibfield  {author} {\bibinfo {author} {\bibfnamefont {H.}~\bibnamefont
  {Delagrange}}, \bibinfo {author} {\bibfnamefont {A.}~\bibnamefont {Fleury}},\
  and\ \bibinfo {author} {\bibfnamefont {J.~M.}\ \bibnamefont {Alexander}},\
  }\href {https://doi.org/10.1103/PhysRevC.17.1706} {\bibfield  {journal}
  {\bibinfo  {journal} {Phys. Rev. C}\ }\textbf {\bibinfo {volume} {17}},\
  \bibinfo {pages} {1706} (\bibinfo {year} {1978})}\BibitemShut {NoStop}%
\bibitem [{\citenamefont {Fleury}\ \emph {et~al.}(1973)\citenamefont {Fleury},
  \citenamefont {Ruddy}, \citenamefont {Namboodiri},\ and\ \citenamefont
  {Alexander}}]{fleury1973excitation}%
  \BibitemOpen
  \bibfield  {author} {\bibinfo {author} {\bibfnamefont {A.}~\bibnamefont
  {Fleury}}, \bibinfo {author} {\bibfnamefont {F.}~\bibnamefont {Ruddy}},
  \bibinfo {author} {\bibfnamefont {M.}~\bibnamefont {Namboodiri}},\ and\
  \bibinfo {author} {\bibfnamefont {J.~M.}\ \bibnamefont {Alexander}},\
  }\href@noop {} {\bibfield  {journal} {\bibinfo  {journal} {Physical Review
  C}\ }\textbf {\bibinfo {volume} {7}},\ \bibinfo {pages} {1231} (\bibinfo
  {year} {1973})}\BibitemShut {NoStop}%
\bibitem [{\citenamefont {Nishio}\ \emph {et~al.}(2004)\citenamefont {Nishio},
  \citenamefont {Ikezoe}, \citenamefont {Nagame}, \citenamefont {Asai},
  \citenamefont {Tsukada}, \citenamefont {Mitsuoka}, \citenamefont {Tsuruta},
  \citenamefont {Satou}, \citenamefont {Lin},\ and\ \citenamefont
  {Ohsawa}}]{PhysRevLett.93.162701}%
  \BibitemOpen
  \bibfield  {author} {\bibinfo {author} {\bibfnamefont {K.}~\bibnamefont
  {Nishio}}, \bibinfo {author} {\bibfnamefont {H.}~\bibnamefont {Ikezoe}},
  \bibinfo {author} {\bibfnamefont {Y.}~\bibnamefont {Nagame}}, \bibinfo
  {author} {\bibfnamefont {M.}~\bibnamefont {Asai}}, \bibinfo {author}
  {\bibfnamefont {K.}~\bibnamefont {Tsukada}}, \bibinfo {author} {\bibfnamefont
  {S.}~\bibnamefont {Mitsuoka}}, \bibinfo {author} {\bibfnamefont
  {K.}~\bibnamefont {Tsuruta}}, \bibinfo {author} {\bibfnamefont
  {K.}~\bibnamefont {Satou}}, \bibinfo {author} {\bibfnamefont {C.~J.}\
  \bibnamefont {Lin}},\ and\ \bibinfo {author} {\bibfnamefont {T.}~\bibnamefont
  {Ohsawa}},\ }\href {https://doi.org/10.1103/PhysRevLett.93.162701} {\bibfield
   {journal} {\bibinfo  {journal} {Phys. Rev. Lett.}\ }\textbf {\bibinfo
  {volume} {93}},\ \bibinfo {pages} {162701} (\bibinfo {year}
  {2004})}\BibitemShut {NoStop}%
\bibitem [{\citenamefont {Sikkeland}\ \emph {et~al.}(1968)\citenamefont
  {Sikkeland}, \citenamefont {Ghiorso},\ and\ \citenamefont
  {Nurmia}}]{PhysRev.172.1232}%
  \BibitemOpen
  \bibfield  {author} {\bibinfo {author} {\bibfnamefont {T.}~\bibnamefont
  {Sikkeland}}, \bibinfo {author} {\bibfnamefont {A.}~\bibnamefont {Ghiorso}},\
  and\ \bibinfo {author} {\bibfnamefont {M.~J.}\ \bibnamefont {Nurmia}},\
  }\href {https://doi.org/10.1103/PhysRev.172.1232} {\bibfield  {journal}
  {\bibinfo  {journal} {Phys. Rev.}\ }\textbf {\bibinfo {volume} {172}},\
  \bibinfo {pages} {1232} (\bibinfo {year} {1968})}\BibitemShut {NoStop}%
\bibitem [{\citenamefont {Nagame}\ \emph {et~al.}(2002)\citenamefont {Nagame},
  \citenamefont {Asai}, \citenamefont {Haba}, \citenamefont {Tsukada},
  \citenamefont {Goto}, \citenamefont {Sakama}, \citenamefont {Nishinaka},
  \citenamefont {Toyoshima}, \citenamefont {Akiyama},\ and\ \citenamefont
  {Ichikawa}}]{nagame2002status}%
  \BibitemOpen
  \bibfield  {author} {\bibinfo {author} {\bibfnamefont {Y.}~\bibnamefont
  {Nagame}}, \bibinfo {author} {\bibfnamefont {M.}~\bibnamefont {Asai}},
  \bibinfo {author} {\bibfnamefont {H.}~\bibnamefont {Haba}}, \bibinfo {author}
  {\bibfnamefont {K.}~\bibnamefont {Tsukada}}, \bibinfo {author} {\bibfnamefont
  {S.}~\bibnamefont {Goto}}, \bibinfo {author} {\bibfnamefont {M.}~\bibnamefont
  {Sakama}}, \bibinfo {author} {\bibfnamefont {I.}~\bibnamefont {Nishinaka}},
  \bibinfo {author} {\bibfnamefont {A.}~\bibnamefont {Toyoshima}}, \bibinfo
  {author} {\bibfnamefont {K.}~\bibnamefont {Akiyama}},\ and\ \bibinfo {author}
  {\bibfnamefont {S.}~\bibnamefont {Ichikawa}},\ }\href@noop {} {\bibfield
  {journal} {\bibinfo  {journal} {Journal of Nuclear and Radiochemical
  Sciences}\ }\textbf {\bibinfo {volume} {3}},\ \bibinfo {pages} {129}
  (\bibinfo {year} {2002})}\BibitemShut {NoStop}%
\bibitem [{\citenamefont {Nishio}\ \emph {et~al.}(2006)\citenamefont {Nishio},
  \citenamefont {Hofmann}, \citenamefont {He{\ss}berger}, \citenamefont
  {Ackermann}, \citenamefont {Antalic}, \citenamefont {Comas}, \citenamefont
  {Gan}, \citenamefont {Heinz}, \citenamefont {Heredia}, \citenamefont {Ikezoe}
  \emph {et~al.}}]{nishio2006measurement}%
  \BibitemOpen
  \bibfield  {author} {\bibinfo {author} {\bibfnamefont {K.}~\bibnamefont
  {Nishio}}, \bibinfo {author} {\bibfnamefont {S.}~\bibnamefont {Hofmann}},
  \bibinfo {author} {\bibfnamefont {F.}~\bibnamefont {He{\ss}berger}}, \bibinfo
  {author} {\bibfnamefont {D.}~\bibnamefont {Ackermann}}, \bibinfo {author}
  {\bibfnamefont {S.}~\bibnamefont {Antalic}}, \bibinfo {author} {\bibfnamefont
  {V.}~\bibnamefont {Comas}}, \bibinfo {author} {\bibfnamefont
  {Z.}~\bibnamefont {Gan}}, \bibinfo {author} {\bibfnamefont {S.}~\bibnamefont
  {Heinz}}, \bibinfo {author} {\bibfnamefont {J.}~\bibnamefont {Heredia}},
  \bibinfo {author} {\bibfnamefont {H.}~\bibnamefont {Ikezoe}}, \emph
  {et~al.},\ }\href@noop {} {\bibfield  {journal} {\bibinfo  {journal} {The
  European Physical Journal A-Hadrons and Nuclei}\ }\textbf {\bibinfo {volume}
  {29}},\ \bibinfo {pages} {281} (\bibinfo {year} {2006})}\BibitemShut
  {NoStop}%
\bibitem [{\citenamefont {Nishio}\ \emph {et~al.}(2010)\citenamefont {Nishio},
  \citenamefont {Hofmann}, \citenamefont {He{\ss}berger}, \citenamefont
  {Ackermann}, \citenamefont {Antalic}, \citenamefont {Aritomo}, \citenamefont
  {Comas}, \citenamefont {D{\"u}llmann}, \citenamefont {Gorshkov},
  \citenamefont {Graeger} \emph {et~al.}}]{nishio2010nuclear}%
  \BibitemOpen
  \bibfield  {author} {\bibinfo {author} {\bibfnamefont {K.}~\bibnamefont
  {Nishio}}, \bibinfo {author} {\bibfnamefont {S.}~\bibnamefont {Hofmann}},
  \bibinfo {author} {\bibfnamefont {F.}~\bibnamefont {He{\ss}berger}}, \bibinfo
  {author} {\bibfnamefont {D.}~\bibnamefont {Ackermann}}, \bibinfo {author}
  {\bibfnamefont {S.}~\bibnamefont {Antalic}}, \bibinfo {author} {\bibfnamefont
  {Y.}~\bibnamefont {Aritomo}}, \bibinfo {author} {\bibfnamefont
  {V.}~\bibnamefont {Comas}}, \bibinfo {author} {\bibfnamefont {C.~E.}\
  \bibnamefont {D{\"u}llmann}}, \bibinfo {author} {\bibfnamefont
  {A.}~\bibnamefont {Gorshkov}}, \bibinfo {author} {\bibfnamefont
  {R.}~\bibnamefont {Graeger}}, \emph {et~al.},\ }\href@noop {} {\bibfield
  {journal} {\bibinfo  {journal} {Physical Review C}\ }\textbf {\bibinfo
  {volume} {82}},\ \bibinfo {pages} {024611} (\bibinfo {year}
  {2010})}\BibitemShut {NoStop}%
\bibitem [{\citenamefont {Oganessian}\ \emph
  {et~al.}(2004{\natexlab{a}})\citenamefont {Oganessian}, \citenamefont
  {Utyonkov}, \citenamefont {Lobanov}, \citenamefont {Abdullin}, \citenamefont
  {Polyakov}, \citenamefont {Shirokovsky}, \citenamefont {Tsyganov},
  \citenamefont {Gulbekian}, \citenamefont {Bogomolov}, \citenamefont {Gikal}
  \emph {et~al.}}]{oganessian2004measurements}%
  \BibitemOpen
  \bibfield  {author} {\bibinfo {author} {\bibfnamefont {Y.~T.}\ \bibnamefont
  {Oganessian}}, \bibinfo {author} {\bibfnamefont {V.}~\bibnamefont
  {Utyonkov}}, \bibinfo {author} {\bibfnamefont {Y.~V.}\ \bibnamefont
  {Lobanov}}, \bibinfo {author} {\bibfnamefont {F.~S.}\ \bibnamefont
  {Abdullin}}, \bibinfo {author} {\bibfnamefont {A.}~\bibnamefont {Polyakov}},
  \bibinfo {author} {\bibfnamefont {I.}~\bibnamefont {Shirokovsky}}, \bibinfo
  {author} {\bibfnamefont {Y.~S.}\ \bibnamefont {Tsyganov}}, \bibinfo {author}
  {\bibfnamefont {G.}~\bibnamefont {Gulbekian}}, \bibinfo {author}
  {\bibfnamefont {S.}~\bibnamefont {Bogomolov}}, \bibinfo {author}
  {\bibfnamefont {B.}~\bibnamefont {Gikal}}, \emph {et~al.},\ }\href@noop {}
  {\bibfield  {journal} {\bibinfo  {journal} {Physical Review C}\ }\textbf
  {\bibinfo {volume} {70}},\ \bibinfo {pages} {064609} (\bibinfo {year}
  {2004}{\natexlab{a}})}\BibitemShut {NoStop}%
\bibitem [{\citenamefont {Oganessian}\ \emph
  {et~al.}(2004{\natexlab{b}})\citenamefont {Oganessian}, \citenamefont
  {Utyonkoy}, \citenamefont {Lobanov}, \citenamefont {Abdullin}, \citenamefont
  {Polyakov}, \citenamefont {Shirokovsky}, \citenamefont {Tsyganov},
  \citenamefont {Gulbekian}, \citenamefont {Bogomolov}, \citenamefont
  {Mezentsev} \emph {et~al.}}]{oganessian2004experiments}%
  \BibitemOpen
  \bibfield  {author} {\bibinfo {author} {\bibfnamefont {Y.~T.}\ \bibnamefont
  {Oganessian}}, \bibinfo {author} {\bibfnamefont {V.}~\bibnamefont
  {Utyonkoy}}, \bibinfo {author} {\bibfnamefont {Y.~V.}\ \bibnamefont
  {Lobanov}}, \bibinfo {author} {\bibfnamefont {F.~S.}\ \bibnamefont
  {Abdullin}}, \bibinfo {author} {\bibfnamefont {A.}~\bibnamefont {Polyakov}},
  \bibinfo {author} {\bibfnamefont {I.}~\bibnamefont {Shirokovsky}}, \bibinfo
  {author} {\bibfnamefont {Y.~S.}\ \bibnamefont {Tsyganov}}, \bibinfo {author}
  {\bibfnamefont {G.}~\bibnamefont {Gulbekian}}, \bibinfo {author}
  {\bibfnamefont {S.}~\bibnamefont {Bogomolov}}, \bibinfo {author}
  {\bibfnamefont {A.}~\bibnamefont {Mezentsev}}, \emph {et~al.},\ }\href@noop
  {} {\bibfield  {journal} {\bibinfo  {journal} {Physical Review C}\ }\textbf
  {\bibinfo {volume} {69}},\ \bibinfo {pages} {021601} (\bibinfo {year}
  {2004}{\natexlab{b}})}\BibitemShut {NoStop}%
\bibitem [{\citenamefont {Khuyagbaatar}\ \emph
  {et~al.}(2013{\natexlab{a}})\citenamefont {Khuyagbaatar}, \citenamefont
  {Yakushev}, \citenamefont {D{\"u}llmann}, \citenamefont {Nitsche},
  \citenamefont {Roberto}, \citenamefont {Ackermann}, \citenamefont
  {Andersson}, \citenamefont {Asai}, \citenamefont {Brand}, \citenamefont
  {Block} \emph {et~al.}}]{khuyagbaatar2013superheavy}%
  \BibitemOpen
  \bibfield  {author} {\bibinfo {author} {\bibfnamefont {J.}~\bibnamefont
  {Khuyagbaatar}}, \bibinfo {author} {\bibfnamefont {A.}~\bibnamefont
  {Yakushev}}, \bibinfo {author} {\bibfnamefont {C.~E.}\ \bibnamefont
  {D{\"u}llmann}}, \bibinfo {author} {\bibfnamefont {H.}~\bibnamefont
  {Nitsche}}, \bibinfo {author} {\bibfnamefont {J.}~\bibnamefont {Roberto}},
  \bibinfo {author} {\bibfnamefont {D.}~\bibnamefont {Ackermann}}, \bibinfo
  {author} {\bibfnamefont {L.-L.}\ \bibnamefont {Andersson}}, \bibinfo {author}
  {\bibfnamefont {M.}~\bibnamefont {Asai}}, \bibinfo {author} {\bibfnamefont
  {H.}~\bibnamefont {Brand}}, \bibinfo {author} {\bibfnamefont
  {M.}~\bibnamefont {Block}}, \emph {et~al.},\ }\href@noop {} {\bibfield
  {journal} {\bibinfo  {journal} {GSI Helmholtzzentrum}\ } (\bibinfo {year}
  {2013}{\natexlab{a}})}\BibitemShut {NoStop}%
\bibitem [{\citenamefont {Ts}\ \emph {et~al.}(2009)\citenamefont {Ts} \emph
  {et~al.}}]{ts2009attempt}%
  \BibitemOpen
  \bibfield  {author} {\bibinfo {author} {\bibfnamefont {O.~Y.}\ \bibnamefont
  {Ts}} \emph {et~al.},\ }\href@noop {} {\bibfield  {journal} {\bibinfo
  {journal} {Physical Rev. C}\ }\textbf {\bibinfo {volume} {79}},\ \bibinfo
  {pages} {024603} (\bibinfo {year} {2009})}\BibitemShut {NoStop}%
\bibitem [{\citenamefont {Hofmann}\ \emph {et~al.}(2008)\citenamefont
  {Hofmann}, \citenamefont {Ackermann}, \citenamefont {Antalic}, \citenamefont
  {Comas},\ and\ \citenamefont {Heinz}}]{Hofmann2008report}%
  \BibitemOpen
  \bibfield  {author} {\bibinfo {author} {\bibfnamefont {S.}~\bibnamefont
  {Hofmann}}, \bibinfo {author} {\bibfnamefont {D.}~\bibnamefont {Ackermann}},
  \bibinfo {author} {\bibfnamefont {S.}~\bibnamefont {Antalic}}, \bibinfo
  {author} {\bibfnamefont {V.~f.}\ \bibnamefont {Comas}},\ and\ \bibinfo
  {author} {\bibfnamefont {S.}~\bibnamefont {Heinz}},\ }\href@noop {}
  {\bibfield  {journal} {\bibinfo  {journal} {GSI Scientific Report}\ ,\
  \bibinfo {pages} {131}} (\bibinfo {year} {2008})}\BibitemShut {NoStop}%
\bibitem [{\citenamefont {Morita}\ \emph {et~al.}(2015)\citenamefont {Morita},
  \citenamefont {Morimoto}, \citenamefont {Kaji}, \citenamefont {Haba} \emph
  {et~al.}}]{Morita2015report}%
  \BibitemOpen
  \bibfield  {author} {\bibinfo {author} {\bibfnamefont {K.}~\bibnamefont
  {Morita}}, \bibinfo {author} {\bibfnamefont {K.}~\bibnamefont {Morimoto}},
  \bibinfo {author} {\bibfnamefont {D.}~\bibnamefont {Kaji}}, \bibinfo {author}
  {\bibfnamefont {H.}~\bibnamefont {Haba}}, \emph {et~al.},\ }\href@noop {}
  {\bibfield  {journal} {\bibinfo  {journal} {RIKEN Accelerator Progress
  Report}\ } (\bibinfo {year} {2015})}\BibitemShut {NoStop}%
\bibitem [{\citenamefont {Hoffman}\ \emph {et~al.}(2000)\citenamefont
  {Hoffman}, \citenamefont {Ghiorso},\ and\ \citenamefont
  {Seaborg}}]{hoffman2000transuranium}%
  \BibitemOpen
  \bibfield  {author} {\bibinfo {author} {\bibfnamefont {D.~C.}\ \bibnamefont
  {Hoffman}}, \bibinfo {author} {\bibfnamefont {A.}~\bibnamefont {Ghiorso}},\
  and\ \bibinfo {author} {\bibfnamefont {G.~T.}\ \bibnamefont {Seaborg}},\
  }\href@noop {} {\emph {\bibinfo {title} {The transuranium people: the inside
  story}}},\ \bibinfo {number} {doi. 10.1142/p074}\ (\bibinfo  {publisher}
  {World Scientific},\ \bibinfo {year} {2000})\BibitemShut {NoStop}%
\bibitem [{\citenamefont {Dmitriev}\ \emph {et~al.}(2016)\citenamefont
  {Dmitriev}, \citenamefont {Itkis},\ and\ \citenamefont
  {Oganessian}}]{dmitriev2016status}%
  \BibitemOpen
  \bibfield  {author} {\bibinfo {author} {\bibfnamefont {S.}~\bibnamefont
  {Dmitriev}}, \bibinfo {author} {\bibfnamefont {M.}~\bibnamefont {Itkis}},\
  and\ \bibinfo {author} {\bibfnamefont {Y.}~\bibnamefont {Oganessian}},\ }in\
  \href@noop {} {\emph {\bibinfo {booktitle} {EPJ Web of Conferences}}},\ Vol.\
  \bibinfo {volume} {131}\ (\bibinfo {organization} {EDP Sciences},\ \bibinfo
  {year} {2016})\ p.\ \bibinfo {pages} {08001}\BibitemShut {NoStop}%
\bibitem [{\citenamefont {Nandi}\ \emph {et~al.}(2021)\citenamefont {Nandi},
  \citenamefont {Swami}, \citenamefont {Gupta}, \citenamefont {Kumar},
  \citenamefont {Chakraborty},\ and\ \citenamefont
  {Manjunatha}}]{nandi2021search}%
  \BibitemOpen
  \bibfield  {author} {\bibinfo {author} {\bibfnamefont {T.}~\bibnamefont
  {Nandi}}, \bibinfo {author} {\bibfnamefont {D.}~\bibnamefont {Swami}},
  \bibinfo {author} {\bibfnamefont {P.}~\bibnamefont {Gupta}}, \bibinfo
  {author} {\bibfnamefont {Y.}~\bibnamefont {Kumar}}, \bibinfo {author}
  {\bibfnamefont {S.}~\bibnamefont {Chakraborty}},\ and\ \bibinfo {author}
  {\bibfnamefont {H.}~\bibnamefont {Manjunatha}},\ }\href@noop {} {\bibfield
  {journal} {\bibinfo  {journal} {arXiv preprint arXiv:2103.06729}\ } (\bibinfo
  {year} {2021})}\BibitemShut {NoStop}%
\bibitem [{\citenamefont {Rauscher}\ \emph {et~al.}(1997)\citenamefont
  {Rauscher}, \citenamefont {Thielemann},\ and\ \citenamefont
  {Kratz}}]{rauscher1997nuclear}%
  \BibitemOpen
  \bibfield  {author} {\bibinfo {author} {\bibfnamefont {T.}~\bibnamefont
  {Rauscher}}, \bibinfo {author} {\bibfnamefont {F.-K.}\ \bibnamefont
  {Thielemann}},\ and\ \bibinfo {author} {\bibfnamefont {K.-L.}\ \bibnamefont
  {Kratz}},\ }\href@noop {} {\bibfield  {journal} {\bibinfo  {journal}
  {Physical Review C}\ }\textbf {\bibinfo {volume} {56}},\ \bibinfo {pages}
  {1613} (\bibinfo {year} {1997})}\BibitemShut {NoStop}%
\bibitem [{\citenamefont {M{\"o}ller}\ \emph {et~al.}(1993)\citenamefont
  {M{\"o}ller}, \citenamefont {Nix}, \citenamefont {Myers},\ and\ \citenamefont
  {Swiatecki}}]{moller1993nuclear}%
  \BibitemOpen
  \bibfield  {author} {\bibinfo {author} {\bibfnamefont {P.}~\bibnamefont
  {M{\"o}ller}}, \bibinfo {author} {\bibfnamefont {J.}~\bibnamefont {Nix}},
  \bibinfo {author} {\bibfnamefont {W.}~\bibnamefont {Myers}},\ and\ \bibinfo
  {author} {\bibfnamefont {W.}~\bibnamefont {Swiatecki}},\ }\href@noop {}
  {\bibfield  {journal} {\bibinfo  {journal} {At. Data and Nucl. Data Tables}\
  }\textbf {\bibinfo {volume} {59}},\ \bibinfo {pages} {185} (\bibinfo {year}
  {1993})}\BibitemShut {NoStop}%
\bibitem [{\citenamefont {Capurro}\ \emph {et~al.}(1997)\citenamefont
  {Capurro}, \citenamefont {DiGregorio}, \citenamefont {Gil}, \citenamefont
  {Abriola}, \citenamefont {Di~Tada}, \citenamefont {Niello}, \citenamefont
  {Macchiavelli}, \citenamefont {Mart{\'\i}}, \citenamefont {Pacheco},
  \citenamefont {Testoni} \emph {et~al.}}]{capurro1997average}%
  \BibitemOpen
  \bibfield  {author} {\bibinfo {author} {\bibfnamefont {O.}~\bibnamefont
  {Capurro}}, \bibinfo {author} {\bibfnamefont {D.~E.}\ \bibnamefont
  {DiGregorio}}, \bibinfo {author} {\bibfnamefont {S.}~\bibnamefont {Gil}},
  \bibinfo {author} {\bibfnamefont {D.}~\bibnamefont {Abriola}}, \bibinfo
  {author} {\bibfnamefont {M.}~\bibnamefont {Di~Tada}}, \bibinfo {author}
  {\bibfnamefont {J.~F.}\ \bibnamefont {Niello}}, \bibinfo {author}
  {\bibfnamefont {A.}~\bibnamefont {Macchiavelli}}, \bibinfo {author}
  {\bibfnamefont {G.}~\bibnamefont {Mart{\'\i}}}, \bibinfo {author}
  {\bibfnamefont {A.}~\bibnamefont {Pacheco}}, \bibinfo {author} {\bibfnamefont
  {J.}~\bibnamefont {Testoni}}, \emph {et~al.},\ }\href@noop {} {\bibfield
  {journal} {\bibinfo  {journal} {Physical Review C}\ }\textbf {\bibinfo
  {volume} {55}},\ \bibinfo {pages} {766} (\bibinfo {year} {1997})}\BibitemShut
  {NoStop}%
\bibitem [{\citenamefont {Manjunatha}\ \emph {et~al.}(2021)\citenamefont
  {Manjunatha}, \citenamefont {Sowmya}, \citenamefont {Damodara~Gupta},
  \citenamefont {Seenappa},\ and\ \citenamefont {Nandi}}]{Manjunath2021PLB}%
  \BibitemOpen
  \bibfield  {author} {\bibinfo {author} {\bibfnamefont {H.~C.}\ \bibnamefont
  {Manjunatha}}, \bibinfo {author} {\bibfnamefont {N.}~\bibnamefont {Sowmya}},
  \bibinfo {author} {\bibfnamefont {P.~S.}\ \bibnamefont {Damodara~Gupta}},
  \bibinfo {author} {\bibfnamefont {L.}~\bibnamefont {Seenappa}},\ and\
  \bibinfo {author} {\bibfnamefont {T.}~\bibnamefont {Nandi}},\ }\href@noop {}
  {\ \textbf {\bibinfo {volume} {arXiv:2103.07177}} (\bibinfo {year}
  {2021})}\BibitemShut {NoStop}%
\bibitem [{\citenamefont {Tanaka}\ \emph {et~al.}(2020)\citenamefont {Tanaka},
  \citenamefont {Morita}, \citenamefont {Morimoto}, \citenamefont {Kaji},
  \citenamefont {Haba}, \citenamefont {Boll}, \citenamefont {Brewer},
  \citenamefont {Van~Cleve}, \citenamefont {Dean}, \citenamefont {Ishizawa}
  \emph {et~al.}}]{tanaka2020study}%
  \BibitemOpen
  \bibfield  {author} {\bibinfo {author} {\bibfnamefont {T.}~\bibnamefont
  {Tanaka}}, \bibinfo {author} {\bibfnamefont {K.}~\bibnamefont {Morita}},
  \bibinfo {author} {\bibfnamefont {K.}~\bibnamefont {Morimoto}}, \bibinfo
  {author} {\bibfnamefont {D.}~\bibnamefont {Kaji}}, \bibinfo {author}
  {\bibfnamefont {H.}~\bibnamefont {Haba}}, \bibinfo {author} {\bibfnamefont
  {R.~A.}\ \bibnamefont {Boll}}, \bibinfo {author} {\bibfnamefont {N.~T.}\
  \bibnamefont {Brewer}}, \bibinfo {author} {\bibfnamefont {S.}~\bibnamefont
  {Van~Cleve}}, \bibinfo {author} {\bibfnamefont {D.~J.}\ \bibnamefont {Dean}},
  \bibinfo {author} {\bibfnamefont {S.}~\bibnamefont {Ishizawa}}, \emph
  {et~al.},\ }\href@noop {} {\bibfield  {journal} {\bibinfo  {journal}
  {Physical Review Letters}\ }\textbf {\bibinfo {volume} {124}},\ \bibinfo
  {pages} {052502} (\bibinfo {year} {2020})}\BibitemShut {NoStop}%
\bibitem [{\citenamefont {Khuyagbaatar}\ \emph
  {et~al.}(2013{\natexlab{b}})\citenamefont {Khuyagbaatar}, \citenamefont
  {Yakushev}, \citenamefont {D¨ullmann}, \citenamefont {Nitsche},
  \citenamefont {Roberto}, \citenamefont {Ackermann}, \citenamefont {Andersson}
  \emph {et~al.}}]{Khuyagbaatar2013report}%
  \BibitemOpen
  \bibfield  {author} {\bibinfo {author} {\bibfnamefont {J.}~\bibnamefont
  {Khuyagbaatar}}, \bibinfo {author} {\bibfnamefont {A.}~\bibnamefont
  {Yakushev}}, \bibinfo {author} {\bibfnamefont {E.}~\bibnamefont
  {D¨ullmann}}, \bibinfo {author} {\bibfnamefont {H.}~\bibnamefont {Nitsche}},
  \bibinfo {author} {\bibfnamefont {J.}~\bibnamefont {Roberto}}, \bibinfo
  {author} {\bibfnamefont {D.}~\bibnamefont {Ackermann}}, \bibinfo {author}
  {\bibfnamefont {L.~L.}\ \bibnamefont {Andersson}}, \emph {et~al.},\
  }\href@noop {} {\bibfield  {journal} {\bibinfo  {journal} {GSI Scientific
  Report}\ ,\ \bibinfo {pages} {131}} (\bibinfo {year}
  {2013}{\natexlab{b}})}\BibitemShut {NoStop}%
\bibitem [{\citenamefont {Khuyagbaatar}\ \emph {et~al.}(2020)\citenamefont
  {Khuyagbaatar}, \citenamefont {Yakushev}, \citenamefont {D{\"u}llmann},
  \citenamefont {Ackermann}, \citenamefont {Andersson}, \citenamefont {Asai},
  \citenamefont {Block}, \citenamefont {Boll}, \citenamefont {Brand},
  \citenamefont {Cox} \emph {et~al.}}]{khuyagbaatar2020search}%
  \BibitemOpen
  \bibfield  {author} {\bibinfo {author} {\bibfnamefont {J.}~\bibnamefont
  {Khuyagbaatar}}, \bibinfo {author} {\bibfnamefont {A.}~\bibnamefont
  {Yakushev}}, \bibinfo {author} {\bibfnamefont {C.~E.}\ \bibnamefont
  {D{\"u}llmann}}, \bibinfo {author} {\bibfnamefont {D.}~\bibnamefont
  {Ackermann}}, \bibinfo {author} {\bibfnamefont {L.-L.}\ \bibnamefont
  {Andersson}}, \bibinfo {author} {\bibfnamefont {M.}~\bibnamefont {Asai}},
  \bibinfo {author} {\bibfnamefont {M.}~\bibnamefont {Block}}, \bibinfo
  {author} {\bibfnamefont {R.}~\bibnamefont {Boll}}, \bibinfo {author}
  {\bibfnamefont {H.}~\bibnamefont {Brand}}, \bibinfo {author} {\bibfnamefont
  {D.}~\bibnamefont {Cox}}, \emph {et~al.},\ }\href@noop {} {\bibfield
  {journal} {\bibinfo  {journal} {Physical Review C}\ }\textbf {\bibinfo
  {volume} {102}},\ \bibinfo {pages} {064602} (\bibinfo {year}
  {2020})}\BibitemShut {NoStop}%
\end{thebibliography}%
\end{document}